\documentclass[11pt]{article}
\pdfmapfile{+bickham.map}
\pdfoutput=1
\parskip \medskipamount

\usepackage{natbib}
\usepackage{har2nat}
\setcitestyle{authoryear,open={(},close={)}}

\usepackage[a4paper, left=1in, right=1in, top=1.8cm, bottom=1.8cm, includehead, includefoot]{geometry} 	
\usepackage{amsfonts}
\usepackage{array}
\usepackage{graphicx} 									
\usepackage[small,bf,hang]{caption}			
\usepackage[labelformat=simple]{subcaption}									

	\usepackage{wrapfig}										
	\usepackage{multirow}
	\usepackage[table]{xcolor}										
	\definecolor{lightblue}{rgb}{0.13, 0.67, 0.8}
	\definecolor{tableblue}{rgb}{250, 250, 255}
	
	\usepackage[latin1]{inputenc}						
	\usepackage{amsmath}										
	\allowdisplaybreaks[1]									
	\usepackage{amssymb}										
	\usepackage{amsthm}	
	\usepackage{leftidx}										
	\usepackage{booktabs}										
	\usepackage{url}                        
	\usepackage{eurosym}                    
	\usepackage{textcomp}                   
	\usepackage{fancyhdr}                   
	\usepackage{listings} 									
	\usepackage{pdfpages}										
	\usepackage{floatrow}
	\usepackage[title,titletoc,toc]{appendix}				
		\usepackage{enumerate}									
		\usepackage[multiple]{footmisc}																		
		\usepackage{enumitem}										
		\usepackage{mathtools}										
		\usepackage{IEEEtrantools}							
		\usepackage{authblk}											
		\usepackage{tikz}
		\usetikzlibrary{snakes}
		\usetikzlibrary{shapes,arrows,decorations.pathmorphing,backgrounds,positioning,fit,petri, matrix, arrows.meta}
		\usepackage[english]{babel}
		\usepackage{grffile}
		\usepackage{amsfonts}
		\usepackage{amssymb}
		\usepackage{relsize}
		\usepackage{rotating}
		\usepackage{siunitx}
		\usepackage{color}	
		\usepackage{etoolbox}
		\usepackage{todonotes}
		\usepackage{longtable}
		\usepackage{spbmark}
		\usepackage[plainpages=false,hyperfootnotes=false]{hyperref}
		\hypersetup{
			colorlinks   = true, 
			urlcolor     = blue, 
			linkcolor    = red, 
			citecolor   = blue 
		}
		\usepackage{kbordermatrix}
		\usepackage[capitalize]{cleveref}
		
		\usepackage{bookmark}
		\usepackage[ruled,vlined,linesnumbered]{algorithm2e}
		\SetAlCapFnt{\small} 
		\SetKw{KwTo}{in}

		\SetKwInput{kwModel}{Model}
		\SetKwInput{kwInit}{Initialization}

		\usetikzlibrary{tikzmark,fit}
		\usetikzlibrary{decorations.pathreplacing,calc}

		\makeatletter
		
		\pretocmd{\NAT@citex}{%
			\let\NAT@hyper@\NAT@hyper@citex
			\def\NAT@postnote{#2}%
			\setcounter{NAT@total@cites}{0}%
			\setcounter{NAT@count@cites}{0}%
			\forcsvlist{\stepcounter{NAT@total@cites}\@gobble}{#3}}{}{}
		\newcounter{NAT@total@cites}
		\newcounter{NAT@count@cites}
		\def\NAT@postnote{}
		
		\def\NAT@hyper@citex#1{%
			\stepcounter{NAT@count@cites}%
			\hyper@natlinkstart{\@citeb\@extra@b@citeb}#1%
			\ifnumequal{\value{NAT@count@cites}}{\value{NAT@total@cites}}
			{\ifNAT@swa\else\if*\NAT@postnote*\else%
				\NAT@cmt\NAT@postnote\global\def\NAT@postnote{}\fi\fi}{}%
			\ifNAT@swa\else\if\relax\NAT@date\relax
			\else\NAT@@close\global\let\NAT@nm\@empty\fi\fi
			\hyper@natlinkend}
		\renewcommand\hyper@natlinkbreak[2]{#1}
		
		\patchcmd{\NAT@citex}
		{\ifNAT@swa\else\if*#2*\else\NAT@cmt#2\fi
			\if\relax\NAT@date\relax\else\NAT@@close\fi\fi}{}{}{}
		\patchcmd{\NAT@citex}
		{\if\relax\NAT@date\relax\NAT@def@citea\else\NAT@def@citea@close\fi}
		{\if\relax\NAT@date\relax\NAT@def@citea\else\NAT@def@citea@space\fi}{}{}
		
		\makeatother
		
		\definecolor{darkgreen}{RGB}{40,150,40}
		
		\DeclarePairedDelimiter{\ev}{\langle}{\rangle}

		\newcommand{\bs}{\boldsymbol}
		\newcommand{\mc}{\mathcal}

		\renewcommand{\leq}{\leqslant}					
		\renewcommand{\geq}{\geqslant}					
		\renewcommand{\epsilon}{\varepsilon}

		\theoremstyle{definition}

		\DeclareMathOperator*{\argminA}{arg\,min} 
		
		\usepackage[scr=bickham, scrscaled=0.95]{mathalpha}
		\makeatletter
		\newcommand{\raisemath}[1]{\mathpalette{\raisem@th{#1}}}
		\newcommand{\raisem@th}[3]{\raisebox{#1}{$#2#3$}}
		\makeatother
		\newcommand{\mb}[1]{\raisemath{1.5pt}{\hspace{-0.75mm}\mathscr{#1}}}
		
		\newcommand{\ignore}[1]{}

		\newcolumntype{K}{>{\centering\arraybackslash}m{1.5cm}}
		\newcolumntype{M}{>{\centering\arraybackslash}m{3.5cm}}
		\newcolumntype{B}{>{\raggedright\arraybackslash}m{5cm}}
		\newcolumntype{W}{>{\raggedright\arraybackslash}p{12cm}}
		\newcolumntype{X}{>{\raggedright\arraybackslash}p{8cm}}
		\newcolumntype{N}{>{\raggedright\arraybackslash}p{9.0cm}}
		\newcolumntype{C}[1]{>{\centering\let\newline\\\arraybackslash\hspace{0pt}}m{#1}}
		\newcolumntype{D}{>{\raggedright\arraybackslash}m{1cm}}
		\newcolumntype{E}{>{\raggedright\arraybackslash}m{1.5cm}}
		\newcolumntype{H}{>{\setbox0=\hbox\bgroup}c<{\egroup}@{}}
		\newcolumntype{I}{>{\centering\arraybackslash}m{2.1cm}}
		\newcolumntype{P}[1]{>{\centering\arraybackslash}p{#1}}
		\newcolumntype{L}[1]{>{\raggedright\arraybackslash}p{#1}}
		
		\newcolumntype{A}{
			>{$}r<{$}
			@{\extracolsep{0pt}}
			>{${}} l <{$}
			@{\extracolsep{\fill}}
		}
		
		\usepackage{threeparttable}
		\usepackage{tabularx}
		\newfloatcommand{capbtabbox}{table}[][\FBwidth]
		\newfloatcommand{ImportTable}{input}[][\FBwidth]
		\newfloatcommand{capthreepart}{threeparttable}[][\FBwidth]
		\DeclareFloatFont{Small}{\fontsize{9}{11}\selectfont}
		\DeclareFloatFont{Small2}{\small}
		\DeclareFloatFont{footnote}{\footnotesize}
		\DeclareFloatFont{scriptsize}{\scriptsize}
		\DeclareMarginSet{hangboth}{\setfloatmargins*{\hskip-4cm}{\hskip-4cm}}
		
		\DeclareNewFloatType{figurea}
		{placement=htbp, name=Figure, within=section, fileext=lofa}
		
		\newfloatcommand{ffigabox}{figurea}[\nocapbeside][]
		\makeatletter
		\def\thickhline{%
			\noalign{\ifnum0=`}\fi\hrule \@height \thickarrayrulewidth \futurelet
			\reserved@a\@xthickhline}
		\def\@xthickhline{\ifx\reserved@a\thickhline
			\vskip\doublerulesep
			\vskip-\thickarrayrulewidth
			\fi
			\ifnum0=`{\fi}}
		\makeatother
		
		\newlength{\thickarrayrulewidth}
		\hypersetup{
			colorlinks = true,
			citecolor  = blue,
			urlcolor   = red,
			breaklinks = true
		}

		\newcolumntype{O}{>{\centering\arraybackslash}m{1.5cm}}
		\renewcommand*{\thesection}{\arabic{section}}
		\let\origappendix\appendix 
		\renewcommand\appendix{\clearpage\pagenumbering{roman}\origappendix}

		\usepackage{subcaption}
		\usepackage{tablefootnote}

		\usepackage{nicematrix}
		
		\graphicspath{{Figures/}}
		
		\author[1]{Bavo D.C. Campo}
		\author[1,2,3,4]{Katrien Antonio}
		\affil[1]{Faculty of Economics and Business, KU Leuven, Belgium.}
		\affil[2]{Faculty of Economics and Business, University of Amsterdam, The Netherlands.}
		\affil[3]{LRisk, Leuven Research Center on Insurance and Financial Risk Analysis, KU Leuven, Belgium.}
		\affil[4]{LStat, Leuven Statistics Research Center, KU Leuven, Belgium.}
		\title{\textbf{On clustering levels of a hierarchical categorical risk factor}}
		\date{}
		
		\allowdisplaybreaks

		\newcommand\Tstrut{\rule{0pt}{3.2ex}}         
		\newcommand\Cstrut{\rule[-1.25ex]{0pt}{0pt}}   
		
		\usepackage[author={Bavo Campo}]{pdfcomment}
		
		\usepackage{afterpage}
		\usepackage{pdflscape}
		
		\newcommand{\killpunct}[1]{}
		
		\usepackage[final]{changes}
		\definechangesauthor[name={Bavo D.C. Campo}, color=red]{B}
		\definecolor{red}{rgb}{0, 0, 0}
		
\begin{document}
			\newcommand{\form}[1]{\scalebox{1.087}{\boldmath{#1}}}
			
			\sloppy
			\maketitle
			\begin{abstract}
				\noindent
				Handling nominal covariates with a large number of categories is challenging for both statistical and machine learning techniques. This problem is further exacerbated when the nominal variable has a hierarchical structure. We commonly rely on methods such as the random effects approach \citep{Campo2023} to incorporate these covariates in a predictive model. Nonetheless, in certain situations, even the random effects approach may encounter estimation problems. We propose the data-driven Partitioning Hierarchical Risk-factors Adaptive Top-down (PHiRAT) algorithm to reduce the hierarchically structured risk factor to its essence, by grouping similar categories at each level of the hierarchy. We work top-down and engineer several features to characterize the profile of the categories at a specific level in the hierarchy. In our workers' compensation case study, we characterize the risk profile of an industry via its observed damage rates and claim frequencies. In addition, we use embeddings \citep{Mikolov, USE} to encode the textual description of the economic activity of the insured company. These features are then used as input in a clustering algorithm to group similar categories. Our method substantially reduces the number of categories and results in a grouping that is generalizable to out-of-sample data. Moreover, we obtain a better differentiation between high-risk and low-risk companies.
				
				\vspace{2mm}
			\end{abstract}
			
			\section{Introduction}
At the heart of a risk-based insurance pricing model is a set of risk factors that are predictive of the loss cost. To model the relation between the risk factors and the loss cost, actuaries rely on statistical and machine learning techniques. Both approaches are able to handle different types of risk factors (i.e. nominal, ordinal, geographical or continuous). In this contribution we put focus on challenges imposed by nominal variables with a hierarchical structure. Such variables may cause estimation problems, due to an exceedingly large number of categories and a limited number of observations for some of the categories. Using default methods to handle these, such as dummy encoding, may result in unreliable parameter estimates in generalized linear models (GLMs) and may cause machine learning methods to become computationally intractable. We refer to this type of risk factor as a hierarchical multi-level factor (MLF) \citep{Ohlsson} or a hierarchical high-cardinality attribute \citep{Micci2001,Pargent2021}. \added[id=B]{Examples of such a nominal variable include provinces and municipalities within provinces, or vehicle brands and models within brands.} Within workers' compensation insurance, a typical example \deleted[id=B]{hereof} is the hierarchical MLF derived from the numerical codes of the NACE system. The NACE system is a hierarchical classification system used in the European Union to group similar companies based on their economic activity \citep{NACE}. A similar example is the Australian and New Zealand Standard Industrial Classification (ANZSIC) system \citep{ANZSIC}, which is closely related to the NACE system \citep{NACE}.

In predictive modelling, such risk factors are potentially a great source of information. In workers' compensation insurance, certain industries (e.g., manufacturing, construction) and occupations (e.g., labouring, roofer) are associated with an increased risk of filing claims \citep{Walters2010, Holizki2008, Wurzel2021}. Furthermore, companies operating in the same industry are exposed to similar risks. This creates a dependency among companies active in the same industry and heterogeneity between companies working in different industries \citep{Campo2023}. Industry classification systems, such as the NACE and ANSZIC, allow to group companies based on their economic activity at varying levels of granularity. Most industry classification systems are hierarchical classifications that work top-down. The classification of a company starts at the highest level in the hierarchy and, from here, proceeds successively to lower levels in the hierarchy. At the top level of the hierarchy, the categories are broad and general, covering a wide range of economic activities. As we move down the hierarchy, the categories are broken down into increasingly specific subcategories that encompass more detailed economic activities. Further, industry classification systems typically provide a textual description for the categories at all levels in the hierarchy. This description explains why companies are grouped in the same category and can be used to judge the similarity of activities among categories.

To incorporate the hierarchical MLF in a predictive model, we can opt for the hierarchical random effects approach \citep{Campo2023}. Here, we specify a random effect at each level in the hierarchy. The random effects capture the unobservable characteristics of the categories at the different levels in the hierarchy. Moreover, random effects models account for the within-category dependency and between-category heterogeneity. To estimate a random effects model, we can either use the hierarchical credibility model \citep{JewellModel}, Ohlsson's combination of the hierarchical credibility model with a GLM \citep{Ohlsson2008} or the mixed models framework \citep{Molenberghs2005}. These estimation procedures rely on the estimation of variance parameters and we require these estimates to be non-negative \citep{Molenberghs2011, Oliveira2017}. In some cases, however, we obtain negative variance estimates and this can occur when there is low variability \citep{Oliveira2017} or when the hierarchical structure of the MLF is misspecified \citep{Pryseley2011}. In these situations, the estimation procedure yields nonsensical results. With the random effects approach we implicitly assume that the risk profiles differ between the different categories \citep{Tutz2017}. However, it is not an unreasonable assumption that certain categories have an identical effect on the response and that these should be grouped into homogeneous clusters. Decreasing the total number of categories leads to sparser models that are easier to interpret and less likely to experience estimation problems or to overfit. Additionally, individual categories will have more observations, leading to more precise estimates of their effect on the response.

To group data into homogeneous clusters, we typically rely on clustering techniques. These techniques partition the data points into clusters such that observations within the same cluster are more similar compared to observations belonging to other clusters \citep[Chapter~14]{Hastie2009}. Within actuarial sciences, clustering methods recently appeared in a variety of applications. In motor insurance, for example, clustering algorithms are employed to group driving styles of policyholders \citep{Wutrich2017, Zhu2021}, to construct tariff classes in an unsupervised way \citep{Yeo2001, Wang2008} and to bin continuous or spatial risk factors \citep{Henckaerts2018}. Also in health insurance we find several examples. \citet{Rosenberg2022} used clustering techniques to identify high-cost health care utilizers who are responsible for a substantial amount of health expenditures.

Our aim is to use clustering algorithms in workers' compensation insurance pricing, to group categories of the hierarchical MLF that are similar in riskiness and economic activity. To characterize the riskiness, we rely on risk statistics such as the average damage rate and the expected claim frequency. Moreover, we also use the textual description of the categories to obtain information on the economic activity. The risk statistics are expressed as numerical values, whereas the textual descriptions are presented as nominal categories. Both types of features are then used as input in a clustering algorithm to group similar categories of the hierarchical MLF. To create clusters, most algorithms rely on distance or (dis)similarity metrics to quantify the degree of relatedness between observations in the feature space \citep{Hastie2009, Foss2019}. These metrics, however, are different for numeric and nominal features which makes it challenging to cluster mixed-type data \citep{Cheung2013,Foss2019,Ahmad2019}. One approach to tackle this problem is to convert the nominal to numeric features. Hereto, we commonly employ dummy encoding \citep{Hsu2006, Cheung2013, Ahmad2019,Foss2019}. This encoding creates binary variables that represent category membership. Hereby, it results in a loss of information when the categories have textual labels. Labels provide meaning to categories and reflect the degree of similarity between different categories. A more suitable encoding is obtained using embeddings, a technique developed within natural language processing (NLP). Embeddings are vector representations of textual data that capture the semantic information \citep{Verma2021,Schomacker2021,Ferrario2020}. In the actuarial literature, \citet{Lee2020} show how embeddings can be used to incorporate textual data into insurance claims modelling. \citet{Xu2022} create embedding based risk factors that are used as features in a claim severity model. \citet{Zappa2021} used embeddings to create risk factors that predict the severity of injuries in road accidents. 

Research on grouping categories of hierarchical MLFs is limited. Most of the research is focused on nominal variables that do not have a hierarchical structure. For example, the Generalized Fused Lasso (GFL) \citep{GFL,Gertheiss2010,Oelker2014} groups MLF categories within a regularized regression framework. Here, categories are merged when there is a small difference between the regression coefficients. Nonetheless, the GFL does not scale well to high-cardinality features since the number of estimated coefficient differences grows exponentially with the number of categories. Another example to fuse non-hierarchically structured nominal variables is the method of \citet{Carrizosa2021}. Here, the authors first specify the order of the categories and the number of clusters. Next, to create the prespecified number of clusters they group consecutive categories. For a specific number of clusters, multiple solutions exist and the solution with the highest out-of-sample accuracy is preferred. The disadvantages of this approach are three-fold. First, it only merges neighbouring categories. Second, there is no procedure to select the optimal number of groups and third, the procedure can not immediately be applied to hierarchical MLFs. To the best of our knowledge, the method described in \citet{Carrizosa2022} is the only approach that puts focus on reducing the number of categories of a hierarchical MLF. Here, the authors propose a bottom-up clustering strategy. This technique begins by considering the categories at the lowest level in the hierarchy. Hereafter, starting from the categories at the lowest level in the hierarchy, they consecutively merge categories at the lowest level into broader categories at higher levels in the hierarchy. As such, categories that are nested within the same category at a higher level in the hierarchy are grouped. Notwithstanding, their proposed optimization strategy is only suited for linear regression models. Further, it is not suitable when we want to maintain the levels in the hierarchical structure. Certain granular categories are replaced the broader categories they are nested in. Hence, it is possible that the optimal solution produces a hierarchical structure with a different depth in its representation.

Our paper contributes to the existing literature in the following ways. Firstly, we present a data-driven approach to reduce an existing granular hierarchical structure to its essence, by grouping similar categories at every level in the hierarchy. We devise a top-down procedure, where we start at the top level, to preserve the hierarchical structure. At a specific level in the hierarchy, we engineer several features to characterize the risk profile of each category. In a case-study with a workers' compensation insurance product, we use \replaced[id = B]{predicted}{estimated} random effects obtained with a generalized linear mixed model for damage rates on the one hand and claim frequencies on the other hand. To extract the textual information contained in the category description, we use embeddings. Next, we use these features as input in a clustering algorithm to group similar categories into clusters. Hereafter, we proceed to grouping the categories at the next hierarchical level. The procedure stops once we grouped the categories at the lowest level in the hierarchy. Secondly, we provide a concise overview of important aspects and algorithms in cluster analysis. Furthermore, we demonstrate that the clustering algorithm and evaluation criterion affect the clustering solution. Thirdly, we show that embeddings can be used to group similar categories of a nominal variable. Contrary to  \citet{Lee2020,Zappa2021,Xu2022}, we do not employ embeddings to create new risk factors in a pricing model. Instead, we use embeddings to extract the textual information of category labels and to cluster categories based on their semantic similarities.

The remainder of this paper is structured as follows. In Section \ref{sec:MotivatingExample}, we use a workers' compensation insurance product as a motivating example with the NACE code as hierarchical MLF. We illustrate the structure of this type of data set, which information is typically available and we explain how to engineer features that characterize the risk profile of the categories. In Section \ref{sec:Clustering}, we define a top-down procedure to cluster similar categories at a specific level in the hierarchy and we discuss several aspects of clustering techniques. The results of applying our procedure to reduce the hierarchical structure to its essence are discussed in Section \ref{sec:CaseStudy}. Moreover, in this section we also compare the use of the original and the reduced structure in a technical pricing model for workers' compensation insurance. Section \ref{sec:Discussion} concludes the article.

			\section{Feature engineering for industrial activities in a workers' compensation insurance product}\label{sec:MotivatingExample}
To illustrate the importance of hierarchical MLFs in insurance pricing, a workers' compensation insurance product is a particularly suitable example. This insurance product compensates employees for lost wages and medical expenses resulting from job-related injury (see \citet{Campo2023} for more information). In this type of insurance, we generally work with an industrial classification system to group companies based on their economic activity. Hereto, we demonstrate and discuss the NACE classification system in Section \ref{subsec:NACE}. We explain the typical structure of a workers' compensation insurance data set and discuss which information is available. In Section \ref{subsec:featureengineering} we show how we use this information to engineer features that characterize the riskiness and economic activity of categories at a specific level in the hierarchy.

\subsection{A hierarchical classification scheme for industrial activities}\label{subsec:NACE}
In a workers' compensation insurance data set, it is common to work with an industrial classification system. Within the European Union, a wide range of organizations (e.g., national statistical institutes, business and trade associations, insurance companies and European national central banks \citep{NACE,Reacfin, ECB}) work with the NACE system \citep{NACE}. This is a hierarchical classification system to group companies based on their economic activity. Each company is assigned a four-digit numerical code, which is used to identify the categories at different levels in the hierarchy. The NACE system works top-down, starts at the highest level in the hierarchy and then proceeds to the lower levels. In our paper, we work with NACE Rev.~1 \citep{NACErev1}, which has five hierarchical levels (in descending order): \texttt{section}, \texttt{subsection}, \texttt{division}, \texttt{group} and \texttt{class}. Most member states of the EU have a national version which follows the same structural and hierarchical framework as the NACE. In Belgium, the national version of NACE Rev.~1 is called the NACE-Bel (2003) \citep{NACEBel2003} and adds one more level to the hierarchy by adding a fifth digit. We refer to this level as \texttt{subclass}. Insurance companies may choose to add a sixth digit to include yet another level, allowing for even more differentiation between companies. This level in the hierarchy is referred to as \texttt{tariff group} and we denote the insurance company's version as NACE-Ins.

\begin{figure}[!htbp]
	\centering
	\caption{\label{fig:NACE}Illustration of the NACE system for a company that manufactures beer. This economic activity is encoded as 1596 in NACE Rev.~1. Based on the NACE code, we assign the company to a certain category at a specific level in the hierarchy. For the purpose of this illustration, we shortened the textual description of the categories.}
	\makebox[\textwidth][c]{\includegraphics[width = \textwidth]{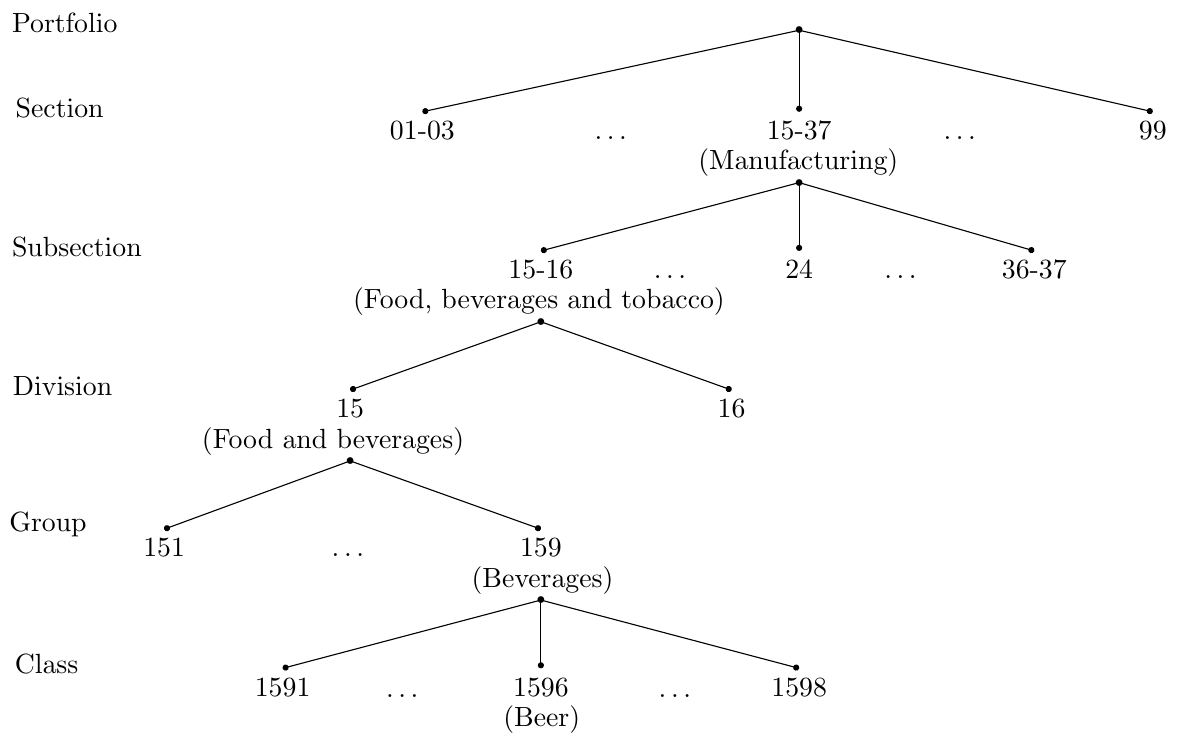}}
\end{figure}

To illustrate how the NACE system works, suppose that we have a company that manufactures beer. Using NACE Rev.~1 \citep{NACErev1}, this company gets the code 1596. At the top level \texttt{section}, the first two digits (i.e. 15) classify this company into \textit{manufacturing} (see \Cref{fig:NACE}). This category contains all NACE codes that start with numbers 15 to 37. Following, the first two digits categorize the company into \textit{manufacture of food products, beverages and tobacco} at the \texttt{subsection} level, which is nested within the \texttt{section} \textit{manufacturing}. The category \textit{manufacture of food products, beverages and tobacco} includes all NACE codes starting with numbers 15 and 16. At the third level in the hierarchy - \texttt{division} - the company is classified into \textit{15 - manufacture of food products and beverages}. At the fourth level \texttt{group}, we use the first three digits to classify the company in \textit{159 - manufacture of beverages}. Finally, at the fifth and lowest level \texttt{class}, the four digit code assigns the company to \textit{1596 - manufacture of beer}.

This example also shows that, at all levels in the hierarchy, the NACE provides a textual description for each category. This text briefly describes the economic activity of a specific category, thereby explaining why certain companies are grouped. We illustrate this in \Cref{tab:TextualNACE}. Herein, we show the textual information that is available for companies with NACE codes 1591, 1596, and 1598. This table displays a separate column for every level in the hierarchy. The corresponding value indicates which category the codes belong to. At the \texttt{section}, \texttt{subsection}, \texttt{division} and \texttt{group} level, the NACE codes 1591, 1596, and 1598 are grouped in the same categories (i.e. \texttt{15-37}, \texttt{15-16}, \texttt{15} and \texttt{159} respectively). At the \texttt{class} level, each of the codes is assigned to a different category. The column description presents the textual information for each corresponding category. For example, at the \texttt{division} level, \texttt{15} has the description \textit{manufacture of food products and beverages}.

\begin{table}[!ht]
	\centering
	\caption{\label{tab:TextualNACE}Illustration of the textual information for NACE codes 1591, 1596, and 1598.}
	\small
	\begin{tabular}{cccccL{6.75cm}}
		\hline
		Section & Subsection & Division & Group & Class & Description\\
		\hline
		15-37 &&&&&Manufacturing\\
		& 15-16&&&&Manufacture of food products, beverages and tobacco\\
		&& 15&&&Manufacture of food products and beverages\\
		&&& 159&&Manufacture of beverages\\
		&&&&1591&Manufacture of distilled potable alcoholic beverages\\
		&&&& $\dots$ & $\dots$\\
		&&&&1596&Manufacture of beer\\
		&&&& $\dots$ & $\dots$\\
		&&&&1598&Production of mineral waters and soft drinks\\
		\hline
	\end{tabular}
\end{table}

The textual description of other categories in the NACE system is similar to the example in \Cref{tab:TextualNACE}. Overall, the description is brief and consists of a single word or phrase. Further, categories at higher levels in the hierarchy have a more concise description using overarching terms such as \textit{manufacturing}. Conversely, at lower levels in the hierarchy, the descriptions are typically more detailed and extensive (e.g. \textit{manufacture of distilled potable alcoholic beverages} and \textit{production of mineral waters and soft drinks}).

\begin{table}[ht]
	\centering
	\caption{\label{tab:DescrNACE}Number of unique categories per level in the hierarchy of the NACE-Bel (2003).}
	\resizebox{\columnwidth}{!}{\begin{tabular}{lcccccc}
		\hline
		& \texttt{Section} & \texttt{Subsection} & \texttt{Division} & \texttt{Group} & \texttt{Class} & \texttt{Subclass}\\ 
		\hline
		Number of categories:\\
		\hspace{1mm} NACE-Bel (2003) &  17 &  31 & 62 &  224 & 515 & 800\\ 
		\hspace{1mm} Portfolio &  17 &  30 &  56 & 197 & 398 & 581 \\
		\hline
	\end{tabular}}
\end{table} 

\subsubsection{Selecting levels in the hierarchy} When working with NACE-Ins, we have seven levels in the hierarchy: \texttt{section}, \texttt{subsection}, \texttt{division}, \texttt{group}, \texttt{class}, \texttt{subclass} and \texttt{tariff group}. This results in an immense amount of categories, some of which contain few to no observations. In this paper, we work with a NACE-Ins that is based on the NACE-Bel (2003) \citep{NACEBel2003}. \Cref{tab:DescrNACE} shows the unique number of categories at each level in the hierarchy. The first row indicates how many categories there are at each level in the NACE-Bel (2003) classification system. The second row specifies how many categories are present at each level in our portfolio. Due to the confidentiality of the data, we do not disclose the number of categories at the \texttt{tariff group} level.

The more levels in the hierarchy, the more complex a tariff model becomes that incorporates this hierarchical MLF in its full granularity. Consequently, in practice, the NACE-Ins may not be used in its entirety. In our database, we have access to a hierarchical MLF designed by the insurance company that offers the workers' compensation insurance product. This structure has been created by merging similar NACE-Ins codes, based on expert judgment. This hierarchical MLF has two levels in the hierarchy, referred to as \texttt{industry} and \texttt{branch} in \citet{Campo2023}. In our paper, we illustrate how such a hierarchical MLF can be constructed using our proposed data-driven approach, as an alternative for the manual grouping by experts. To demonstrate our method we put focus on two selected levels in the hierarchy of NACE-Ins. This is merely for illustration purposes. Our approach is applicable with any number of levels in the hierarchy.

To align with the insurance company's hierarchical MLF, we select the \texttt{subsection} and \texttt{tariff group} level (see \Cref{fig:NACE}) and aim to reduce the hierarchical structure consisting of these levels. We use $l = (1, \dots, L)$ to index the levels in the hierarchy, where $L$ denotes the total number of levels. In our illustration, \texttt{subsection} corresponds to the highest level $l = 1$ in the hierarchy. At $l = 1$, we index the categories using $j = (1, \dots, J)$ where $J$ denotes the total number of categories. The \texttt{tariff group} level represents the second level in the hierarchy. Here, we use $jk = (j1, \dots, jK_j)$ to index the categories nested within \texttt{subsection} $j$. We refer to $k$ as the child category that is nested within parent category $j$ and $K_j$ denotes the total number of categories nested within $j$. Due to confidentiality of the classification system and data, no comparisons between the company's hierarchical MLF and our proposed clustering solutions will be provided in the paper.

\subsection{Feature engineering}\label{subsec:featureengineering}
We start at the highest level in the hierarchy, \texttt{subsection}, and engineer a set of features that capture the riskiness and the economic activity of the categories. Using these features, the clustering algorithms can identify and group similar categories at the \texttt{subsection} level. Hereafter, we proceed to the \texttt{tariff group} level and engineer the same type of features for the categories at this level in the hierarchy.

We assume that we have a workers' compensation insurance data set with historical, claim related information of the companies in our portfolio. For each company $i$, we have the NACE-Ins code in year $t$. We use this code to categorize company $i$ in \texttt{subsection} $j$ and \texttt{tariff group} $k$. In our database, we have the total claim amount $Z_{ijkt}$, the number of claims $N_{ijkt}$ and the salary mass $w_{ijkt}$ for year $t$ and company $i$, a member of \texttt{subsection} $j$ and \texttt{tariff group} $k$.

\subsubsection{Riskiness} We express the riskiness of the \texttt{subsection} and \texttt{tariff group} in terms of their damage rate and their claim frequency. The higher the damage rate and claim frequency, the riskier the category. For an individual company $i$, the damage rate in year $t$ is calculated as
\begin{equation}\label{eq:DamageRate}
	\begin{aligned}
		Y_{ijkt} = \frac{Z_{ijkt}}{w_{ijkt}}.
	\end{aligned}
\end{equation}
\noindent
To capture the \texttt{subsection}- and \texttt{tariff group}-specific effect on the damage rate and claim frequency, we use random effects models \citep{Campo2023}. In this approach, the \replaced[id=B]{prediction}{estimate} of the category-specific random effect is dependent on how much information is available. The random effect \replaced[id=B]{predictions}{estimates} are shrunk towards zero for categories with high variability, a low number of observations or when the variability between categories is small \citep{Breslow1993, Gelman2017, Brown2006, Pinheiro2009}. 

We start at the \texttt{subsection} level and model the damage rate as a function of the \texttt{subsection} using a (Tweedie generalized) linear mixed model
\begin{equation}\label{eq:REDR1}
	\begin{aligned}
		g(E[Y_{ijkt} | U_j^d]) = \mu^d + U_j^{d}.
	\end{aligned}
\end{equation}
\noindent
Here, $\mu^d$ denotes the intercept and $U_j^d$ the random effect of \texttt{subsection} $j$. We include the salary mass $w_{ijkt}$ as weight. As discussed in \citet{Campo2023}, we typically model the damage rate by assuming either a Gaussian or Tweedie distribution for the response. The $U_j^d$'s represent the between-\texttt{subsection} variability and enable us to discern between low- and high-risk profiles. The higher $U_j^d$, the higher the expected damage rate for \texttt{subsection} $j$ and vice versa. To engineer the first feature, we extract the $\widehat{U}_{j}^d$'s from the fitted damage rate model \eqref{eq:REDR1}. Hereafter, we fit a Poisson generalized linear mixed model 
\begin{equation}\label{eq:RECF1}
	\begin{aligned}
		g(E[N_{ijkt}| U_j^f]) = \mu^f + U_j^{f} + \log(w_{ijkt})
	\end{aligned}
\end{equation}
\noindent
to assess the \texttt{subsection}'s effect on the claim frequency. $\mu^f$ denotes the intercept and $U_j^f$ the random effect of \texttt{subsection} $j$. We include the log of the salary mass as an offset variable. We extract the $\widehat{U}_j^f$'s from the fitted claim frequency model \eqref{eq:RECF1} to engineer the second feature. The $\widehat{U}_j^d$'s and $\widehat{U}_j^f$'s will be combined with the features representing the economic activity to group similar categories at the \texttt{subsection} level. We index the resulting grouped categories at the \texttt{subsection} level using $j' = (1, \dots, J')$.

Next, we engineer the features at the \texttt{tariff group} level. To capture the \texttt{tariff group}-specific effect on the damage rate, we fit a (Tweedie generalized) linear mixed model
\begin{equation}\label{eq:REDR2}
	\begin{aligned}
		g(E[Y_{ij'kt} | U_{j'}^d, U_{j'k}^d]) &= \mu^d + U_{j'}^{d} + U_{j'k}^d
	\end{aligned}
\end{equation}
\noindent
where the random effect $U_{j'k}^d$ represents the \texttt{tariff group}-specific deviation from $\mu^d + U_{j'}^{d}$. As before, we include the $w_{ij'kt}$ as weight. The $U_{j'k}^d$'s reflect the between-\texttt{tariff group} variability, after having accounted for the variability between the grouped \texttt{subsections}. $U_{j'k}^d$ quantifies the \texttt{tariff group}-specific effect of \texttt{tariff group} $k$, in addition to the (grouped) \texttt{subsection}-specific effect $U_{j'}^d$, on the expected damage rate. To characterize the riskiness of the categories at the \texttt{tariff group} level in terms of the claim frequency, we extend \eqref{eq:RECF1} to
\begin{equation}\label{eq:RECF2}
	\begin{aligned}
		g(E[N_{ij'kt}| U_{j'}^f, U_{j'k}^f]) &= \mu^f + U_{j'}^{f} + U_{j'k}^f  + \log(w_{ij'kt}).
	\end{aligned}
\end{equation}
\noindent
In this model, $U_{j'k}^f$ represents the \texttt{tariff group}-specific deviation from $\mu^d + U_{j'}^{f}$. We extract the $\widehat{U}_{j'k}^d$'s and $\widehat{U}_{j'k}^f$'s from the fitted damage rate \eqref{eq:REDR2} and claim frequency model \eqref{eq:RECF2} to engineer the features at the \texttt{tariff group} level. 

\added[id = B]{We do not include any additional covariates in the random effects models (see equations \cref{eq:REDR1,eq:REDR2,eq:RECF1,eq:RECF2}) to fully capture the variability that is due to heterogeneity between categories. If, however, there are covariates available at the \texttt{subsection} or \texttt{tariff group} level, these can be incorporated in the model specification in equations \cref{eq:REDR1,eq:REDR2,eq:RECF1,eq:RECF2} and used further on in the construction of the feature matrix at the \texttt{subsection} or \texttt{tariff group} level.}

\added[id=B]{The approach to construct features based on the random effect predictions is closely related to target encoding \citep{Micci2001}. In target encoding, the numerical value of a category is the weighted average of the category-specific average of the response variable and the response variable's average at the higher levels in the hierarchy. \citet{Micci2001} presents various approaches to determine the weights. A linear mixed model can be seen as a special case of the weights specification as it has a closed-form solution for the random effects predictions. For most GLMMs, however, there is no available analytical expression. In these cases, there is no strict equivalence with target encoding.}

\subsubsection{Economic activity} Next, we require a feature that expresses similarity in economic activity. Given that industry codes of similar activities tend to be closer, we might rely on their numerical values. The closer the numerical value, the more overlap there will be between categories at a specific level in the hierarchy. Hence, we could use the four-digit NACE codes as discussed in Section \ref{subsec:NACE}. Notwithstanding, not all categories can readily be converted to a numerical format. At the higher levels in the hierarchy we have categories that encompass various NACE codes, such as \textit{manufacture of pulp, paper and paper products: publishing and printing} at the \texttt{subsection} level. This specific \texttt{subsection} consists of all NACE codes that start with numbers 21 to 22. To obtain a numerical representation of this \texttt{subsection} we can, for example, take the mean of these NACE codes. We then obtain the encodings as illustrated in \Cref{tab:NumericalEncodingNACE}. The column \texttt{subsection} indicates which category the NACE codes are appointed to at the \texttt{subsection} level and the column description gives the textual description of this category. The examples in this table highlight two issues with this approach. Firstly, the numerical distance may not reflect the similarity between economic activities. The difference between \textit{manufacture of wood and wood products} and \textit{manufacture of pulp, paper and paper products: publishing and printing} is the same as the difference between \textit{manufacture of pulp, paper and paper products: publishing and printing} and \textit{manufacture of coke, refined petroleum products and nuclear fuel}. Hence, using this encoding would imply that \textit{manufacture of pulp, paper and paper products: publishing and printing} is as similar to \textit{manufacture of wood and wood products} as it is to \textit{manufacture of coke, refined petroleum products and nuclear fuel}. Secondly, there might be gaps between consecutive codes in the classification system. In NACE Rev.~1, for example, there are no codes that start with 43 or 44. Gaps such as these can be present at various levels in the hierarchy and generally exist to allow for future additional categories \citep{NACE}.

\begin{table}[!htbp]
	\centering
	\caption{\label{tab:NumericalEncodingNACE}Illustration of a possible encoding for the categories at the \texttt{subsection} level of the NACE Rev.~1.}
	\small
	\begin{tabular}{ccL{9.5cm}}
		\hline
		\texttt{Subsection} & Encoding & Description\\
		\hline
		$\dots$ & $\dots$ & $\dots$\\
		20 & 20 & Manufacture of wood and wood products\\
		21-22 & 21.5 & Manufacture of pulp, paper and paper products: publishing and printing\\
		23 & 23 & Manufacture of coke, refined petroleum products and nuclear fuel\\
		24 & 24 & Manufacture of chemicals, chemical products and mad-made fibres\\
		$\dots$ & $\dots$ & $\dots$\\
		36-37 & 36.5 & Manufacturing not elsewhere classified\\
		40-41 & 40.5 & Electricity, gas and water supply\\
		45 & 45 & Construction\\
		$\dots$ & $\dots$ & $\dots$\\
		\hline
	\end{tabular}
\end{table}

Alternatively, we use embeddings to encode the economic activity of a category \citep{Mikolov}. Embeddings map the textual information to a continuous vector. Hence, we use embeddings to map the description for \texttt{subsection} $j$ to a vector $\bs{e}_j = (e_{j1}, e_{j2}, \dots, e_{jE})$. For \texttt{tariff group} $k$ within \texttt{subsection} $j$, we denote the embedding vector as $\bs{e}_{jk} = (e_{jk1}, e_{jk2}, \dots, e_{jkE})$. Here, $E$ is the dimension of the vector which depends on the encoder. The textual information from similar categories is expected to lie closer in the vector space and this enables us to group semantically similar texts (i.e. texts that are similar in meaning). 

Consequently, we group categories with comparable economic activities based on the embeddings of their textual labels. To engineer these embeddings, we may train an encoder on a large corpus of text. Within the area of Natural Language Processing (NLP), researchers commonly use neural networks (NN) as encoders. By training the NN on large amounts of unstructured text data, it is able to learn high-quality vector representations \citep{Mikolov}. In addition, the encoders can be trained to either learn vector representations for words, phrases or paragraphs. The disadvantage of these models is that they generally need a large amount of data \citep{Simran2020, Troxler2022}. Furthermore, words that do not appear often are poorly represented \citep{Luong2013}. We therefore prefer pre-trained encoders, which are trained on large text corpuses, to encode the textual description of a \texttt{subsection}. For example, the universal sentence encoder of \citet{USE} is trained on Wikipedia, web news, web question-answer pages and discussion forums. This encoder is publicly available via TensorFlow Hub (\url{https://tfhub.dev/}).

\subsubsection{Feature matrix} After engineering the features at level $l$ in the hierarchy, we assemble them into a feature matrix $\mc{F}_l$. \Cref{tab:FeatureMatrixIndustry} shows an example of the feature matrix $\mc{F}_1$ at the \texttt{subsection} level. Herein, $_1\widehat{\bs{U}}^d = (\widehat{U}_1^d, \dots, \widehat{U}_j^d, \dots, \widehat{U}_J^d)$ represents the vector with the \replaced[id=B]{predicted}{estimated} random effects from the fitted damage rate GLMM (see \eqref{eq:REDR1}). $_1\widehat{\bs{U}}^f = (\widehat{U}_1^f, \dots, \widehat{U}_j^f, \dots, \widehat{U}_J^f)$ denotes the vector with the \replaced[id=B]{predicted}{estimated} random effects of the claim frequency GLMM (see \eqref{eq:RECF1}) and we use the notation $\bs{e}_{\star 1} = (e_{11}, \dots, e_{j1}, \dots, e_{J1})$ for the embeddings. Each row in $\mathcal{F}_1$ corresponds to a numerical representation of a specific \texttt{subsection} $j$, with $j = (1, \dots, J)$. We denote the feature vector of \texttt{subsection} $j$ by $\boldsymbol{x}_{j} = (\widehat{U}_j^d, \widehat{U}_j^f, \bs{e}_j)$.

\begin{table}[!htbp]
	\renewcommand{\arraystretch}{1.6}
	\caption{\label{tab:FeatureMatrixIndustry}Feature matrix $\mathcal{F}_1$, consisting of the engineered features for the categories at $l = 1$ in the hierarchy. The columns $_1\widehat{\bs{U}}^d$ and $_1\widehat{\bs{U}}^f$ contain the \replaced[id=B]{predicted}{estimated} random effects of the damage rate and claim frequency GLMM, respectively. The embedding vector is represented by the values in columns $\bs{e}_{\star 1}, \bs{e}_{\star 2}, \dots, \bs{e}_{\star E}$.}
	\begin{tabular}{cccccccc}
		\hline
		\texttt{Subsection} & $_1\widehat{\bs{U}}^d$ & $_1\widehat{\bs{U}}^f$ & $\bs{e}_{\star 1}$ & $\bs{e}_{\star 2}$ & $\bs{e}_{\star 3}$ & $\dots$ & $\bs{e}_{\star E}$\\
		\hline
		1 &-1.25 & -0.25 & -2.13 & 1.25 & 0.15 & \dots & -0.05\\
		\dots & \dots & \dots & \dots & \dots & \dots & \dots & \dots\\
		$J$ & 0.75 & 0.15 & 1.79 & -2.13 & 0.5 & \dots & 1.07\\
		\hline
	\end{tabular}
\end{table}

At the \texttt{tariff group} level, we gather the features in $\mc{F}_2$. An example hereof is given in \Cref{tab:FeatureMatrixOccupation}. $_2\widehat{\bs{U}}^d = (\widehat{U}_{11}^d, \dots, \widehat{U}_{jk}^d, \dots, \widehat{U}_{JK_J}^d)$ and $_2\widehat{\bs{U}}^f = (\widehat{U}_{11}^f, \dots, \widehat{U}_{jk}^f, \dots, \widehat{U}_{JK_J}^f)$ denote the vectors of the \texttt{tariff group}-specific random effects. To denote the embeddings, we use $\bs{e}_{\star \star 1} = (e_{111}, \dots, e_{jk1}, \dots, e_{JK_{J}1})$. $\boldsymbol{x}_{jk} = (\widehat{U}_{jk}^d, \widehat{U}_{jk}^f, \bs{e}_{jk})$ denotes the feature vector of \texttt{tariff group} $k$, nested within \texttt{subsection} $j$.
\begin{table}[!htbp]
	\caption{\label{tab:FeatureMatrixOccupation}Feature matrix $\mathcal{F}_2$, consisting of the engineered features for the categories at $l = 2$ in the hierarchy. The columns $_2\widehat{\bs{U}}^d$ and $_2\widehat{\bs{U}}^f$ contain the \replaced[id=B]{predicted}{estimated} random effects of the damage rate and claim frequency GLMM, respectively. The embedding vector is represented by the values in columns $\bs{e}_{\star \star 1}, \bs{e}_{\star \star 2}, \dots, \bs{e}_{\star \star E}$.}
	\begin{tabular}{ccccccccc}
		\hline
		\texttt{Subsection} & \texttt{Tariff group} & $_2\widehat{\bs{U}}^d$ & $_2\widehat{\bs{U}}^f$ & $\bs{e}_{\star \star 1}$ & $\bs{e}_{\star \star 2}$ & $\bs{e}_{\star \star 3}$ & $\dots$ & $\bs{e}_{\star \star E}$\Tstrut\Cstrut\\
		\hline
		1 & 11 & -1.55 & -0.01 & -0.54 & 1.08 & 2.12 & \dots & 0.10\Tstrut\\
		\dots & \dots & \dots & \dots & \dots & \dots & \dots & \dots & \dots\\
		1 & $1K_1$ & 0.15 & 0.96 & -0.37 & -0.26 & 0.58 & \dots & -0.99\\
		\dots & \dots & \dots & \dots & \dots & \dots & \dots & \dots & \dots\\
		$j$ & $j$1 & -0.29 & -0.41 & -0.05 & -0.72 & 0.41 & \dots & -0.73\\
		\dots & \dots & \dots & \dots & \dots & \dots & \dots & \dots & \dots\\
		$j$ & $jK_j$ & 0.11 & 0.26 & 0.16 & 0.69 & 0.87 & \dots & 1.59\\
		\dots & \dots & \dots & \dots & \dots & \dots & \dots & \dots & \dots\\
		$J$ & $J$1 & 0.11 & 0.26 & 0.16 & 0.69 & 0.87 & \dots & 1.59\\
		\dots & \dots & \dots & \dots & \dots & \dots & \dots & \dots & \dots\\
		$J$ & $JK_J$ & 0.67 & -1.74 & 0.19 & 0.45 & 0.5 & \dots & -0.61\\
		\hline
	\end{tabular}
\end{table}

			\section{Clustering levels in a hierarchical categorical risk factor}\label{sec:Clustering}
\subsection{Partitioning Hierarchical Risk-factors Adaptive Top-down}\label{sec:Algo}
To group similar categories at each level in the hierarchy, we devise the PHiRAT to \textbf{P}artition \textbf{Hi}erarchical \textbf{R}isk-factors in an \textbf{A}daptive \textbf{T}op-down way, see Algorithm \ref{algo:GeneralProcedure}. We introduce some additional notation to explain how PHiRAT works. $\mc{J}_l$ denotes the set of categories at a specific level $l$ in the hierarchy. Hence, when the total number of levels $L = 2$, $\mc{J}_1 = (1, \dots, J)$ and $\mc{J}_2 = (11, \dots, 1K_1, \dots, j1, \dots, jK_j, \dots, J1, \dots, JK_J)$ as illustrated in \Cref{tab:FeatureMatrixIndustry,tab:FeatureMatrixOccupation}, respectively. We use $\pi(c) = p$ to indicate that child category $c$ has parent category $p$ and $\{c : \pi(c) = p\}$ denotes the set of all child categories $c$ nested within parent category $p$. $\mc{F}_{l, \{c : \pi(c) = p\}}$ represents the subset of feature matrix $\mc{F}_l$, which contains only those rows of $\mc{F}_l$ that correspond to the child categories nested in parent category $p$. For example, when $L = 2$, $\mc{F}_{2, \{c : \pi(c) = j\}}$ contains only the rows corresponding to the $(j1, \dots, jK_j)$ child categories of $j$ (see \Cref{tab:FeatureMatrixOccupation}). Further, $\mc{K}_l$ denotes the number of clusters at level $l$ in the hierarchy. For $l > 1$, we extend the notation to $\mc{K}_{l, \{c : \pi(c) = p\}}$ to indicate that, at level $l$ in the hierarchy, we group the child categories of $p$ into $\mc{K}_{l, \{c : \pi(c) = p\}}$ clusters. The subscript $l$ indicates at which level in the hierarchy we are. We use the additional subscript $\{c : \pi(c) = p\}$ to specify that we only consider the set of child categories of parent category $p$.

PHiRAT works top-down, starting from the highest level ($l = 1$) and working its way down to the lowest level ($l = L$) in the hierarchy (see Algorithm \ref{algo:GeneralProcedure}). Every iteration consists of three steps. First, we engineer features for the categories in $\mc{J}_l$. Second, we combine these features in a feature matrix $\mc{F}_l$. Third, we employ clustering techniques to group the categories in $\mc{J}_l$ using (a subset of) $\mc{F}_l$ as input. \added[id=B]{In most clustering methods, we define a tuning grid and perform a grid search to determine the optimal number of clusters. Consequently, the minimum value within the tuning grid sets the lower bound for the number of grouped categories.} \replaced[id=B]{Further, t}{T}he third step differs slightly when $l = 1$. At $l = 1$, we use the full feature matrix as input in the clustering algorithm. Conversely, when $l > 1$, we loop over the parent categories in $\mc{J}_{l - 1}$ and in every loop, we use a different subset of $\mc{F}_l$ as input. For parent category $p$, we only consider $\{c : \pi(c) = p\}$ and use $\mc{F}_{l, \{c : \pi(c) = p\}}$ as input in the clustering algorithm. Hereby, we ensure that we only group child categories nested within parent category $p$. Further, we remove a specific level $l$ from the hierarchy if $\mc{J}_{l}$ is reduced to $\mc{J}_{l - 1}$ (i.e.
all child categories, of each parent category in $\mc{J}_{l - 1}$, are merged into a single group). The algorithm stops when the clustering at level $l = L$ is done.

\vspace{2mm}
\IncMargin{1em}
\LinesNotNumbered
\begin{algorithm}[!htbp]
	\SetAlgoLined
	\caption{\label{algo:GeneralProcedure}PHiRAT Pseudo-code}
	\For{$l = 1$ to $L$}{
		Engineer features that characterize the categories\;
		Combine the features in a feature matrix $\mc{F}_l$\;
		\uIf{$l = 1$}{
			Use a clustering algorithm to group the $(1, \dots, J)$ categories into $\mc{K}_1$ clusters, with $\mc{F}_1$ as input\;
		}
		\uElse{
			\ForEach{$p$ in $\mc{J}_{l - 1}$}{
				Use a clustering algorithm to group the $\{c : \pi(c) = p\}$ child categories of parent category $p$ into $\mc{K}_{l, \{c : \pi(c) = p\}}$ clusters, using $\mc{F}_{l, \{c : \pi(c) = p\}}$ as input\;
			}
		}
		
	}
\end{algorithm}

\begin{figure}[!htbp]
	\centering
	\caption{A fictive example illustrating how the PHiRAT algorithm clusters categories at the \texttt{subsection} and \texttt{tariff group} level. The textual labels of the categories are shortened for the purpose of this illustration.}
	\label{fig:ExamplePHiRAT}
	\begin{subfigure}[b]{1\textwidth}
		\centering
		\caption{\label{fig:ExamplePHiRAT1}Visualization of the hierarchically structured categories at the \texttt{subsection} and \texttt{tariff group} level before clustering. The blue rectangles depict which categories are grouped when employing the PHiRAT algorithm.}
		\includegraphics[width=1\textwidth]{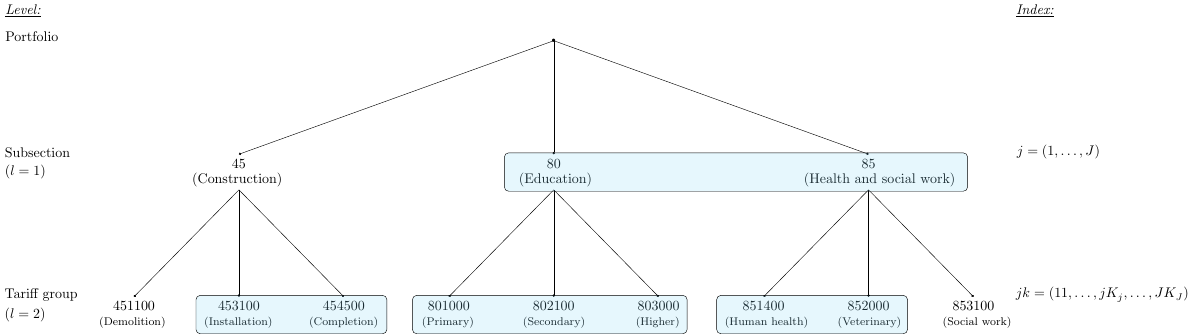}
	\end{subfigure}\\
	\vspace{1em}
	\begin{subfigure}[b]{1\textwidth}
		\centering
		\caption{\label{fig:ExamplePHiRAT2}Visualization of the reduced hierarchical structure after clustering with the PHiRAT algorithm.}
		\includegraphics[width=1\textwidth]{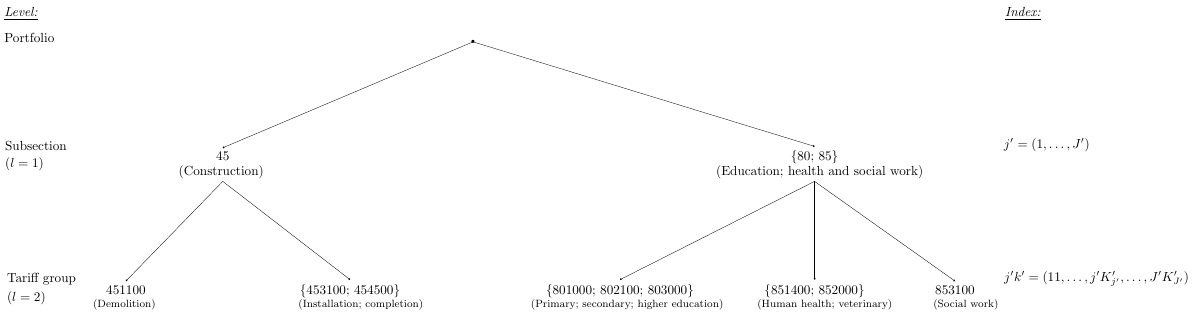}
	\end{subfigure}%
\end{figure}

We visualize how the procedure works in \Cref{fig:ExamplePHiRAT}. In this fictive example, we focus on the \texttt{subsection} and \texttt{tariff group} level. At the \texttt{subsection} level, there are three unique categories $\mc{J}_1 = (45, 80, 85)$ and at the \texttt{tariff group} level, we have nine unique categories $\mc{J}_2 = (451100, 453100, 454500, 801000, 802100, 803000, 851400, 852000, 853100)$ (see \Cref{fig:ExamplePHiRAT1}). We use the index $j = (1, \dots, J)$ at $l = 1$. At $l = 2$, we use $jk = (j1, \dots, jK_j)$ to index the categories nested within $j$. Using PHiRAT, we first group the categories 80 and 85 at $l = 1$. This is depicted in \Cref{fig:ExamplePHiRAT1} by the blue rectangle. Consequently, $\mc{J}_1 = (45, \{80;85\})$ where $\{80;85\}$ denotes that categories 80 and 85 are merged. The grouped categories at $l = 1$ are now indexed by $j' = (1, \dots, J')$. Hereafter, the algorithm iterates over the (fused) categories in $\mc{J}_1$ and clusters the child categories at $l = 2$. Within \texttt{subsection} 45, it groups the categories 453100 and 454500. Within $\{80; 85\}$, the clustering results in three groups of categories: (1) $\{801000; 802100; 803000\}$; (2) $\{851400; 852000\}$ and (3) 853100. We index the fused categories at $l = 2$ and nested within $j'$ using $j'k' = (j'1, \dots, j'K'_{j'})$. At this point the algorithm stops and \Cref{fig:ExamplePHiRAT2} depicts the clustering solution.

We opt for a top-down approach for several reasons. Firstly, at the highest level we have more observations available for the categories. The more data, the more precise the category-specific risk estimates will be. Conversely, categories at more granular levels have fewer observations, leading to less precise risk estimates. Secondly, we preserve the original hierarchical structure and maintain the parent-child relationship between categories at different levels in the hierarchy. Hereby, we divide the grouping of categories, at a specific level in the hierarchy, into smaller and more specific separate clustering problems.

Alternatively, it might be interesting to construct a similar algorithm that works bottom-up and groups child categories that have different parent categories. We leave this as a topic for future research.

\subsection{Clustering analysis}\label{sec:ClusteringAlgorithms}
To partition the set of categories $\mc{J}_l$ into homogeneous groups, we rely on clustering algorithms using (a subset of) $\mc{F}_l$ as input. At $l = 1$, for example, we employ clustering to divide the rows in $\mc{F}_1$ into $\mc{K}_1$ homogeneous groups such that categories in each cluster $j'$ are more similar to each other compared to categories of other clusters $\mb{j'} \neq j'$. In the remainder of this section, we continue with the example of grouping the $(1, \dots, J)$ categories at $l = 1$ to explain and illustrate the key concepts of the clustering methods used in this paper.

Most clustering algorithms rely on distance or (dis)similarity metrics to quantify the proximity between observations. \Cref{tab:DistanceMeasures} gives an overview of a selected set of (dis)similarity measures relevant to our paper. A dissimilarity measure $d(\bs{x}_j, \bs{x}_{\mb{j}})$ expresses how different two observations $\bs{x}_j$ and $\bs{x}_{\mb{j}}$ are (the higher the value, the more they differ). Dissimilarity metrics that satisfy the triangle inequality $d(\bs{x}_j, \bs{x}_{\mb{j}}) \leq d(\bs{x}_j, \bs{x}_{z}) + d(\bs{x}_z, \bs{x}_{\mb{j}})$ for any $z$ \citep{Schubert2021, Phillips2021} are considered proper distance metrics. Most dissimilarity metrics can easily be converted to a similarity measure $s(\cdot, \cdot)$ which expresses how comparable two observations are, with similar observations obtaining higher values for these measures. The most commonly used dissimilarity metric is the squared Euclidean distance $\|\bs{x}_j - \bs{x}_{\mb{j}}\|^2_2$. Here, $\| \bs{x}_j \|_2 \coloneqq \sqrt{x_{j1}^2 + \dots + x_{jn_f}^2}$ and $n_f$ denotes the number of features considered. Euclidean based (dis)similarity measures, however, are not appropriate to capture the similarities between embeddings \citep{Kogan2006}. Within NLP, the cosine similarity is therefore most often used to measure the similarity between embeddings \citep{Mohammed2012,Schubert2021}. Notwithstanding, the cosine similarity ranges from -1 to 1 and in cluster analysis we generally require the (dis)similarity measure to be non-negative \citep{Everitt2011, Kogan2006, Hastie2009}. In this case, we can convert the cosine similarity to the angular similarity which is restricted to [0, 1]. The angular similarity can be converted to the angular distance, which is a proper distance metric. Conversely, the cosine dissimilarity is not a distance measure since it does not satisfy the triangle inequality. In our study, we therefore rely on the angular similarity and angular distance to group comparable categories.

\begin{table}[!htbp]
	\caption{\label{tab:DistanceMeasures}Overview of existing (dis)similarity metrics to quantify the proximity between observations. We select the angular similarity and angular distance, as they are better suited to measure the similarity between embeddings and also compatible with clustering algorithms.}
	\begin{threeparttable}
	\centering
	\small
	\renewcommand{\arraystretch}{1.6}
		\begin{tabular}{lcccc}
			\hline
			& Dissimilarity & Similarity\\
			\hline
			Euclidean\tnote{a} & \(\displaystyle \|\bs{x}_j - \bs{x}_{\mb{j}}\|^2_2 \) & \(\displaystyle \exp\left( \frac{-\|\bs{x}_j - \bs{x}_{\mb{j}}\|^2_2}{\sigma^2} \right) \) \rule{0pt}{5ex}\rule[-2ex]{0pt}{0pt}\\
			Cosine & \(1 - \displaystyle \frac{\bs{x}_{j}^\top \bs{x}_{\mb{j}}}{\|\bs{x}_{j}\|_2 \cdot \|\bs{x}_{\mb{j}}\|_2} \) & \(\displaystyle \frac{\bs{x}_{j}^\top \bs{x}_{\mb{j}}}{\|\bs{x}_{j}\|_2 \cdot \|\bs{x}_{\mb{j}}\|_2} \)\rule{0pt}{6.5ex}\rule[-2ex]{0pt}{0pt}\\
			Angular & \textcolor{red}{\(\displaystyle \pi^{-1} \cos^{-1}\left(\frac{\bs{x}_{j}^\top \bs{x}_{\mb{j}}}{\|\bs{x}_{j}\|_2 \cdot \|\bs{x}_{\mb{j}}\|_2}\right) \)} & \textcolor{red}{\(\displaystyle 1 - \pi^{-1} \cos^{-1}\left(\frac{\bs{x}_{j}^\top \bs{x}_{\mb{j}}}{\|\bs{x}_{j}\|_2 \cdot \|\bs{x}_{\mb{j}}\|_2}\right) \)} \rule{0pt}{7.5ex}\\[2ex]
			\hline
		\end{tabular}
	\begin{tablenotes}
		\small
		\item[a] $\sigma$ is a scaling parameter set by the user \citep{Ng2001, Poon2012}
	\end{tablenotes}
	\end{threeparttable}
\end{table}

Using a selected (dis)similarity metric, we compute the proximity measure between all pairwise observations in the matrix with input features. We combine all values in a $J \times J$ similarity matrix $S$ or dissimilarity matrix $D$, which is used as input in a clustering algorithm. Most algorithms require these matrices to be symmetric \citep{Hastie2009}. In literature on unsupervised learning algorithms, clustering methods are typically divided into three different types: combinatorial algorithms, mixture modelling and mode seeking \citep{Hastie2009}. Both the mixture modelling and mode seeking algorithms rely on probability density functions. Conversely, the combinatorial algorithms do not rely on an underlying probability model and work directly on the data. In our paper we opt for a distribution-free approach and therefore focus on the combinatorial algorithms summarized in \Cref{tab:ClusteringAlgorithms}. For the interested reader, a detailed overview of these algorithms is given Appendix \ref{App:ClusteringAlgorithms}.

\begin{table}[!htbp]
	\caption{\label{tab:ClusteringAlgorithms}Overview of clustering algorithms, together with their strengths and drawbacks.}
	\begin{threeparttable}
		\centering
		\footnotesize
		\begin{NiceTabular}{l|L{4.4cm}L{4.4cm}}[
			code-before = \rowcolor[HTML]{FFFFFF}{1,6,7,8,11,12}
			\rowcolor[HTML]{FAFAFF}{2,3,4,5,8,9,10}
			]
			\toprule
			\textbf{Algorithm} & \textbf{Strengths} & \textbf{Drawbacks}\\
			\midrule
			k-means & - Well-known & - Only suited for numeric\\
			\citep{kmeans}&& features\\
			& - Simple and easy to implement & - Sensitive to outliers and the initialization\\
			& - Computationally efficient & - Local optima\\[0.5em]
			\midrule
			
			k-medoids & - Applicable to any feature type & - Sensitive to the initialization\\
			\citep{PAM}& - Less sensitive to outliers & - Local optima\\[0.5em]
			\midrule
			
			Spectral clustering & - Applicable to any feature type & - Computationally expensive\\
			\citep{Hastie2009} & - Less sensitive to outliers, initialization and local optima & - Sensitive to the employed similarity metric\\
			& - Able to identify non-convex clusters\tnote{a }&\\[0.5em]
			\midrule
			
			HCA\tnote{b}  & - Applicable to any feature type & - Computationally expensive  \\ 
			\citep{Hastie2009} & - Less sensitive to outliers\tnote{c}, initialization and local optima & - Static; divisions or fusions of clusters are irrevocable\\
			\bottomrule
		\end{NiceTabular}
		\begin{tablenotes}
			\tiny
			\item[a] For every pair of points inside a convex cluster, the connecting straight line segment is within this cluster
			\item[b] Hierarchical clustering analysis
			\item[c] When using the single-linkage criterion
		\end{tablenotes}
	\end{threeparttable}
\end{table}

In most clustering techniques, the number of clusters $\mc{K}$ can be considered a tuning parameter that needs to be carefully chosen from a range of possible (integer) values. Hereto, we require a cluster validation index to select the value for $\mc{K}$ which results in the most optimal clustering solution. We divide the cluster validation indices into two groups: internal and external \citep{Liu2013, Everitt2011, Wierzchon2019, Halkidi2001}. Using external validation indices, we evaluate the clustering criterion with respect to the true partitioning (i.e. the actual assignment of the observations to different groups is known). Conversely, we rely on internal validation indices when we do not have the true cluster label at our disposal. With such indices, we evaluate the compactness and separation of a clustering solution. The compactness indicates how dense the clusters are and compact clusters are characterized by observations that are similar and close to each other. Clusters are well separated when observations belonging to different clusters are dissimilar and far from each other. Consequently, we employ internal validation indices to choose the value for $\mc{K}$ which results in compact clusters that are well separated \citep{Liu2013, Everitt2011, Wierzchon2019}. 

Several internal validation indices exist and each index formalizes the compactness and separation of the clustering solution differently. An extensive overview of internal (and external) validation indices is given in \citet{Liu2013}, \citet{Wierzchon2019} and in the benchmark study of \citet{Vendramin2010}. The authors concluded that the silhouette and Cali{\'n}ski-Harabasz (CH) indices are superior compared to other validation criteria. While these indices are well-known within cluster analysis \citep{Wierzchon2019, Govender2020, Vendramin2010}, the results of \citet{Vendramin2010} do not necessarily generalize to our data set. We therefore include two additional, commonly used criteria: the Dunn index and Davies-Bouldin index. \Cref{tab:EvaluationCriteria} shows how these four indices are calculated. For the Cali{\'n}ski-Harabasz, Dunn and silhouette index, higher values are associated with a better clustering solution. Conversely, for the Davies-Bouldin index, we arrive at the best partition by minimizing this criterion. A more in-depth discussion of these criteria can be found in Appendix \ref{App:EvaluationCriteria}.

Numerous papers compared the performance of different clustering algorithms using different data sets and various evaluation criteria \citep{Mangiamelo1996, Costa2004, DeSouto2008, Kinnunen2011, Jung2014, Kou2014, Rodriguez2019, Murugesan2021}. They concluded that none of the considered algorithms consistently outperforms the others and that the performance is dependent on the type of data \citep{Hennig2015, McNicholas2016a, McNicholas2016b, Murugesan2021}. The authors advise to compare different clustering methods using more than one performance measure. Consequently, we test the PHiRAT algorithm (see Algorithm \ref{algo:GeneralProcedure}) with all possible combinations of the selected clustering methods (i.e. k-medoids, spectral clustering and HCA, see \Cref{tab:ClusteringAlgorithms}) and internal validation criteria (i.e. Cali{\'n}ski-Harabasz index, Davies-Bouldin index, Dunn index and silhouette index, see \Cref{tab:EvaluationCriteria}). 

\begin{table}[!htbp]
	\caption{\label{tab:EvaluationCriteria}Internal clustering validation criteria used in this paper.}
	\begin{threeparttable}
		\centering
		\footnotesize
			\begin{NiceTabular}{L{4cm}L{9cm}}
				\hline
				Criterion & Definition\\
				\hline
				Cali{\'n}ski-Harabasz index \hspace{0.75cm}\citep{Calinski1974} & \scalebox{0.85}{\(\displaystyle \frac{\sum_{j' = 1}^{J'} n_{j'} \|\bs{c}_{j'} - \bs{c}\|^2_2 / (J' - 1)}{\sum_{j' = 1}^{J'} \sum_{\bs{x}_j \in C_{j'}} \|\bs{x}_j - \bs{c}_{j'}\|^2_2 / (J - J')} \)} where \scalebox{0.85}{\(\displaystyle \bs{c} = \frac{1}{J} \sum_{j = 1}^{J} \bs{x}_j \)} \rule{0pt}{5ex}\rule[-4ex]{0pt}{0pt}\\
				
				Davies-Bouldin index \hspace{0.75cm}\citep{Davies1979} & \scalebox{0.85}{\(\displaystyle \frac{1}{J'} \sum_{j' = 1}^{J'} \max_{\mb{j}', \ \mb{j}' \neq j'} \left( \frac{\frac{1}{n_{j'}} \sum_{\bs{x}_j \in C_{j'}} d(\bs{x}_j, \bs{c}_{j'}) + \frac{1}{n_{\mb{j}'}} \sum_{\bs{x}_{\mb{j}} \in C_{\mb{j}'}} d(\bs{x}_{\mb{j}}, \bs{c}_{\mb{j}'})}{d(\bs{c}_{j'}, \bs{c}_{\mb{j}'})} \right) \)} \rule{0pt}{5ex}\rule[-4ex]{0pt}{0pt}\\
				
				Dunn index \hspace{2cm}\citep{Dunn1974} & \scalebox{0.85}{\(\displaystyle \min_{1 \leq j' \leq J'} \left(\min_{\substack{1 \leq \mb{j}' \leq J'\\ j \neq \mb{j}}} \left( \frac{\min\limits_{\substack{\bs{x}_j \in C_{j'}\\\bs{x}_{\mb{j}} \in C_{\mb{j}'}}} d(\bs{x}_{j}, \bs{x}_{\mb{j}})}{\max\limits_{1 \leq \kappa \leq J'} \left\lbrace \max\limits_{\substack{\bs{x}_j, \bs{x}_{\mb{j}} \in C_{\kappa}}} d(\bs{x}_{j}, \bs{x}_{\mb{j}}) \right\rbrace } \right) \right) \)} \rule{0pt}{5ex}\rule[-4ex]{0pt}{0pt}\\
				
				Silhouette index \hspace{1.5cm}\citep{Rousseeuw1987} & \scalebox{0.85}{\(\displaystyle\tilde{s} = \frac{1}{J} \sum_{j = 1} s(\bs{x}_j) \)} where \scalebox{0.85}{\(\displaystyle s(\bs{x}_j) = \frac{b(\bs{x}_j) - a(\bs{x}_j)}{\max(a(\bs{x}_j), b(\bs{x}_j))} \)}, \rule{0pt}{5ex}\\
				& \scalebox{0.85}{\(\displaystyle b(\bs{x}_j) = \min_{C_{j'} \neq C_{\mb{j}'}} \frac{1}{n_{\mb{j}'}} \sum_{\bs{x}_{\mb{j}} \in C_{\mb{j}'}} d(\bs{x}_{j}, \bs{x}_{\mb{j}}) \)} and \scalebox{0.85}{\(\displaystyle a(\bs{x}_j) = \frac{1}{n_{j'} - 1} \sum_{\bs{x}_{\mb{j}} \in C_{j'}, \ \mb{j} \neq j} d(\bs{x}_{j}, \bs{x}_{\mb{j}}). \)} \rule{0pt}{2ex}\\[2ex]
				\hline
			\end{NiceTabular}
			\begin{tablenotes}
			\tiny
			\item $\bs{c}_{j'}$ denotes the cluster centre or centroid of cluster $C_{j'}$; $n_{j'}$ denotes the number of observations in cluster $C_{j'}$
		\end{tablenotes}
	\end{threeparttable}
\end{table}

			\section{Clustering NACE codes in a workers' compensation insurance product}\label{sec:CaseStudy}
We use a workers' compensation insurance data set from a Belgian insurer to illustrate the PHiRAT procedure. The portfolio consists of Belgian companies that are active in various industries and occupations. The database contains claim-related information, such as the number of claims and the claim sizes, over a course of eight years. Additionally, for each of the companies we have the corresponding NACE-Ins code (see \Cref{sec:MotivatingExample}). In this section, we demonstrate how to reduce the NACE-Ins to its essence using PHiRAT. Our end objective is to incorporate the reduced version of the hierarchical MLF as a risk factor in a technical pricing model as discussed in \citet{Campo2023}. We therefore also assess the effect of clustering on the predictive accuracy. Furthermore, if the reduced structure properly captures the essence, the predictive accuracy should generalize to out-of-sample and out-of-time data. We employ PHiRAT using the training data set, which contains data from the first seven years, to construct the reduced hierarchical risk factor. To examine the generalizability of the reduced structure to out-of-sample and out-of-time data, we use the test data set which contains data from the eight and most recent year.

\subsection{Exploring the workers' compensation insurance database}

\subsubsection{Claim-related information at the company level} 
\begin{figure}[!htbp]
	\centering
	\caption{\label{fig:EmpDistr}Empirical distribution of the individual companies' (a) damage rates $Y_{it}$; (b) log transformed damage rates $Y_{it}$ for $Y_{it} > 0$; (c) number of claims $N_{it}$; (d) log transformed $N_{it}$ for $N_{it} > 0$. This figure depicts the $Y_{it}$'s and $N_{it}$'s of all available years in our data set.}
	\makebox[\textwidth][c]{\includegraphics[scale = 0.5]{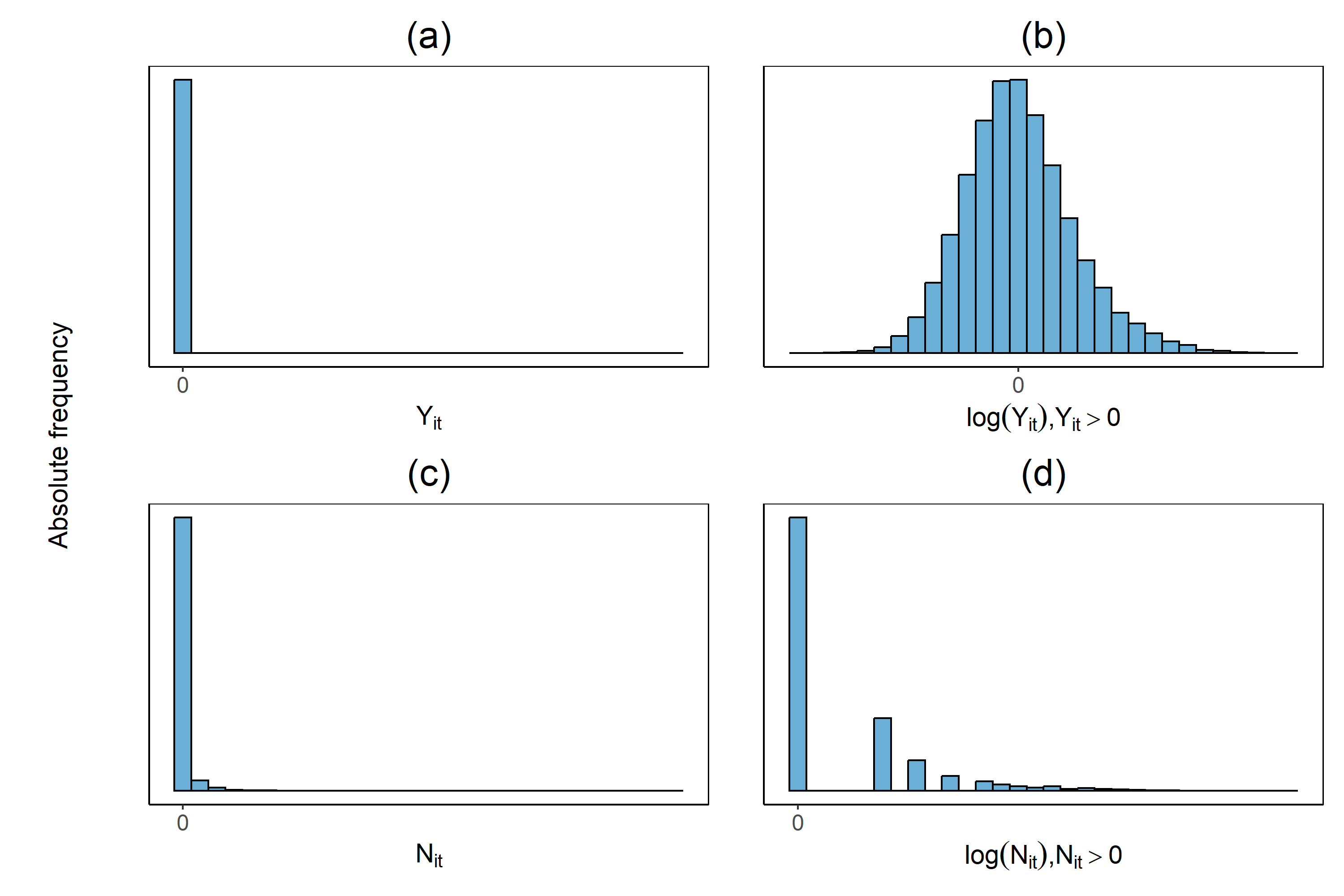}}
\end{figure}

We first explore the distribution of the claim-related information in our database. We consider the damage rate $Y_{it}$ and the number of claims $N_{it}$ of company $i$ in year $t$. When calculating $Y_{it} = \mathcal{Z}_{it} / w_{it}$ (see equation \eqref{eq:DamageRate}), we use the capped claim amount $\mathcal{Z}_{it}$ for company $i$ in year $t$ to prevent that large losses have a disproportionate impact on the results (see \citet{Campo2023} for details on the capping procedure). To account for inflation and for the size of the company, we express the damage rate per unit of company- and year-specific salary mass $w_{it}$. \Cref{fig:EmpDistr} depicts the empirical distribution of the damage rates $Y_{it}$ and number of claims $N_{it}$  of the individual companies. A strong right skew is visible in the empirical distribution of the $Y_{it}$ (\Cref{fig:EmpDistr}(a)) and the $N_{it}$ (\Cref{fig:EmpDistr}(c)). Moreover, this right skewness persists in the log transformed counterparts (see panels (b) and (d) in \Cref{fig:EmpDistr}).

\subsubsection{Claim-related information at different levels in the NACE-Ins hierarchy} The NACE-Ins partitions the companies according to their economic activity, at varying levels of granularity. We compute the category-specific weighted average damage rate and claim frequency at all levels in the hierarchy. Considering the top level in the hierarchy (as explained in \Cref{sec:MotivatingExample}), we calculate the weighted average damage rate for category $j = (1, \dots, J)$ as
\begin{equation}
	\begin{aligned}
		\bar{Y}_{j} = \frac{\sum_{i, k, t} w_{ijkt} Y_{ijkt}}{\sum_{i, k, t} w_{ijkt}}
	\end{aligned}
\end{equation}
\noindent
and the expected claim frequency as
\begin{equation}
	\begin{aligned}
		\bar{\mathcal{C}}_j = \frac{\sum_{i, k, t} N_{ijkt}}{\sum_{i, k, t} w_{ijkt}}.
	\end{aligned}
\end{equation}
We then calculate the weighted average damage rate and expected claim frequency for child-category $jk = (j1, \dots, jK_j)$ within parent-category $j$ as
\begin{equation}
	\begin{aligned}
		\bar{Y}_{j k} = \frac{\sum_{i, t} w_{ijkt} Y_{ijkt}}{\sum_{i, t} w_{ijkt}} \mspace{37.5mu} \text{and} \mspace{37.5mu} 
		\bar{\mathcal{C}}_{jk} = \frac{\sum_{i, t} N_{ijkt}}{\sum_{i, t} w_{ijkt}}.
	\end{aligned}
\end{equation}
\noindent
When we consider more granular levels in the hierarchy, the computation is similar. To calculate these quantities for a specific category, we only use observations that are classified herein.

At different levels in the hierarchy, the empirical distribution of the category-specific weighted average damage rates and expected claims frequencies show a strong right skew (see Appendix \Cref{App:EmpiricalDistrDRCF}). The large range in values hinders the visual comparison of the weighted average damage rates and expected claim frequencies across categories at different levels in the hierarchy. We therefore apply a transformation solely for the purpose of visual comparison. Illustrating the procedure with $\bar{Y}_j$, we first apply the following transformation
\begin{equation}
	\begin{aligned}
		\bar{\mathcal{Y}}_j = \log(\bar{Y}_j + 0.0001)
	\end{aligned}
\end{equation}
\noindent
since we have categories for which $\bar{Y}_j = 0$. Hereafter, we cap $\bar{\mathcal{Y}}_j$ using
\begin{equation}
	\begin{aligned}
		\bar{\mathcal{Y}}_j^c = \max(\min(\bar{\mathcal{Y}}_j, \ Q_3(\bar{\mathcal{Y}}_j) + 1.5 \ \text{IQR}(\bar{\mathcal{Y}}_j)), \ Q_1(\bar{\mathcal{Y}}_j) - 1.5 \ \text{IQR}(\bar{\mathcal{Y}}_j))
	\end{aligned}
\end{equation}
\noindent
where $Q_n(\bar{\mathcal{Y}}_j)$ denotes the $n^{th}$ quantile of $\bar{\mathcal{Y}}_j$ and IQR$(\bar{\mathcal{Y}}_j) = Q_3(\bar{\mathcal{Y}}_j) - Q_1(\bar{\mathcal{Y}}_j)$ denotes the interquartile range of $\bar{\mathcal{Y}}_j$. The lower and upper bound of $\bar{\mathcal{Y}}_j^c$ correspond to the inner fences of a boxplot \citep{Schwertman2004}. By transforming and capping the quantities, we focus on the pattern seen in the majority of the categories. Additionally, this facilitates the visual comparison across different categories.

\afterpage{%
	\begin{landscape}
		\begin{figure}[htbp]
			\centering
			\caption{\label{fig:AvgDR}Category-specific weighted average damage rates: (a) at all levels in the hierarchy; (b) of the \texttt{section} $\lambda$ and $\phi$, including those of their child categories at all levels in the hierarchy; (c) at the \texttt{subsection} and \texttt{tariff group} level. (b) is a close-up of the top right part of (a). In this close-up, the width of $\phi$ at the \texttt{section} level and its child categories is increased by a factor 10 to allow for better visual inspection.}
			\makebox[\textwidth][c]{\includegraphics[width = \textwidth]{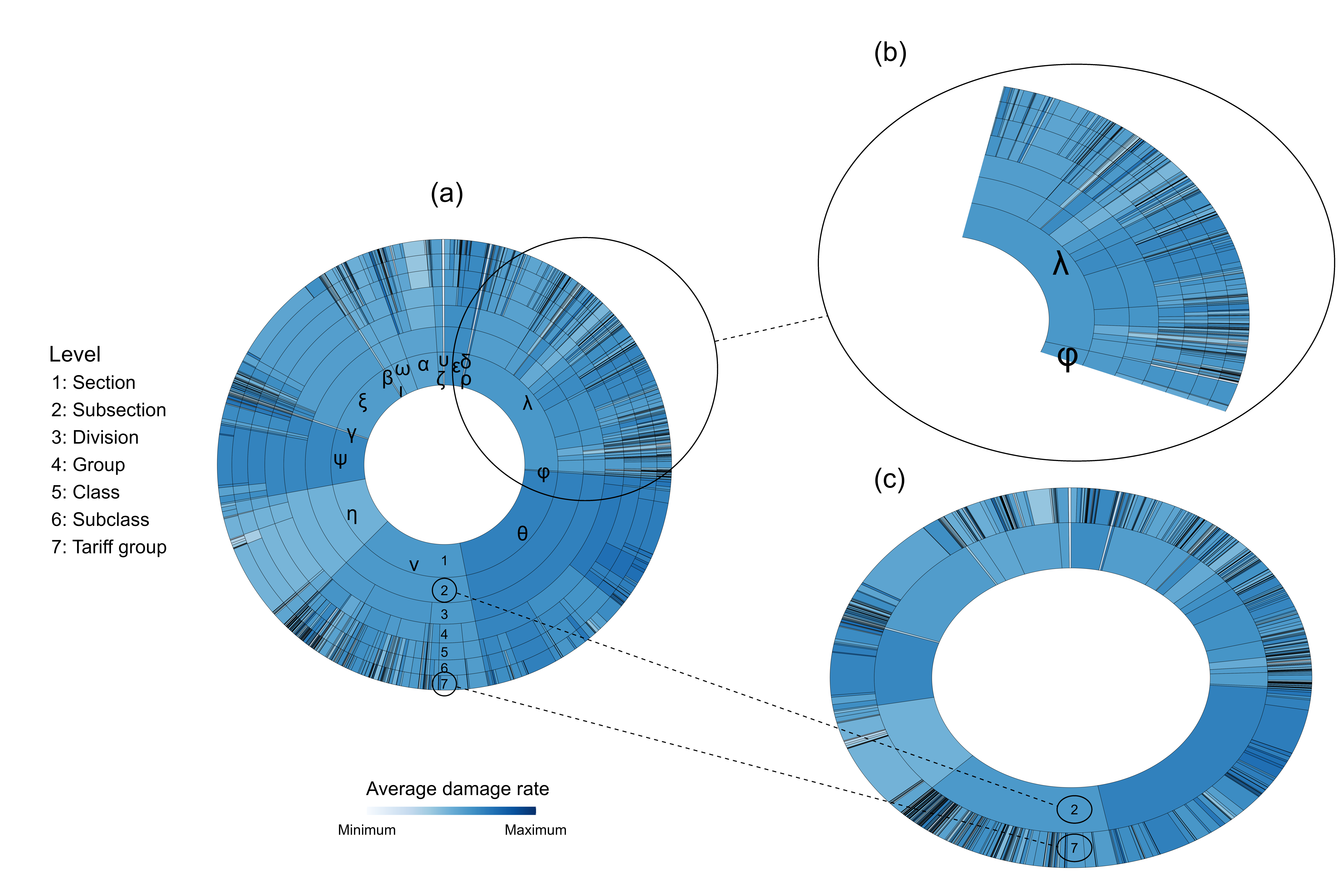}}
		\end{figure}
		\begin{figure}[htbp]
			\centering
			\caption{\label{fig:AvgCF}Category-specific expected claim frequencies: (a) at all levels in the hierarchy; (b) at the \texttt{subsection} and \texttt{tariff group} level.}
			\makebox[\textwidth][c]{\includegraphics[width = \textwidth]{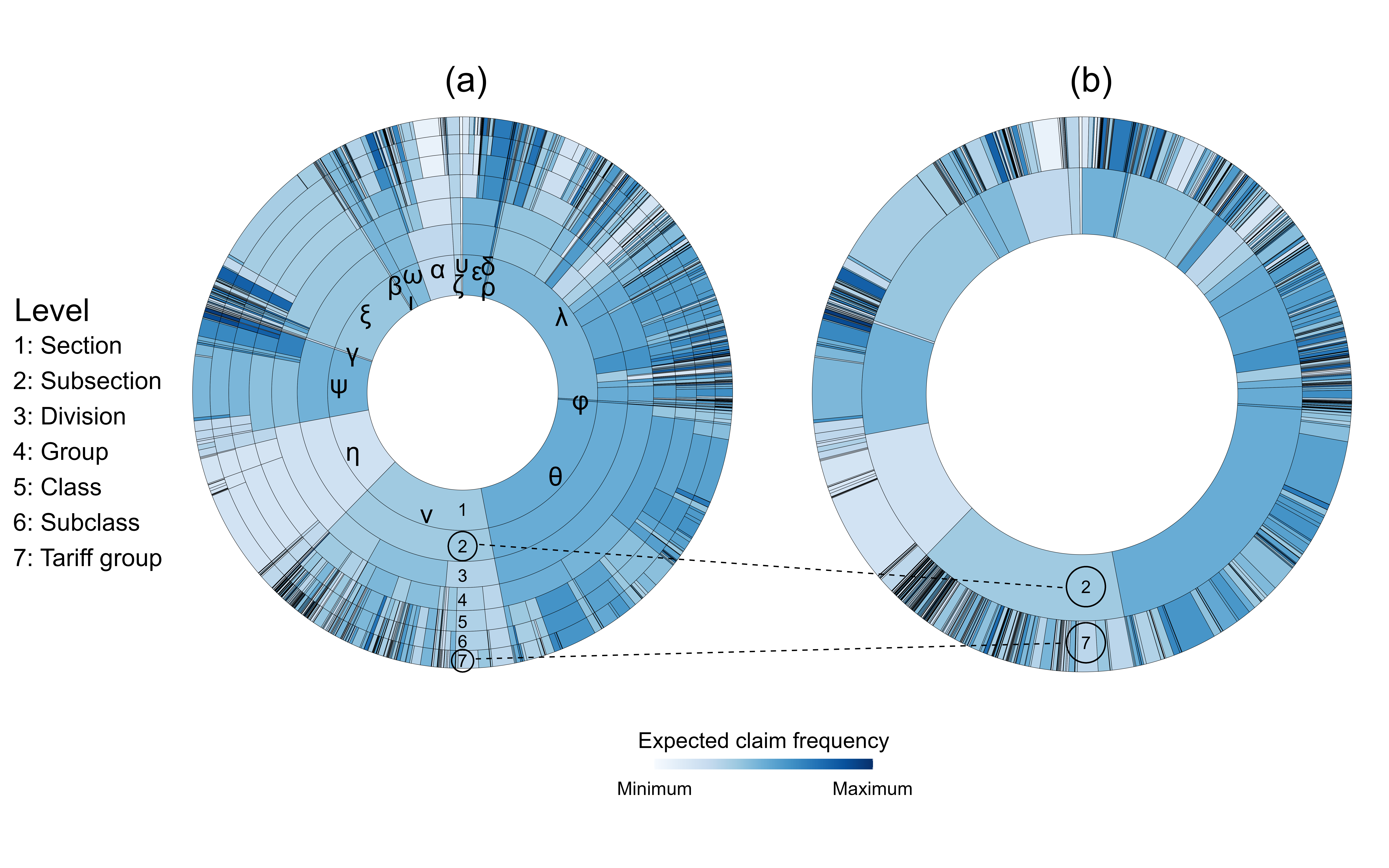}}
		\end{figure}
	\end{landscape}
}

\Cref{fig:AvgDR} visualizes the category-specific weighted average damage rates and salary masses. Panel (a) depicts the category-specific weighted averages at all levels in the hierarchy, panel (b) is close-up of the top right portion of panel (a) and panel (c) represents the averages at the \texttt{subsection} and \texttt{tariff group} level. We use a colour gradient for the weighted average damage rate. The darker the colour, the higher the weighted average damage rate. Further, each ring in the circle corresponds to a specific level in the hierarchy and the level is depicted by the number in the ring. To represent the categories at a specific level, the rings are split into slices proportional to the corresponding summed salary mass. The bigger the slice, the larger the summed salary mass that corresponds to a specific category at the considered level in the hierarchy. To preserve the confidentiality of the data, we randomly assign Greek letters to each of the categories at the \texttt{section} level. 

At all levels in the hierarchy, there is variation in the category-specific weighted average damage rates. This is demonstrated in \Cref{fig:AvgDR}(b), displaying a magnified view of the top right section of \Cref{fig:AvgDR}(a). Here, the blue tone varies between the child categories of parent category $\lambda$ at the \texttt{section} level. Additionally, at all levels in the hierarchy, there are categories with a low summed salary mass (e.g. $\gamma, \beta, \upsilon, \delta, \rho, \phi$ at the \texttt{section} level). In \Cref{fig:AvgDR}, this is depicted by the thin slices. These categories represent only a small part of our portfolio. Furthermore, most of the categories with a low salary mass have a less granular representation in the NACE system, having very few child categories compared to other categories. For example, the parent category $\phi$ at the \texttt{section} level, has a low salary mass and only a limited number of child categories across all levels in the hierarchy of the NACE system.

\Cref{fig:AvgCF} depicts the category-specific expected claim frequency and salary masses. Similarly to \Cref{fig:AvgDR}, we use a colour gradient for the claim frequencies and the size of a slice is again proportional to the summed salary mass. Overall, the findings are comparable to \Cref{fig:AvgDR}. The category-specific expected claim frequency varies between the categories at all levels in the hierarchy.

Following, we focus only on the \texttt{subsection} and \texttt{tariff group} level. The category-specific weighted average damage rates and expected claim frequencies at the \texttt{subsection} and \texttt{tariff group} level are shown in \Cref{fig:AvgDR}(c) and \Cref{fig:AvgCF}(b), respectively. To illustrate PHiRAT, we group similar categories at these levels in the hierarchy and hereby, reduce the granularity of the hierarchical risk factor.

\subsection{Engineering features to improve clustering results}\label{subsec:Embeddings}
Feature engineering is crucial to obtain a reliable clustering solution through PHiRAT. As discussed in \Cref{subsec:featureengineering}, we therefore engineer a set of features to capture the riskiness and the economic activity of the categories. The \replaced[id=B]{predicted}{estimated} random effect from the damage rate and claim frequency model expresses the category-specific riskiness at the \texttt{subsection} and \texttt{tariff group} level. Further, we use LMMs to fit the damage rate random effects models in \eqref{eq:REDR1} and \eqref{eq:REDR2}. LMMs are less complex, computationally more efficient and are less likely to experience convergence problems compared to GLMMs. Alternatively, we can consider Tweedie GLMMs to model the damage rate. We refer the reader to \citet{Campo2023} for a discussion on the effect of the distributional assumption on the response. We use embeddings to encode the category's textual labels that describe the economic activity. In our database, we have a description for every category at the \texttt{subsection} and \texttt{tariff group} level.

To encode the textual information, we rely on pre-trained encoders. The first pre-trained encoder we use is a Word2Vec model \citep{Mikolov} trained on part of the Google News data set that contains approximately 100 billion words (\url{https://code.google.com/archive/p/word2vec/}). This encoder is only able to give vector representations of words. As a result, when encoding a sentence for example, we get a separate vector for each word in the sentence. To obtain a single embedding for a sentence, we first remove stop words (e.g. the, and, $\dots$) and then take the element-wise average of the embedding vectors of the individual words \citep{Troxler2022}. We also use the Universal Sentence Encoder (USE) trained on Wikipedia, web news, web question-answer pages and discussion forums \citep{USE} (\url{https://tfhub.dev/google/collections/universal-sentence-encoder/1}). There are two different versions, v4 and v5, which are specifically designed to encode greater-than-word length text (i.e. sentences, phrases or short paragraphs). In our paper, we use both versions.

\begin{figure}[!htbp]
	\centering
	\caption{\label{fig:tSNE}Low-dimensional visualization of all embedding vectors at the \texttt{subsection} level, resulting from the pre-trained Word2Vec model, encoding the textual labels of the categories (see \Cref{fig:AvgDR}). The text boxes display the textual labels. The blue dots connected to the boxes depict the position in the low-dimensional representation of the embeddings.}
	\makebox[\textwidth][c]{\includegraphics[width = \textwidth]{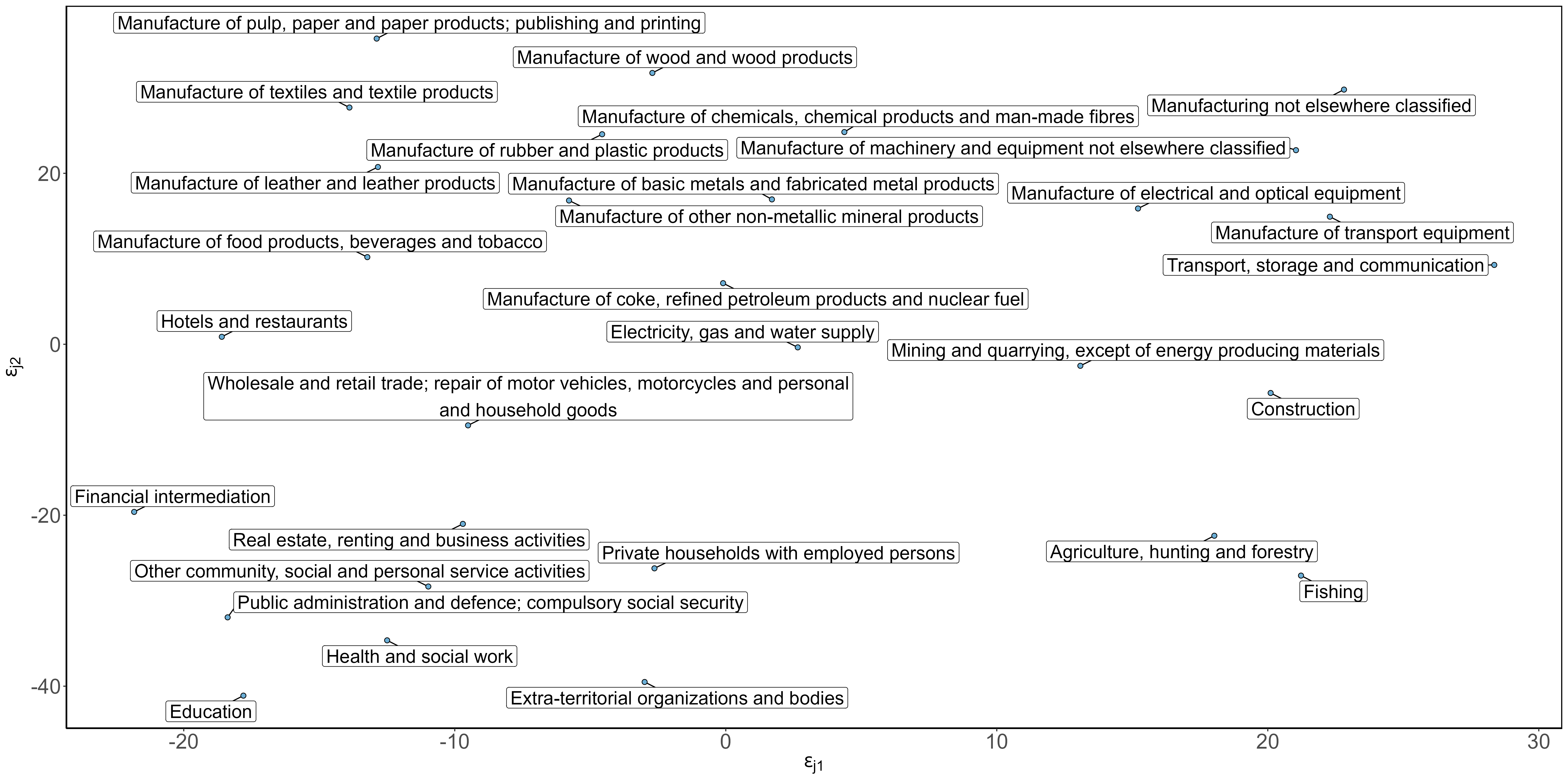}}
\end{figure}

The pre-trained Word2Vec encoder outputs a 300-dimensional embedding vector and the USEs a 512-dimensional vector. To assess the quality of the resulting embeddings, we inspect whether embeddings of related economic activities lie close to each other in the vector space. We employ the dimension reduction technique t-distributed stochastic neighbour embedding (t-SNE) \citep{tSNE} to obtain a two-dimensional visualization of the embeddings constructed for different categories at the \texttt{subsection} level (see \Cref{fig:AvgCF}). We opt to reduce the dimensionality to two dimensions, since it allows for an easy representation of the information in a scatterplot. t-SNE maps the high-dimensional embedding vector $\bs{e}_j$ for category $j$ into a lower dimensional representation $\bs{\epsilon}_j = (\epsilon_{j1}, \epsilon_{j2})$ whilst preserving its structure, enabling the visualization of relationships and the identification of patterns and groups. \Cref{fig:tSNE} visualizes the low-dimensional representation of the embeddings resulting from the pre-trained Word2Vec encoder (see Appendix \ref{App:tSNEFig} for similar figures of USE v4 and v5). 

In \Cref{fig:tSNE} we see that the embeddings of economically similar activities lie close to each other. All manufacturing related activities are situated at the top of the plot and in the left bottom corner, we have mostly activities that have a social component (e.g., education). In the right bottom corner, we have activities related to the exploitation of natural resources. Further, the more unrelated the activities are, the bigger the distance between their embedding vectors. For example, \textit{financial intermediation} ($\ev{\epsilon_{j1}, \epsilon_{j2}} \approxeq \ev{-22.5, -20}$) and \textit{transport, storage and communication} ($\ev{\epsilon_{j1}, \epsilon_{j2}} \approxeq \ev{28, 10}$) lie at opposite sides of the plot.

\subsection{Clustering subsections and tariff groups using PHiRAT}\label{subsec:ClusteringExpl}
We illustrate the effectiveness of PHiRAT by applying it to cluster categories at the \texttt{subsection} and \texttt{tariff group} level, using the training set. We slightly adjust Algorithm \ref{algo:GeneralProcedure} to ensure that each of the resulting categories at the \texttt{subsection} and \texttt{tariff group} level has sufficient salary mass. The minimum salary mass is based on previous analyses of the insurance company and we need to adhere to this minimum during clustering. Further, as discussed at the end of \Cref{sec:Clustering}, we run PHiRAT with all possible combinations of the clustering methods (i.e. k-medoids, spectral clustering and HCA) and internal validation criteria (i.e.~Cali{\'n}ski-Harabasz index, Davies-Bouldin index, Dunn index and silhouette index). Per run, a specific combination is used to cluster the categories at all levels in the hierarchy.

\begin{figure}[!htbp]
	\centering
	\caption{\label{fig:SelectClustersIllustration}Visualization of the search grid. The $x$-axis depicts the tuning grid $\mc{K}_1$ and the $y$-axis the encoder-specific feature matrix that is used. Here, we use k-medoids for clustering and the CH index as internal validation measure. The CH index is highest ($= 11.913$) for $\mc{K}_1 = 19$ in combination with $\mc{F}_1^{\text{USEv5}}$, indicating that this combination results in the most optimal clustering solution.}
	\makebox[\textwidth][c]{\includegraphics[width = \textwidth]{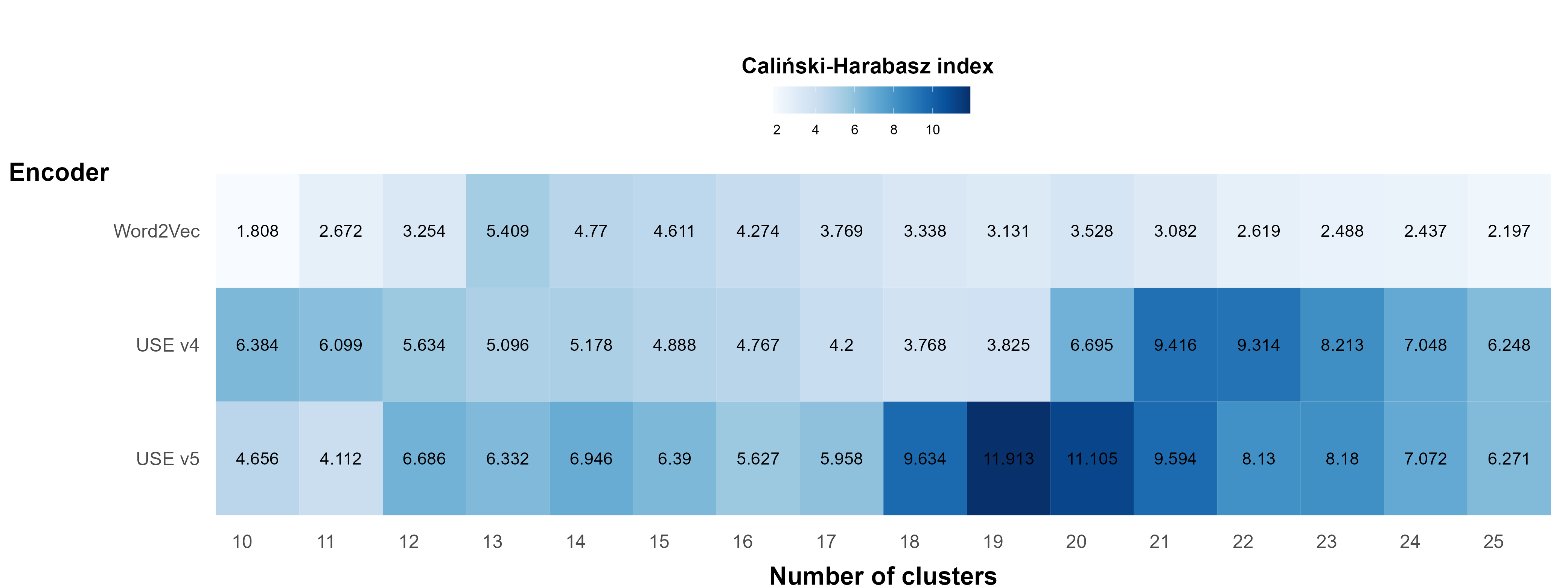}}
\end{figure}

\Cref{fig:ExamplePHiRAT} visualizes how PHiRAT works its way down the hierarchy. We start at the \texttt{subsection} level and construct the feature matrix $\mc{F}_1$. Part of $\mc{F}_1$ is built using an encoder to capture the textual information (see \Cref{subsec:featureengineering} and \Cref{subsec:Embeddings}). In our paper, we consider three different pre-trained encoders: (a) Word2Vec; (b) USE v4 and (c) USE v5. We use these encoders to obtain the embeddings and construct three separate feature matrices: (a) $\mc{F}_1^{\text{Word2Vec}}$; (b) $\mc{F}_1^{\text{USEv4}}$ and (c) $\mc{F}_1^{\text{USEv5}}$. All three feature matrices contain the same \replaced[id=B]{predicted}{estimated} random effects vectors $_1\widehat{\bs{U}}^d$ and $_1\widehat{\bs{U}}^f$. The difference between the feature matrices is that the embeddings are encoder-specific. Next, we define a tuning grid for the number of clusters $\mc{K}_1 \in (10, 11, \dots, 24, 25)$. \added[id = B]{Hence, the lower and upper bound to the number of grouped categories is determined by the minimum and maximum value of the tuning grid, respectively.} In combination with the encoder-specific feature matrices, this results in a search grid of three (i.e. the encoder-specific feature matrices) by 16 (i.e. the possible values for the number of clusters). For every combination in this grid, we run the clustering algorithm to obtain a clustering solution and calculate the value of the internal validation criterion. We select the combination that results in the most optimal clustering solution according to the validation measure. \Cref{fig:SelectClustersIllustration} provides a visualization of this search grid. In this figure, we use k-medoids as clustering algorithm, the CH index as internal validation measure and we use a colour gradient for the value of the validation criterion. The darker the colour, the higher the value and the better the clustering solution. The $x$-axis depicts the tuning grid for $\mc{K}_1$ and the $y$-axis the encoder-specific feature matrices. The CH index is highest ($= 11.913$) for $\mc{K}_1 = 19$ in combination with $\mc{F}_1^{\text{USEv5}}$. According to the selected validation measure, this specific combination results in the most optimal clustering solution. Hence, we use this clustering solution to group the categories $j = (1, \dots, J)$ into clusters $j' = (1, \dots, J')$ (see \Cref{fig:ExamplePHiRAT}). Hereafter, we merge clusters $j' = (1, \dots, J')$ with neighbouring clusters until each cluster has sufficient salary mass.

Next, we proceed to cluster the categories at the \texttt{tariff group} level. Within each cluster $j'$ at the \texttt{section} level, we first group consecutive categories (i.e. those with consecutive NACE codes, see \Cref{subsec:NACE}) at the \texttt{tariff group} level to ensure that the salary mass is sufficient for every category. As before, we construct three encoder-specific versions of $\mc{F}_2$. We then iterate over every parent category $j'$ to group the child categories. At iteration $j'$, we use $\mc{F}_{2, \{c : \pi(c) = j'\}}$ as input in a clustering algorithm. At the \texttt{tariff group} level, we use the tuning grid $\mc{K}_{2, \{c : \pi(c) = j'\}} \in (5, 6, \dots, \ \min(K_{j'}, 25))$. Here $K_{j'}$ denotes the total number of child categories of parent category $j'$. \added[id = B]{Similarly, the lower and upper bound to the number of grouped child categories is dependent on the minimum and maximum value in the tuning grid.} We select the combination of the encoder-specific feature matrix and $\mc{K}_{2, \{c : \pi(c) = j'\}}$ that results in the most optimal clustering solution to group the categories $j'k = (j'1, \dots, j'K_{j'})$, nested within $j'$, into subclusters $j'k' = (j'1, \dots, j'K'_{j'})$ (see \Cref{fig:ExamplePHiRAT}).

\subsubsection{Distance measures and evaluation metrics} For k-medoids clustering and HCA we use the angular distance (see Section \ref{sec:ClusteringAlgorithms}), as both algorithms require a distance or dissimilarity measure. For spectral clustering we use the angular similarity. We also need to define a distance measure $d(\cdot, \cdot)$ for the selected internal evaluation criteria, except for the CH index. Hereto, we define $d(\cdot, \cdot)$ as the angular distance. We calculate the evaluation criteria using all engineered features. However, since a large part of the feature vector consists of the high-dimensional embedding vector, too much weight might be given to the similarity in economic activity when choosing the optimal cluster solution. Therefore, we also evaluate a second variation of the internal evaluation criteria. When calculating the criteria, we remove the embedding vectors from the feature matrices. As such, we focus on constructing a clustering solution that is most optimal in terms of riskiness. In this evaluation we use the Euclidean distance for $d(\cdot, \cdot)$. Hence, at the \texttt{subsection} level, we define $d(\bs{x}_j, \bs{x}_{\mb{j}}) = \| \bs{x}_j - \bs{x}_{\mb{j}} \|^2_2$ where $\bs{x}_{j} = (\widehat{U}^d_j, \widehat{U}^f_j)$. Similarly, at the \texttt{tariff group} level, $d(\bs{x}_{jk}, \bs{x}\sub[sbvmove = -1pt]{j\mathscr{k}}) = \| \bs{x}_{jk} - \bs{x}\sub[sbvmove = -1pt]{j\mathscr{k}} \|^2_2$ and $\bs{x}_{jk} = (\widehat{U}^d_{jk}, \widehat{U}^f_{jk})$. In what follows we discuss the results obtained with this approach. Results obtained with the complete feature vector, including the embedding, are given in Appendix \ref{App:PredPerfAngular}.

\subsubsection{Implementation} We perform the main part of the analysis using the statistical software R \citep{Rsoftware}. For k-medoids and HCA, we rely on the \verb|cluster| \citep{clusterPackage} and \verb|stats| package, respectively. For spectral clustering, we follow the implementation of \citet{Ng2001} and developed our own code. Spectral clustering partially depends on the k-means algorithm to obtain a clustering solution (see Appendix \ref{App:ClusteringAlgorithms}). However, k-means is sensitive to the initialization and can get stuck in an inferior local minimum \citep{Franti2019}. One way to alleviate this issue is by repeating k-means with different initializations and to select the most optimal clustering solution. Hence, to prevent that spectral clustering results in a suboptimal clustering solution, we repeat the k-means step a 100 times.

\subsection{Evaluating the clustering solution}
The main aim of our procedure is to reduce the cardinality of a hierarchically structured categorical variable, which can then be incorporated as a risk factor in a predictive model to underpin the technical price list \citep{Campo2023}. Ideally, we maintain the predictive accuracy whilst reducing the granularity of the risk factor. As discussed in \Cref{subsec:ClusteringExpl}, we employ PHiRAT to group similar categories at the \texttt{subsection} and \texttt{tariff group} level. We refer to the less granular version of the hierarchical risk factor as the reduced risk factor.

To assess the predictive accuracy of a clustering solution, we fit an LMM
\begin{equation}\label{eq:EvaluationClustering}
	\begin{aligned}
		E[Y_{ij'k't} | U_{j'}^d, U_{j'k'}^d] = \mu^d + U_{j'}^{d} + U_{j'k'}^{d}\
	\end{aligned}
\end{equation}
\noindent
where the reduced risk factor enters the model through the random effects $U_{j'}$ and $U_{j'k'}$. We include the salary mass $w_{ij'k't}$ as weight. We opt for an LMM given its simplicity and computational efficiency \citep{Campo2023}. Next, we calculate the predicted damage rate as $\widehat{Y}_{ij'k't} = \hat{\mu}^d + \widehat{U}_{j'}^d + \widehat{U}_{j'k'}^d$ for the training and test set as introduced in the beginning of \Cref{sec:CaseStudy}.

\subsubsection{Benchmark clustering solution} To evaluate the clustering solution obtained with the proposed data-driven approach, we need to compare it to a benchmark where we do not rely on PHiRAT. One possibility is to fit \eqref{eq:EvaluationClustering} with the nominal variable composed of the original categories at the \texttt{subsection} and \texttt{tariff group} level. \replaced[id = B]{In our example, fitting this LMM}{However, this} results in negative variance estimates and yields incorrect random effect \replaced[id=B]{predictions}{estimates}. Consequently, to obtain a benchmark clustering solution, we start at the \texttt{subsection} level and merge consecutive categories until the salary mass is sufficient (e.g. \textit{01 agriculture} and \textit{02 forestry}). \added[id=B]{Similarly, we index the merged categories using $j' = (1, \dots, J')$.} Hereafter, within each of the (grouped) categories at the \texttt{subsection} level, we group consecutive categories (e.g. \textit{142121} and \textit{142122}) at the \texttt{tariff group} level. Again, we use the minimum salary mass as defined by the insurance company \added[id=B]{and we use $k' = (1, \dots, K'_{j'})$ as an index for the grouped categories. This results in a hierarchical MLF in which we merged adjacent categories with insufficient salary mass. To construct the benchmark model, we fit an LMM with the same specification as in \eqref{eq:EvaluationClustering}.}

\subsubsection{Performance measures} Using the predicted damage rates on the training and test set, we compute the Gini-index \citep{Gini1921} and loss ratio. The Gini-index assesses how well a model is able to distinguish high-risk from low-risk companies and is considered appropriate for the comparison of competing pricing models \citep{Denuit2019, Campo2023}. The higher the value, the better the model can differentiate between risks. The Gini-index has a maximum theoretical value of 1. The loss ratio measures the overall accuracy of the fitted damage rates and is defined as $\mathcal{Z}^{tot}_t / \widehat{\mathcal{Z}}^{tot}_t$, where $\mathcal{Z}^{tot}_t = \sum_{i, j', k'} \mathcal{Z}_{ij'k't}$ denotes the total capped claim amount and $\widehat{\mathcal{Z}}^{tot}_t = \sum_{i, j', k'} \widehat{\mathcal{Z}}_{ij'k't}$ the total fitted claim amount.
To obtain $\widehat{\mathcal{Z}}_{ij'k't}$, we transform the individual predictions $\widehat{Y}_{ij'k't}$  as (see equation \eqref{eq:DamageRate})
\begin{equation}
	\begin{aligned}
		\widehat{\mathcal{Z}}_{ij'k't} = \widehat{Y}_{ij'k't} \cdot w_{ij'k't}.
	\end{aligned}
\end{equation}
\noindent
Due to the balance property \citep{Campo2023}, the loss ratio is one when calculated using the training data set. We therefore compute the loss ratio only for the test set.

\subsubsection{Predictive performance} \Cref{tab:PredPerf} depicts the performance of the different clustering solutions on the training and test sets. In this table, $J'$ denotes the total number of grouped categories at the \texttt{subsection} level and $\sum_{j' = 1}^{J'} K'_{j'}$ denotes the total number of grouped categories at the \texttt{tariff group} level. With the benchmark clustering solution, we end up with a large number of different categories at both hierarchical levels ($18 + 641 = 659$ separate categories in total). Conversely, with our data-driven approach the maximum is 221 (14 + 207; k-medoids with Davies-Bouldin index) and the minimum is 97 (10 + 87; k-medoids with the CH index). Consequently, PHiRAT substantially reduces the number of categories. Moreover, we are able to retain the predictive performance. On both the training and test set, the Gini-indices of nearly all clustering solutions are higher than the Gini-index of the benchmark solution. Hence, we are better able to differentiate between high- and low-risk companies with the clustering solutions. On the training data set, the highest Gini-indices are seen for the k-medoids algorithm using the silhouette index and the spectral clustering algorithm using the Dunn index. On the test set, spectral clustering using the CH index and silhouette index result in the highest Gini-index whereas the benchmark has the lowest Gini-index. However, the loss ratio of most clustering solutions is higher than the loss ratio of the benchmark clustering solution (= 1.006). This indicates that, when we use the reduced risk factor in an LMM, we generally underestimate the total damage. One exception is the clustering solution resulting from HCA using the Davies-Bouldin index, which has the same loss ratio as the benchmark (= 1.006). In addition, for this solution the Gini-index is among the highest (0.675 on the training and 0.616 on the test set). Using the result from HCA with the Davies-Bouldin index, we are able to reduce the total number of categories to 207 (= 14 + 193). Compared to the benchmark solution, we obtain the same overall predictive accuracy (i.e. the loss ratio is approximately equal) and we can better differentiate between high- and low-risk companies.

\begin{table}[!htbp]
	\centering
	\caption{Predictive performance on the training and test set.} 
	\label{tab:PredPerf}
	\begin{tabular}{@{\extracolsep{4pt}}lccccc@{}}
		\hline
		& & & Training & \multicolumn{2}{c}{Test}\\
		\cline{4-4} \cline{5-6}
		& $J'$ & $\sum_{j' = 1}^{J'} K'_{j'}$ & Gini-index & Gini-index & Loss ratio\\ 
		\hline
		Benchmark &   18 &  641 & 0.658 & 0.585 & 1.006\\ 
		HCA:&&&&&\\
		\hspace{1mm} Silhouette index &   13 &  100 & 0.667 &  0.596 & 1.007\\ 
		\hspace{1mm} Dunn index &   15 &  202 & 0.656 &  0.599 & 1.011 \\ 
		\hspace{1mm} Davies-Bouldin index &   14 &  193 & 0.675 &  0.616 & 1.006\\ 
		\hspace{1mm} CH index &   15 &  140 & 0.667 &  0.614 & 1.010\\ 
		
		k-medoids:&&&&&
		\\\hspace{1mm} Silhouette index &   12 &  107 & 0.678 & 0.613 &1.008  \\ 
		\hspace{1mm} Dunn index &   10 &  130 & 0.657 &  0.598 &  1.010\\ 
		\hspace{1mm} Davies-Bouldin index &   14 &  207 & 0.670 & 0.619 &  1.010 \\ 
		\hspace{1mm} CH index &   10 &   87 & 0.676 & 0.605 & 1.011 \\ 
		
		Spectral clustering:&&&&&\\
		\hspace{1mm} Silhouette index & 12 & 86 & 0.673 & 0.628 & 1.012 \\ 
		\hspace{1mm} Dunn index & 12 &  143 & 0.677 & 0.624 &  1.010 \\  
		\hspace{1mm} Davies-Bouldin index & 12 &  166 &0.669 & 0.618 & 1.010 \\  
		\hspace{1mm} CH index &  12 &  102 & 0.670 & 0.628 & 1.013  \\  
		\hline
	\end{tabular}
\end{table}

\begin{figure}[!htbp]
	\centering
	\caption{\label{fig:HierarchicalStructureDR_ClusterSolution}Cluster-specific weighted average damage rates at the \texttt{subsection} and \texttt{tariff group} level, when employing PHiRAT with spectral clustering and the silhouette index.}
	\makebox[\textwidth][c]{\includegraphics[width = \textwidth]{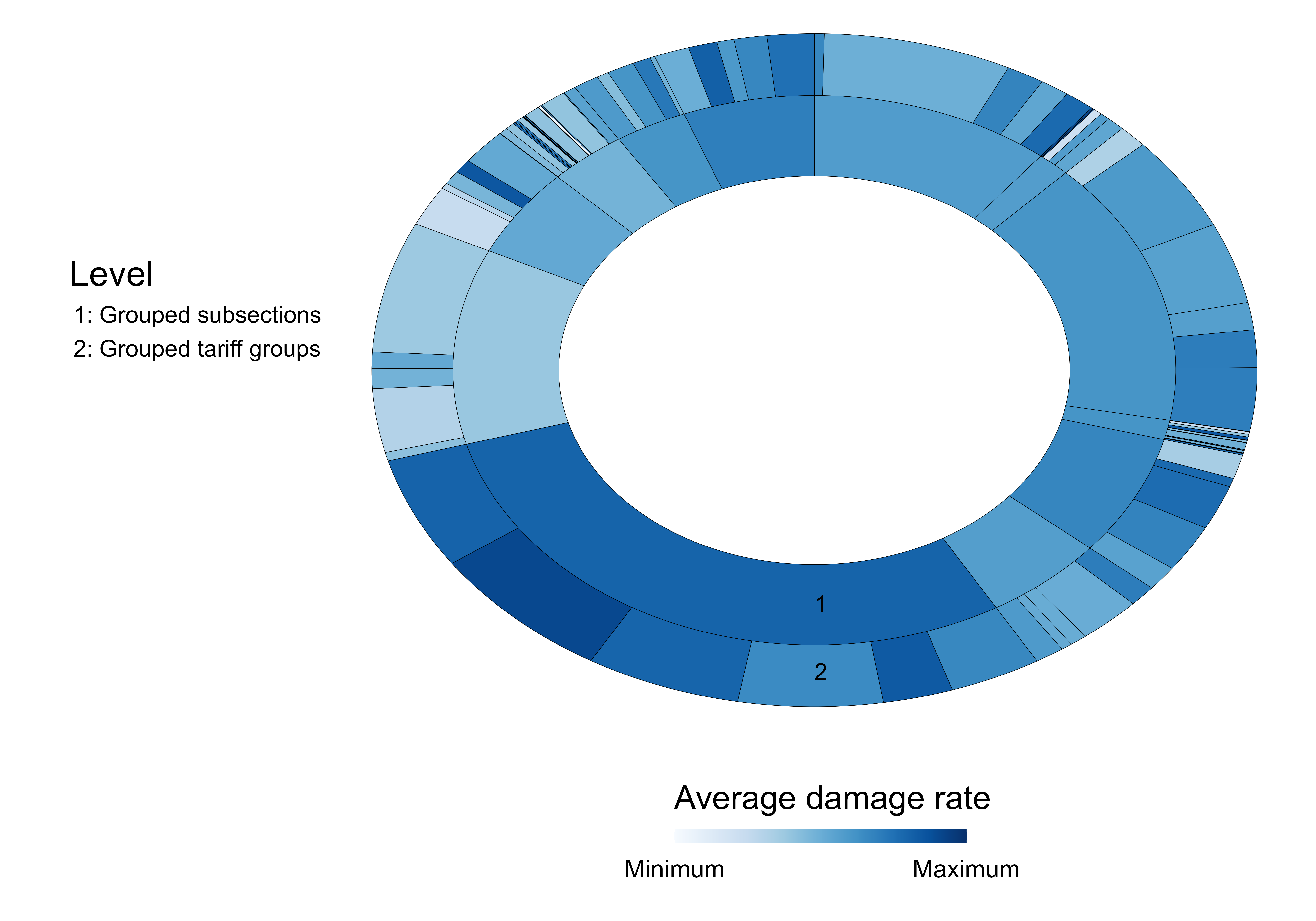}}
\end{figure}

The clustering solution resulting from HCA with the Davies-Bouldin index offers a good balance between reducing granularity and enhancing differentiation whilst preserving predictive accuracy. If, however, a sparse representation and good differentiation is more important than the overall predictive accuracy, the result from spectral clustering with the silhouette index is a better option. The latter clustering solution is visualized in \Cref{fig:HierarchicalStructureDR_ClusterSolution}. This figure is similar to \Cref{fig:AvgDR}(a) and depicts the cluster-specific weighted average damage rates at the \texttt{subsection} and \texttt{tariff group} level. The two figures show that PHiRAT substantially reduces the total number of categories (12 + 86 = 98). Furthermore, spectral clustering with the silhouette index is one of the combinations that results in the best differentiation between high- and low-risk companies (Gini index is 0.673 and 0.628 on the training and test set, respectively).

\subsubsection{Interpretability} After clustering, it is imperative to evaluate the interpretability of the solution. The grouping should provide us with meaningful insights into the underlying structure of the data. Building on the strong predictive performance of the solution resulting from spectral clustering with the silhouette index, we examine this specific clustering solution in more detail.

\begin{figure}[!htbp]
	\centering
	\caption{\label{fig:ClusteringSolutionl1}Visualization of the grouped categories at the \texttt{subsection} level (see \Cref{fig:ExamplePHiRAT}): a) the clustered categories and the original random effect \replaced[id=B]{predictions}{estimates} $\widehat{U}_j^d$ and $\widehat{U}_j^f$; b) the description of the economic activity of the categories.}
	\makebox[\textwidth][c]{\includegraphics[width = \textwidth]{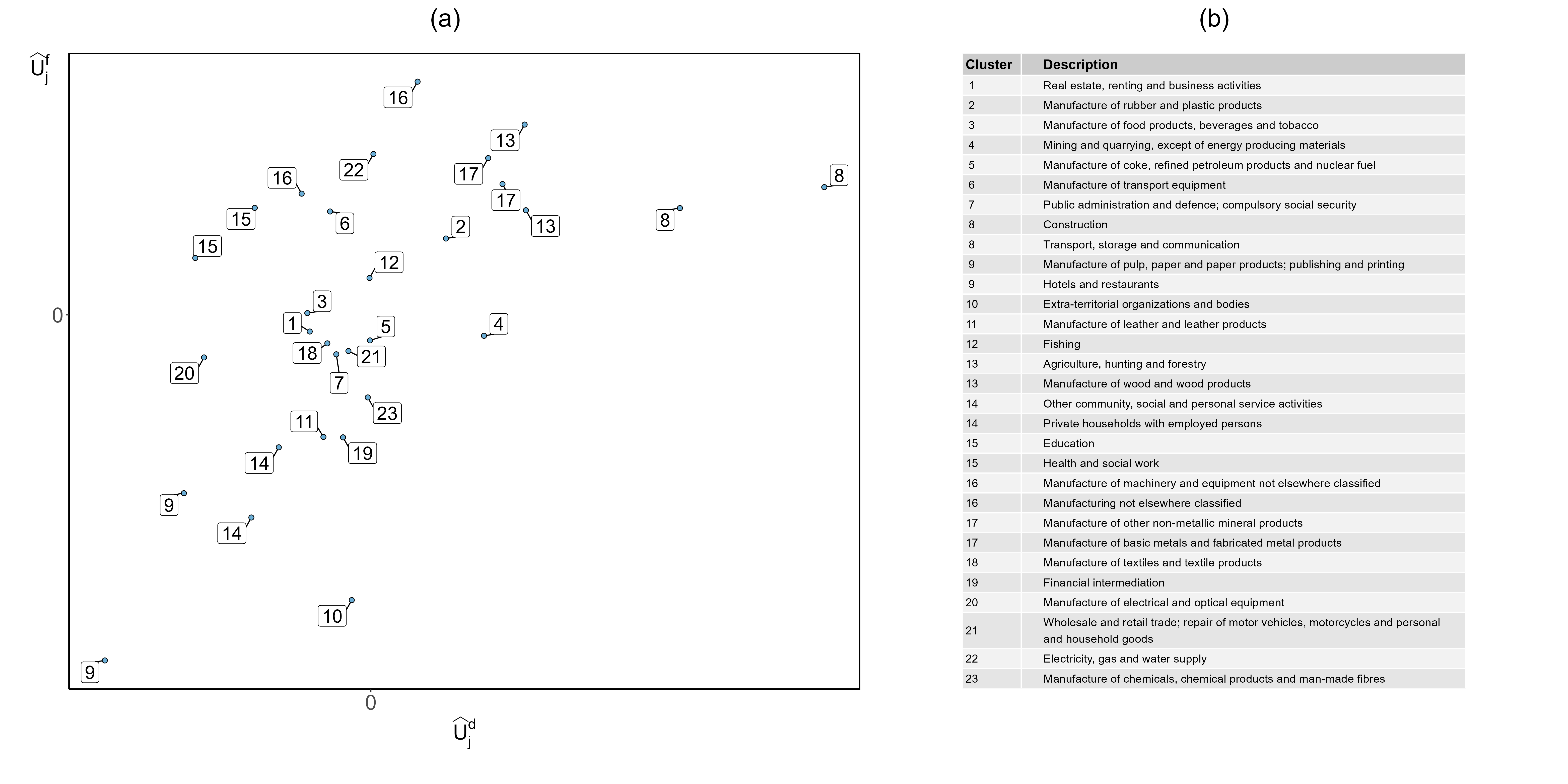}}
\end{figure}

\Cref{fig:ClusteringSolutionl1} visualizes which categories are clustered at the \texttt{subsection} level. At this point we have not yet merged neighbouring clusters to ensure sufficient salary mass (see \Cref{subsec:ClusteringExpl}). \Cref{fig:ClusteringSolutionl1}(a) shows the $\widehat{U}_j^d$ and $\widehat{U}_j^f$ of the fitted damage rate and claim frequency random effects models (see \eqref{eq:REDR1} and \eqref{eq:RECF1}). The number in the plot indicates which cluster the categories are appointed to. The description of the category's economic activity is given in \Cref{fig:ClusteringSolutionl1}(b). In total we have 24 clusters, 19 of which consist of a single category. Inspecting the clusters in detail, we find that the riskiness and economic activity of grouped categories are similar. Moreover, categories that are similar in terms of riskiness (i.e. those with similar random effect \replaced[id=B]{predictions}{estimates}) but different in economic activity are assigned to different clusters. For example, the $\widehat{U}_j^d$'s and $\widehat{U}_j^f$'s are similar for: a) \textit{manufacture of rubber and plastic products}; b) \textit{agriculture, hunting and forestry}; c) \textit{manufacture of wood and wood products}; d) \textit{manufacture of other non-metallic mineral product} and e) \textit{manufacture of basic metals and fabricated metal products}. These industries are partitioned into three clusters. The category in cluster 2 (i.e. \textit{manufacture of rubber and plastic products}) manufactures organic products. Conversely, the categories in cluster 17 (i.e. \textit{manufacture of other non-metallic mineral product} and \textit{manufacture of basic metals and fabricated metal products}) use inorganic materials and the categories in cluster 13 (i.e. \textit{agriculture, hunting and forestry} and \textit{manufacture of wood and wood products}) utilize products derived from plants and animals.

\begin{figure}[!htbp]
	\centering
	\caption{\label{fig:ClusteringSolutionl2}Visualization of the clustered child categories at the \texttt{tariff group} level, with parent category \textit{manufacture of chemicals, chemical products and man-made fibres} at the \texttt{subsection} level (see \Cref{fig:ExamplePHiRAT}): a) the clustered categories and the original random effect \replaced[id=B]{predictions}{estimates} $\widehat{U}_{j'k}^d$ and $\widehat{U}_{j'k}^f$; b) the description of the economic activity of the categories.}
	\makebox[\textwidth][c]{\includegraphics[width = \textwidth]{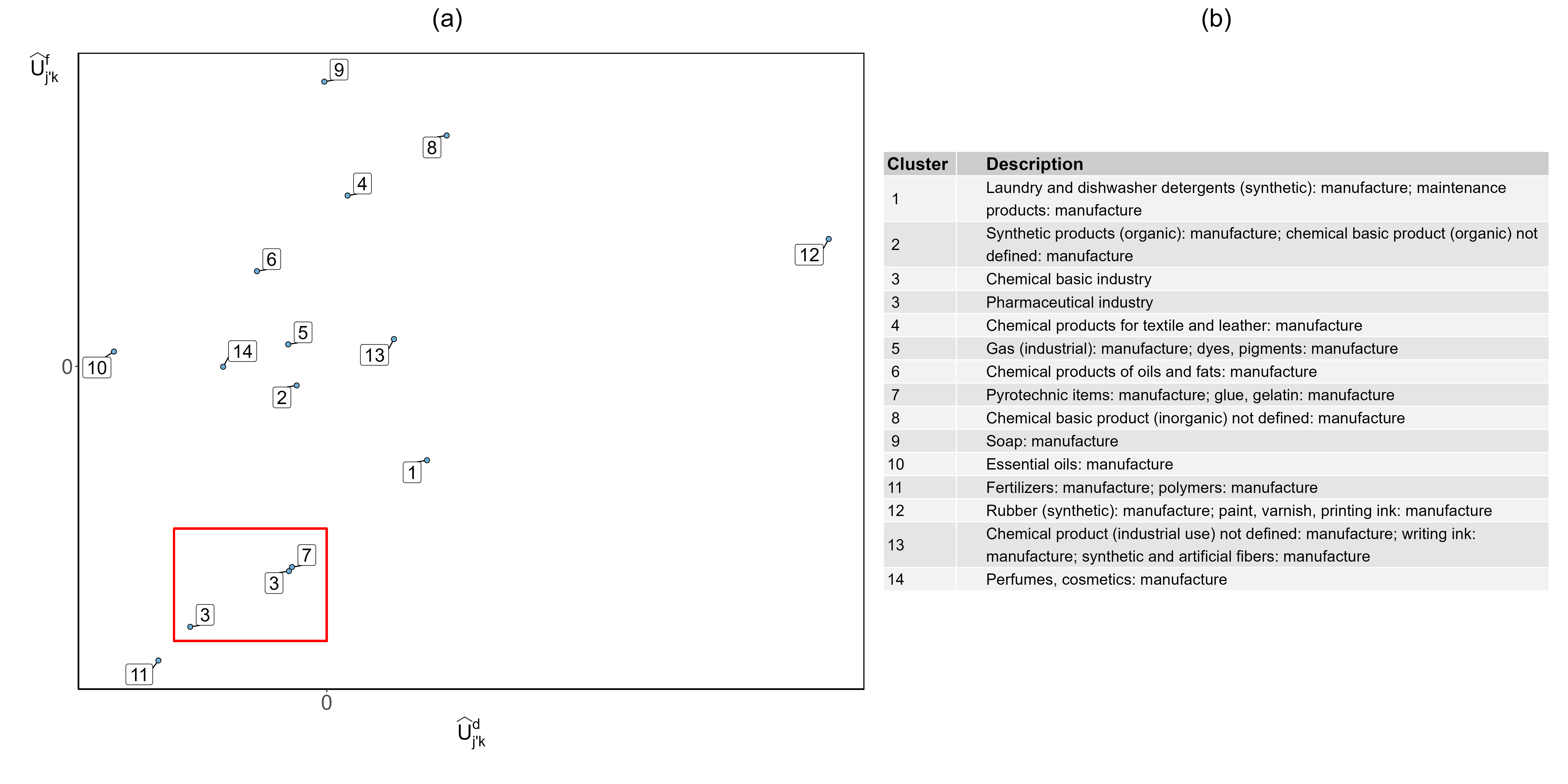}}
\end{figure}

\Cref{fig:ClusteringSolutionl2} depicts the clustered categories $k'$ at the \texttt{tariff group} level, nested within category $j' =$ \textit{manufacture of chemicals, chemical products and man-made fibres} at the \texttt{subsection} level. The $\widehat{U}_{j'k}^d$ and $\widehat{U}_{j'k}^f$ of the fitted damage rate and claim frequency random effects model (see \eqref{eq:REDR2} and \eqref{eq:RECF2}) are shown in the left panel of \Cref{fig:ClusteringSolutionl2} and the textual information in the right panel. When the text is separated by a semicolon, this indicates that categories are merged before clustering to ensure sufficient salary mass (see \Cref{subsec:ClusteringExpl}). Similar to \Cref{fig:ClusteringSolutionl1}, both the riskiness and economic activity are taken into account when grouping categories. This is best illustrated for the categories in the red rectangle in \Cref{fig:ClusteringSolutionl2}(a). Herein, we have three categories: (a) \textit{pyrotechnic items: manufacture; glue, gelatin: manufacture}; (b) \textit{chemical basic industry} and (c) \textit{pharmaceutical industry}. Of these, \textit{pyrotechnic items: manufacture; glue, gelatin: manufacture} (number 7 in the top right corner of the red rectangle) and \textit{chemical basic industry} (number 3 in the top right corner of the red rectangle) have nearly identical random effect \replaced[id=B]{predictions}{estimates}. Notwithstanding, both categories are not grouped. Instead, the category \textit{chemical basic industry} is merged with \textit{pharmaceutical industry} (number 3 in the bottom left corner). The latter two categories manufacture chemical compounds. In addition, companies in \textit{chemical basic industry} mainly produce chemical compounds that are used as building blocks in other products (e.g. polymers). Building blocks (or raw materials) that are needed by companies involved in manufacturing pyrotechnic items, glue and gelatin (e.g. glue is typically made from polymers \citep{Glue}).

			\section{Discussion}\label{sec:Discussion}
This paper presents the data-driven PHiRAT approach to reduce a hierarchically structured categorical variable with a large number of categories to its essence, by grouping similar categories at every level in the hierarchy. PHiRAT is a top-down procedure that preserves the hierarchical structure. It starts with grouping categories at the highest level in the hierarchy and proceeds to lower levels. At a specific level in the hierarchy, we first engineer several features to characterize the profile and specificity of each category. Using these features as input in a clustering algorithm, we group similar categories. The procedure stops once we grouped the categories at the lowest level in the hierarchy. When deployed in a predictive model, the reduced structure leads to a more parsimonious model that is easier to interpret, less likely to experience estimation problems or to overfit. Further, the increased number of observations of grouped categories leads to more precise estimates of the category's effect on the response.

Using a workers' compensation insurance portfolio from a Belgian insurer, we illustrate how to employ PHiRAT to reduce the granular structure of the NACE code. Using PHiRAT, we are able to substantially reduce the dimensionality of this hierarchical risk factor whilst maintaining its predictive accuracy. The reduced risk factor allows for better differentiation, has the same overall precision as the original risk factor and the grouping seems to generalize well to out-of-sample data. Moreover, the resulting clusters consist of categories that are similar in terms of riskiness and economic activity. Furthermore, the results show that embeddings are an efficient and effective method to capture textual information.

\added[id = B]{Our approach results in a clustering solution that can provide insurers with improved insights into the underlying risk structure of a hierarchical covariate. By capturing the key characteristics of the original hierarchically structured risk factor, the reduced risk factor offers a more informative and concise representation of the various risk profiles. Insurers can incorporate the reduced version into their pricing algorithms to better assess the riskiness associated with each category in said hierarchical covariate.
}

\added[id = B]{Negative variance estimates in the random effects model and computational limitations with the generalized fused lasso penalty prevent us from constructing a different benchmark model. Hence, future research can examine the robustness of our approach by examining its performance in other data sets or applications and by comparing it with alternative approaches. Additionally,} \replaced[id=B]{f}{F}uture research can aim to improve our proposed solution in multiple directions. The final clustering solution depends on the engineered features, the clustering algorithm and the cluster evaluation criterion. Constructing appropriate and reliable features when employing PHiRAT is crucial to obtain a good clustering solution. In our paper, we rely on random effects models to capture the riskiness of the categories. Alternatively, we can characterize the risk profile using entity embeddings \citep{Guo2016}. Here, we train a neural network to create entity embeddings that map the categorical values to a continuous vector, ensuring that categories with similar response values are located closer to each other in the embedding space. Additionally, we advise to run PHiRAT using different clustering algorithms and cluster evaluation criteria, since no algorithm consistently outperforms the others \citep{Hennig2015, McNicholas2016a, McNicholas2016b, Murugesan2021}. \added[id = B]{Moreover, the robustness of the clustering solution largely depends on the stability of the selected clustering method.} Further, NLP is a rapidly evolving field and we only considered three pre-trained encoders. We did not include the pre-trained Bidirectional Encoder Representations from Transformers (BERT) model \citep{BERT}, an encoder that is being increasingly used by actuarial researchers (see, for example, \citet{Xu2022, Troxler2022}). Subsequent studies can assess whether BERT-based models result in higher quality embeddings, which in turn leads to better clustering solutions. In addition, researchers can employ alternative approaches to represent the parent-child relationships between categories in hierarchically structured data. Such as \citet{Argyrou2009}, for example, who formalizes the hierarchical relations using graph theory and applies a self-organizing map \citep{Kohonen1995} to obtain a reduced representation of the hierarchical data.

			\section*{Acknowledgements}
\textcolor{red}{We sincerely thank the associate editor and referees for their invaluable feedback and constructive comments, which significantly improved this manuscript. We also express our appreciation to the editor for their guidance and support throughout the review process. We are especially grateful to W.S., who provided us with valuable insights and expertise that helped shape our research. Katrien Antonio gratefully acknowledges funding from the FWO and Fonds De La Recherche Scientifique - FNRS (F.R.S.-FNRS) under the Excellence of Science (EOS) program, project ASTeRISK Research Foundation Flanders [grant number 40007517]. The authors gratefully acknowledge support from the Ageas research chair on insurance analytics at KU Leuven, from the Chaire DIALog sponsored by CNP Assurances and the FWO network W001021N.}

			\section*{References}
			\begingroup
			\renewcommand{\section}[2]{}%
			\renewcommand\harvardurl{}
			\bibliographystyle{agsmAdj}
			\bibliography{ReferencesAdj}

@book{Ohlsson,
series = {EAA Lecture Notes},
issn = {3-642-10790-7},
publisher = {Springer Berlin Heidelberg : Imprint: Springer},
isbn = {1280391626},
year = {2010},
title = {Non-Life Insurance Pricing with Generalized Linear Models},
language = {eng},
address = {Berlin, Heidelberg},
author={Ohlsson, E. and Johansson, B.},
}

@article{Ohlsson2008,
issn = {0346-1238},
journal = {Scandinavian Actuarial Journal},
pages = {301--314},
volume = {2008},
publisher = {Taylor \& Francis Group},
number = {4},
year = {2008},
title = {Combining generalized linear models and credibility models in practice},
author = {Ohlsson, Esbj{\"o}rn},
}

@article{Gertheiss2010,
issn = {1932-6157},
journal = {Ann. Appl. Stat.},
pages = {2150--2180},
volume = {4},
publisher = {The Institute of Mathematical Statistics},
number = {4},
year = {2010},
title = {Sparse modeling of categorial explanatory variables},
language = {eng},
author = {Gertheiss, Jan and Tutz, Gerhard},
}

@techreport{Reacfin,
year = {2017},
author = {Stassen, B. and Denuit, M. and Mahy, S. and Mar{\'e}chal, X. and J. Trufin},
title = {A unified approach for the modelling of rating factors in workers compensation insurance},
institution = {Reacfin},
type = {White paper},
note = {\url{https://www.reacfin.com/wp-content/uploads/2016/12/170131-Reacfin-White-Paper-A-Unified-Approach-for-the-Modeling-of-Rating-Factors-in-Work-ers\%E2\%80\%99-Compensation-Insurance.pdf}}
}

@article{Haberman1996,
issn = {0039-0526},
journal = {Journal of the Royal Statistical Society: Series D (The Statistician)},
pages = {407--436},
volume = {45},
number = {4},
year = {1996},
title = {Generalized Linear Models and Actuarial Science},
author = {Haberman, Steven and Renshaw, Arthur E.},
}

@article{Oelker2014,
issn = {1471-082X},
journal = {Statistical Modelling},
pages = {157--177},
volume = {14},
publisher = {SAGE Publications},
number = {2},
year = {2014},
title = {Regularization and model selection with categorical predictors and effect modifiers in generalized linear models},
language = {eng},
address = {New Delhi, India},
author = {Oelker, Margret-Ruth and Gertheiss, Jan and Tutz, Gerhard},
}

@book{Molenberghs2005,
series = {Springer Series in Statistics},
publisher = {Springer New York},
isbn = {9780387251448},
year = {2005},
title = {Models for Discrete Longitudinal Data},
language = {eng},
address = {New York, NY},
author = {Molenberghs, Geert and Verbeke, Geert},
}

@article{Gini1921,
issn = {00130133},
journal = {The Economic Journal},
pages = {124--126},
volume = {31},
publisher = {MacMillan and Co. Limited},
number = {121},
year = {1921},
title = {Measurement of Inequality of Incomes},
language = {eng},
author = {Gini, Corrado},
}

@Manual{Rsoftware,
    title = {R: A Language and Environment for Statistical Computing},
    author = {{R Core Team}},
    organization = {R Foundation for Statistical Computing},
    address = {Vienna, Austria},
    year = {2022},
    url = {https://www.R-project.org/},
  }

@article{JewellModel,
journal = {Giornale dell'Istituto Italiano degli Attuari},
title = {The use of collateral data in credibility theory : a hierarchical model},
volume = {38},
pages = {1-16},
year = {1975},
author = {Jewell, William S},
}

@article{Henckaerts2018,
issn = {0346-1238},
journal = {Scandinavian Actuarial Journal},
pages = {681--705},
volume = {2018},
publisher = {Taylor \& Francis},
number = {8},
year = {2018},
title = {A data driven binning strategy for the construction of insurance tariff classes},
copyright = {2018 Informa UK Limited, trading as Taylor \&amp; Francis Group 2018},
language = {eng},
author = {Henckaerts, Roel and Antonio, Katrien and Clijsters, Maxime and Verbelen, Roel},
}

@article{Denuit2019,
issn = {0167-6687},
journal = {Insurance, Mathematics \& Economics},
pages = {128--139},
volume = {89},
publisher = {Elsevier B.V},
year = {2019},
title = {Model selection based on {Lorenz} and concentration curves, {Gini} indices and convex order},
copyright = {2019 Elsevier B.V.},
language = {eng},
author = {Denuit, Michel and Sznajder, Dominik and Trufin, Julien},
}

@article{Micci2001,
issn = {1931-0145},
journal = {SIGKDD Explorations},
pages = {27--32},
volume = {3},
publisher = {ACM},
number = {1},
year = {2001},
title = {A preprocessing scheme for high-cardinality categorical attributes in classification and prediction problems},
language = {eng},
author = {Micci-Barreca, Daniele},
}

@article{NACE,
author = {Eurostat},
series = {Methodologies and working papers},
issn = {1977-0375},
publisher = {Publications Office},
isbn = {9279047418},
year = {2008},
title = {{NACE Rev. 2}: statistical classification of economic activities in the {European} Community},
journal = {Eurostat: Methodologies and Working Papers},
language = {eng},
address = {Luxembourg},
}

@book{Gelman2017,
publisher = {Cambridge University Press},
isbn = {9780521686891},
year = {2017},
title = {Data analysis using regression and multilevel/hierarchical models},
edition = {17th pr.},
address = {Cambridge},
author = {Gelman, Andrew and Hill, Jennifer},
}

@misc{Mikolov,
      title={Efficient Estimation of Word Representations in Vector Space}, 
      author={Tomas Mikolov and Kai Chen and Greg Corrado and Jeffrey Dean},
      year={2013},
note = {arXiv: 1301.3781. Available at: \url{https://arxiv.org/abs/1301.3781}},
      eprint={1301.3781},
      archivePrefix={arXiv},
      primaryClass={cs.CL}
}

@misc{USE,
      title={{Universal Sentence Encoder}}, 
      author={Daniel Cer and Yinfei Yang and Sheng-yi Kong and Nan Hua and Nicole Limtiaco and Rhomni St. John and Noah Constant and Mario Guajardo-Cespedes and Steve Yuan and Chris Tar and Yun-Hsuan Sung and Brian Strope and Ray Kurzweil},
      year={2018},
      eprint={1803.11175},
note = {arXiv: 1803.11175. Available at: \url{https://arxiv.org/abs/1803.11175}},
      archivePrefix={arXiv},
      primaryClass={cs.CL}
}

@book{Kogan2006,
publisher = {Springer Berlin / Heidelberg},
booktitle = {Grouping Multidimensional Data},
isbn = {354028348X},
year = {2005},
title = {Grouping Multidimensional Data: Recent Advances in Clustering},
copyright = {Springer-Verlag Berlin Heidelberg 2006},
language = {eng},
address = {Berlin, Heidelberg},
author = {Kogan, Jacob and Nicholas, Charles and Teboulle, Marc},
}

@misc{Poon2012,
  doi = {10.48550/ARXIV.1210.4883},
note = {arXiv: 1210.4883. Available at: \url{https://arxiv.org/abs/1210.4883}},
  author = {Poon, Leonard K. M. and Liu, April H. and Liu, Tengfei and Zhang, Nevin Lianwen},
  title = {A Model-Based Approach to Rounding in Spectral Clustering},
  publisher = {arXiv},
  year = {2012},
  copyright = {arXiv.org perpetual, non-exclusive license}
}

@misc{Mohammed2012,
  doi = {10.48550/ARXIV.1203.1858},
  
note = {arXiv: 1203.1858. Available at: \url{https://arxiv.org/abs/1203.1858}},
  
  author = {Mohammad, Saif M. and Hirst, Graeme},
  
  title = {Distributional Measures of Semantic Distance: A Survey},
  
  publisher = {arXiv},
  
  year = {2012},
  
  copyright = {arXiv.org perpetual, non-exclusive license}
}

@incollection{Schubert2021,
series = {Lecture Notes in Computer Science},
issn = {0302-9743},
pages = {32--44},
publisher = {Springer International Publishing},
booktitle = {Similarity Search and Applications},
isbn = {9783030896560},
year = {2021},
title = {A Triangle Inequality for Cosine Similarity},
copyright = {Springer Nature Switzerland AG 2021},
language = {eng},
address = {Cham},
author = {Schubert, Erich},
}

@book{Phillips2021,
series = {Springer Series in the Data Sciences},
issn = {2365-5674},
publisher = {Springer International Publishing},
isbn = {3030623408},
title = {Mathematical Foundations for Data Analysis},
copyright = {Springer Nature Switzerland AG 2021},
language = {eng},
address = {Cham},
author = {Phillips, Jeff M},
year = {2021},
}

@article{Ng2001,
  title={On spectral clustering: Analysis and an algorithm},
  author={Ng, Andrew and Jordan, Michael and Weiss, Yair},
  journal={Advances in Neural Information Processing Systems},
  volume={14},
  year={2001}
}

@article{Luxburg2007,
  doi = {10.48550/ARXIV.0711.0189},
  
note = {arXiv: 0711.0189. Available at: \url{https://arxiv.org/abs/0711.0189}},
  
  author = {von Luxburg, Ulrike},
  
  title = {A Tutorial on Spectral Clustering},
  
  publisher = {arXiv},
  
  year = {2007},
  
  copyright = {Assumed arXiv.org perpetual, non-exclusive license to distribute this article for submissions made before January 2004}
}

@inproceedings{kmeans,
  title={Some methods for classification and analysis of multivariate observations},
  author={MacQueen, James and others},
  booktitle={Proceedings of the fifth Berkeley symposium on mathematical statistics and probability},
  volume={1},
  pages={281--297},
  year={1967},
  organization={Oakland, CA, USA}
}

@misc{Rentzmann2019,
  doi = {http://dx.doi.org/10.2139/ssrn.3439358 },
  author = {Rentzmann, Simon and Wuthrich, Mario V.},
  title = {Unsupervised Learning: What is a Sports Car? },
  note = {Available at SSRN: https://ssrn.com/abstract=3439358 or http://dx.doi.org/10.2139/ssrn.3439358},
  publisher = {SSRN},
  year = {2019},
}

@article{Struyf1997,
issn = {1548-7660},
journal = {Journal of Statistical Software},
pages = {1--30},
volume = {1},
publisher = {Foundation for Open Access Statistics},
number = {1},
year = {1997},
title = {Clustering in an Object-Oriented Environment},
language = {eng},
author = {Anja Struyf and Mia Hubert and Peter Rousseeuw},
}

@inbook{PAM,
publisher = {John Wiley \& Sons, Ltd},
author = {Kaufman, L. and Rousseeuw, P.J.},
isbn = {9780470316801},
title = {Partitioning Around Medoids (Program PAM)},
booktitle = {Finding Groups in Data},
chapter = {2},
pages = {68-125},
doi = {https://doi.org/10.1002/9780470316801.ch2},
eprint = {https://onlinelibrary.wiley.com/doi/pdf/10.1002/9780470316801.ch2},
year = {1990}
}

@book{Hastie2009,
series = {Springer series in statistics},
publisher = {Springer},
booktitle = {Elements of Statistical Learning},
isbn = {0387848576},
year = {2009},
title = {Elements of Statistical Learning: Data Mining, Inference, and Prediction},
language = {eng},
address = {New York},
author = {Hastie, Trevor and Tibshirani, Robert and Friedman, Jerome},
}

@article{Liu2013,
issn = {2168-2267},
journal = {IEEE Transactions on Cybernetics},
pages = {982--994},
volume = {43},
publisher = {IEEE},
number = {3},
year = {2013},
title = {Understanding and Enhancement of Internal Clustering Validation Measures},
copyright = {Copyright 2014 Elsevier B.V., All rights reserved.},
language = {eng},
address = {PISCATAWAY},
author = {Liu, Yanchi and Li, Zhongmou and Xiong, Hui and Gao, Xuedong and Wu, Junjie and Wu, Sen},
}

@article{Calinski1974,
author = {Cali{\'n}ski, T.  and Harabasz, J.},
title = {A dendrite method for cluster analysis},
journal = {Communications in Statistics},
volume = {3},
number = {1},
pages = {1-27},
year  = {1974},
publisher = {Taylor & Francis},
doi = {10.1080/03610927408827101},
eprint = { 
        https://www.tandfonline.com/doi/pdf/10.1080/03610927408827101
}
}

@article{Rousseeuw1987,
title = {Silhouettes: A graphical aid to the interpretation and validation of cluster analysis},
journal = {Journal of Computational and Applied Mathematics},
volume = {20},
pages = {53-65},
year = {1987},
issn = {0377-0427},
doi = {https://doi.org/10.1016/0377-0427(87)90125-7},
author = {Peter J. Rousseeuw}
}

@article{Davies1979,
issn = {0162-8828},
journal = {IEEE transactions on pattern analysis and machine intelligence},
pages = {224--227},
volume = {PAMI-1},
publisher = {IEEE},
number = {2},
year = {1979},
title = {A Cluster Separation Measure},
language = {eng},
address = {United States},
author = {Davies, David L and Bouldin, Donald W},
}

@article{Pargent2021,
author = {Pargent, Florian and Pfisterer, Florian and Thomas, Janek and Bischl, Bernd},
address = {Berlin/Heidelberg},
copyright = {The Author(s) 2022},
issn = {0943-4062},
journal = {Computational Statistics},
language = {eng},
number = {5},
pages = {2671-2692},
publisher = {Springer Berlin Heidelberg},
title = {Regularized target encoding outperforms traditional methods in supervised machine learning with high cardinality features},
volume = {37},
year = {2022},
}

@article{Pryseley2011,
title = {Estimating negative variance components from Gaussian and non-Gaussian data: A mixed models approach},
journal = {Computational Statistics \& Data Analysis},
volume = {55},
number = {2},
pages = {1071-1085},
year = {2011},
issn = {0167-9473},
doi = {https://doi.org/10.1016/j.csda.2010.09.002},
author = {Assam Pryseley and Clotaire Tchonlafi and Geert Verbeke and Geert Molenberghs}
}

@article{Molenberghs2011,
issn = {1471-082X},
journal = {Statistical Modelling},
pages = {389--408},
volume = {11},
publisher = {SAGE Publications},
number = {5},
year = {2011},
title = {A note on a hierarchical interpretation for negative variance components},
language = {eng},
author = {Molenberghs, Geert and Verbeke, Geert},
}

@article{Oliveira2017,
issn = {0266-4763},
journal = {Journal of Applied Statistics},
pages = {1047--1063},
volume = {44},
publisher = {Sheffield City Polytechnic},
number = {6},
year = {2017},
title = {Negative variance components for non-negative hierarchical data with correlation, over-, and/or underdispersion},
language = {eng},
author = {Oliveira, IRC and Molenberghs, Geert and Verbeke, Geert and Demetrio, CGB and Dias, CTS},
}

@article{Tutz2017,
issn = {0306-7734},
journal = {International Statistical Review},
pages = {204--227},
volume = {85},
number = {2},
year = {2017},
title = {Modelling Clustered Heterogeneity: Fixed Effects, Random Effects and Mixtures},
author = {Tutz, Gerhard and Oelker, Margret‐Ruth},
}

@article{Campo2023,
author = {Bavo D.C. Campo and Katrien Antonio},
title = {Insurance pricing with hierarchically structured data an illustration with a workers' compensation insurance portfolio},
journal = {Scandinavian Actuarial Journal},
year  = {2023},
publisher = {Taylor & Francis},
doi = {10.1080/03461238.2022.2161413},
note={Epub ahead of print, https://doi.org/10.1080/03461238.2022.2161413},
URL = {https://doi.org/10.1080/03461238.2022.2161413},
eprint = {https://doi.org/10.1080/03461238.2022.2161413}
}

@Book{ANZSIC,
author = {{Australian Bureau of Statistics and New Zealand}},
title = { Australian and New Zealand Standard Industrial Classification, (ANZSIC) 2006 },
publisher = { Australian Bureau of Statistics : Statistics New Zealand},
pages = { 507 p. ; },
year = { 2006 },
type = { Book, Online },
copyright = { Commonwealth of Australia, 2006. Crown Copyright New Zealand, 2006. },
url = { https://www.ausstats.abs.gov.au/ausstats/subscriber.nsf/0/5718D13F2E345B57CA257B9500176C8F/$File/12920_2006.pdf},
language = { English },
subjects = { Industrial surveys -- Australia -- Classification.; Industrial surveys -- New Zealand -- Classification. },
life-dates = { 2006 -  },
catalogue-url = { https://nla.gov.au/nla.cat-vn3707266 },
}

@article{Yeo2001,
author = {Yeo, Ai Cheo and Smith, Kate A. and Willis, Robert J. and Brooks, Malcolm},
title = {Clustering technique for risk classification and prediction of claim costs in the automobile insurance industry},
journal = {Intelligent Systems in Accounting, Finance and Management},
volume = {10},
number = {1},
pages = {39-50},
doi = {https://doi.org/10.1002/isaf.196},
eprint = {https://onlinelibrary.wiley.com/doi/pdf/10.1002/isaf.196},
year = {2001}
}

@incollection{Wang2008,
series = {Studies in Fuzziness and Soft Computing},
issn = {1434-9922},
pages = {113--127},
volume = {230},
publisher = {Springer Berlin Heidelberg},
booktitle = {Soft Computing Applications in Business},
isbn = {9783540790044},
year = {2008},
title = {A Clustering Analysis for Target Group Identification by Locality in Motor Insurance Industry},
copyright = {Springer-Verlag Berlin Heidelberg 2008},
language = {eng},
address = {Berlin, Heidelberg},
author = {Wang, Xiaozhe and Keogh, Eamonn},
}

@article{Zhu2021, 
  title={Clustering driving styles via image processing}, 
  volume={15}, 
  DOI={10.1017/S1748499520000317}, 
  number={2}, journal={Annals of Actuarial Science}, 
  publisher={Cambridge University Press}, 
  author={Zhu, Rui and W{\"u}thrich, Mario V.},
  year={2021}, 
  pages={276--290}
}

@article{Rosenberg2022,
author = {Marjorie Rosenberg and Fanghao Zhong},
title = {Using Clusters Based on Social Determinants to Identify the Top 5\% Utilizers of Health Care},
journal = {North American Actuarial Journal},
volume = {26},
number = {3},
pages = {456-469},
year  = {2022},
publisher = {Routledge},
doi = {10.1080/10920277.2021.2000876},
}

@article{Wutrich2017,
issn = {2190-9733},
journal = {European Actuarial Journal},
pages = {89--108},
volume = {7},
publisher = {Springer Berlin Heidelberg},
number = {1},
year = {2017},
title = {Covariate selection from telematics car driving data},
copyright = {EAJ Association 2017},
language = {eng},
address = {Berlin/Heidelberg},
author = {W{\"u}thrich, Mario V},
}

@misc{Troxler2022,
  doi = {10.48550/ARXIV.2206.02014},
note = {arXiv: 2206.02014. Available at: \url{https://arxiv.org/abs/2206.02014}},
  author = {Troxler, Andreas and Schelldorfer, J{\"u}rg},
  title = {Actuarial Applications of Natural Language Processing Using Transformers: Case Studies for Using Text Features in an Actuarial Context},
  publisher = {arXiv},
  year = {2022},  
  copyright = {Creative Commons Attribution 4.0 International}
}

@article{Zappa2021,
	journal={Variance},
	number=1,
	title={Text Mining in Insurance: From Unstructured Data to Meaning},
	volume=14,
	author={Zappa, Diego and Borrelli, Mattia and Clemente, Gian Paolo and Savelli, Nino},
	date={2021-09-21},
	year=2021,
	month=9,
	day=21,
}

@article{Ferrario2020,
author = {Ferrario, Andrea and Naegelin, Mara},
year = {2020},
month = {03},
pages = {},
title = {The Art of Natural Language Processing: Classical, Modern and Contemporary Approaches to Text Document Classification},
note = {Available at SSRN: \url{https://ssrn.com/abstract=3547887}},
doi = {10.2139/ssrn.3547887}
}

@article{Lee2020,
issn = {0515-0361},
journal = {{ASTIN Bulletin: The Journal of the IAA}},
pages = {1--24},
volume = {50},
number = {1},
year = {2020},
title = {ACTUARIAL APPLICATIONS OF WORD EMBEDDING MODELS},
language = {eng},
author = {Lee, Gee Y and Manski, Scott and Maiti, Tapabrata},
}

@article{Schomacker2021,
issn = {1099-4300},
journal = {Entropy (Basel, Switzerland)},
pages = {1422},
volume = {23},
publisher = {MDPI AG},
number = {11},
year = {2021},
title = {Language Representation Models: An Overview},
copyright = {2021 by the authors. Licensee MDPI, Basel, Switzerland. This article is an open access article distributed under the terms and conditions of the Creative Commons Attribution (CC BY) license (https://creativecommons.org/licenses/by/4.0/). Notwithstanding the ProQuest Terms and Conditions, you may use this content in accordance with the terms of the License.},
language = {eng},
address = {Basel},
author = {Schomacker, Thorben and Tropmann-Frick, Marina},
}

@article{Verma2021,
issn = {0972-0529},
journal = {Journal of Discrete Mathematical Sciences \& Cryptography},
pages = {1509--1515},
volume = {24},
publisher = {Taylor \& Francis},
number = {5},
year = {2021},
title = {Dissecting word embeddings and language models in natural language processing},
copyright = {2021 Taru Publications 2021},
language = {eng},
author = {Verma, Vivek Kumar and Pandey, Mrigank and Jain, Tarun and Tiwari, Pradeep Kumar},
}

@article{Xu2022,
title = {{BERT}-based {NLP} techniques for classification and severity modeling in basic warranty data study},
journal = {Insurance: Mathematics and Economics},
volume = {107},
pages = {57-67},
year = {2022},
issn = {0167-6687},
doi = {https://doi.org/10.1016/j.insmatheco.2022.07.013},
author = {Shuzhe Xu and Chuanlong Zhang and Don Hong}
}

@article{Breslow1993,
pages = {9-25},
publisher = {Taylor & Francis Group},
title = {Approximate Inference in Generalized Linear Mixed Models},
volume = {88},
year = {1993},
author = {Breslow, N. E. and Clayton, D. G.},
address = {Alexandria, VA},
copyright = {Copyright Taylor & Francis Group, LLC 1993},
issn = {0162-1459},
journal = {Journal of the American Statistical Association},
language = {eng},
number = {421},
}

@book{Brown2006,
  title={Applied Mixed Models in Medicine},
  author={Brown, H. and Prescott, R.},
  isbn={9780470023570},
  lccn={2005036661},
  series={Statistics in Practice},
  url={https://books.google.be/books?id=7BdjFUUwTzgC},
  year={2006},
  publisher={Wiley}
}

@inproceedings{Luong2013,
    title = "Better Word Representations with Recursive Neural Networks for Morphology",
    author = "Luong, Thang  and
      Socher, Richard  and
      Manning, Christopher",
    booktitle = "Proceedings of the Seventeenth Conference on Computational Natural Language Learning",
    month = aug,
    year = "2013",
    address = "Sofia, Bulgaria",
    publisher = "Association for Computational Linguistics",
    url = "https://aclanthology.org/W13-3512",
    pages = "104--113",
}

@misc{Simran2020,
  doi = {10.48550/ARXIV.2005.09117},
  
note = {arXiv: 2005.09117. Available at: \url{https://arxiv.org/abs/2005.09117}},
  
  author = {Arora, Simran and May, Avner and Zhang, Jian and R{\'e}, Christopher},
  
  title = {Contextual Embeddings: When Are They Worth It?\killpunct},
  
  publisher = {arXiv},
  
  year = {2020},
  
  copyright = {arXiv.org perpetual, non-exclusive license}
}

@book{Everitt2011,
  title={Cluster Analysis},
  edition = {Fifth},
  author={Everitt, B.S. and Landau, S. and Leese, M.},
  isbn={9780340761199},
  lccn={93168652},
  series={A Hodder Arnold Publication},
  url={https://books.google.be/books?id=htZzDGlCnQYC},
  year={2011},
  publisher={Wiley}
}

@article{vonLuxburg2008,
 ISSN = {00905364},
 author = {Ulrike {von Luxburg} and Mikhail Belkin and Olivier Bousquet},
 journal = {The Annals of Statistics},
 number = {2},
 pages = {555--586},
 publisher = {Institute of Mathematical Statistics},
 title = {Consistency of Spectral Clustering},
 urldate = {2022-09-15},
 volume = {36},
 year = {2008}
}

@inproceedings{vonLuxburg2004,
pages = {457-471},
publisher = {Springer Berlin Heidelberg},
series = {Lecture Notes in Computer Science},
title = {On the Convergence of Spectral Clustering on Random Samples: The Normalized Case},
volume = {3120},
year = {2004},
author = {von Luxburg, Ulrike and Bousquet, Olivier and Belkin, Mikhail},
address = {Berlin, Heidelberg},
booktitle = {LEARNING THEORY, PROCEEDINGS},
copyright = {Springer-Verlag Berlin Heidelberg 2004},
isbn = {9783540222828},
issn = {0302-9743},
language = {eng},
}

@article{Dunn1974,
author = {J. C. Dunn},
title = {Well-Separated Clusters and Optimal Fuzzy Partitions},
journal = {Journal of Cybernetics},
volume = {4},
number = {1},
pages = {95-104},
year  = {1974},
publisher = {Taylor & Francis},
doi = {10.1080/01969727408546059},
eprint = {https://doi.org/10.1080/01969727408546059},
}

@book{NACErev1,
  title={{NACE Rev. 1}: Statistical Classification of Economic Activities in the European Community},
  author={{Statistical Office of the European Communities}},
  isbn={9789282687673},
  series={Eurostat: Series E, Methods},
  year={1996},
  publisher={Office for Official Publications of the European Communities}
}

@book{NACEBel2003,
  title={{NACE-Bel}: Activiteitennomenclatuur},
  author={{FOD Economie}},
  publisher={Algemene Directie Statistiek en Economische Informatie},
  year={2004}
}

@article{Schwertman2004,
title = {A simple more general boxplot method for identifying outliers},
journal = {Computational Statistics \& Data Analysis},
volume = {47},
number = {1},
pages = {165-174},
year = {2004},
issn = {0167-9473},
doi = {https://doi.org/10.1016/j.csda.2003.10.012},
author = {Neil C Schwertman and Margaret Ann Owens and Robiah Adnan}
}

@article{Walters2010,
author = {Walters, Jaime K. and A. Christensen, Kari and K. Green, Mandy and E. Karam, Lauren and D. Kincl, Laurel},
address = {Hoboken},
issn = {0271-3586},
journal = {American Journal of Industrial Medicine},
language = {eng},
number = {10},
pages = {984-994},
publisher = {Wiley Subscription Services, Inc., A Wiley Company},
title = {Occupational injuries to oregon workers 24 years and younger: An analysis of workers' compensation claims, 2000-2007},
volume = {53},
year = {2010},
}

@article{Holizki2008,
author = {Holizki, Theresa and McDonald, Rose and Foster, Valerie and Guzmicky, Michael},
address = {Hoboken},
copyright = {Copyright © 2008 Wiley‐Liss, Inc.},
issn = {0271-3586},
journal = {American Journal of Industrial Medicine},
language = {eng},
number = {5},
pages = {357-363},
publisher = {Wiley Subscription Services, Inc., A Wiley Company},
title = {Causes of work-related injuries among young workers in {British Columbia}},
volume = {51},
year = {2008},
}

@article{Wurzel2021,
title = {Workers' compensation claim counts and rates by injury event/exposure among state-insured private employers in Ohio, 2007-2017},
journal = {Journal of Safety Research},
volume = {79},
pages = {148-167},
year = {2021},
issn = {0022-4375},
doi = {https://doi.org/10.1016/j.jsr.2021.08.015},
author = {Steven J. Wurzelbacher and Alysha R. Meyers and Michael P. Lampl and P. {Timothy Bushnell} and Stephen J. Bertke and David C. Robins and Chih-Yu Tseng and Steven J. Naber}
}

@article{Cheung2013,
author = {Cheung, Yiu-ming and Jia, Hong},
address = {OXFORD},
copyright = {2013 Elsevier Ltd},
issn = {0031-3203},
journal = {Pattern Recognition},
language = {eng},
number = {8},
pages = {2228-2238},
publisher = {Elsevier Ltd},
title = {Categorical-and-numerical-attribute data clustering based on a unified similarity metric without knowing cluster number},
volume = {46},
year = {2013},
}

@incollection{Ahmad2019,
author = {Ahmad, Amir and Ray, Santosh Kumar and Aswani Kumar, Ch},
address = {Cham},
booktitle = {Sustainable Communication Networks and Application},
copyright = {Springer Nature Switzerland AG 2020},
isbn = {9783030345143},
issn = {2367-4512},
language = {eng},
pages = {478-485},
publisher = {Springer International Publishing},
series = {Lecture Notes on Data Engineering and Communications Technologies},
title = {Clustering Mixed Datasets by Using Similarity Features},
year = {2019},
}

@article{Foss2019,
author = {Foss, Alexander H. and Markatou, Marianthi and Ray, Bonnie},
address = {Hoboken},
copyright = {2018 The Authors. International Statistical Review © 2018 International Statistical Institute},
issn = {0306-7734},
journal = {International Statistical Review},
language = {eng},
number = {1},
pages = {80-109},
publisher = {Wiley Subscription Services, Inc},
title = {Distance Metrics and Clustering Methods for Mixed-type Data},
volume = {87},
year = {2019},
}

@article{Hsu2006,
author = {Hsu, Chung-Chian},
address = {PISCATAWAY},
copyright = {Copyright 2008 Elsevier B.V., All rights reserved.},
issn = {1045-9227},
journal = {IEEE Transactions on Neural Networks},
language = {eng},
number = {2},
pages = {294-304},
publisher = {IEEE},
title = {Generalizing self-organizing map for categorical data},
volume = {17},
year = {2006},
}

@article{Carrizosa2021,
author = {Carrizosa, Emilio and Galvis Restrepo, Marcela and Romero Morales, Dolores},
address = {OXFORD},
copyright = {2021 Elsevier Ltd},
issn = {0957-4174},
journal = {Expert Systems with Applications},
language = {eng},
pages = {115245},
publisher = {Elsevier Ltd},
title = {On clustering categories of categorical predictors in generalized linear models},
volume = {182},
year = {2021},
}

@article{Carrizosa2022,
author = {Carrizosa, Emilio and Mortensen, Laust Hvas and Romero Morales, Dolores and Sillero-Denamiel, M. Remedios},
copyright = {2022 The Author(s)},
issn = {0957-4174},
journal = {Expert Systems with Applications},
language = {eng},
pages = {117423},
publisher = {Elsevier Ltd},
title = {The tree based linear regression model for hierarchical categorical variables},
volume = {203},
year = {2022},
}

@misc{GFL,
year = {2010},
archivePrefix = {arXiv},
eprint = {1011.6409},
title = {A coordinate-wise optimization algorithm for the Fused Lasso},
author = {H{\"o}fling, Holger and Binder, Harald and Schumacher, Martin},
note = {arXiv: 1011.6409. Available at: \url{https://arxiv.org/abs/1011.6409}},
}

@book{Pinheiro2009,
  title={Mixed-Effects Models in S and S-PLUS},
  author={Pinheiro, J.C. and Pinheiro, J. and Bates, D.},
  isbn={9781441903174},
  lccn={99053566},
  series={Statistics and Computing},
  url={https://books.google.be/books?id=y54QDUTmvDcC},
  year={2009},
  publisher={Springer}
}

@article{tSNE,
author = {Van Der Maaten, Laurens and Hinton, Geoffrey},
address = {BROOKLINE},
copyright = {Copyright 2008 Elsevier B.V., All rights reserved.},
issn = {1532-4435},
journal = {Journal of Machine Learning Research},
language = {eng},
pages = {2579-2625},
publisher = {Microtome Publ},
title = {Visualizing data using {t-SNE}},
volume = {9},
year = {2008},
}

@article{Ostrovsky2012,
author = {Ostrovsky, Rafail and Rabani, Yuval and Schulman, Leonard and Swamy, Chaitanya},
address = {NEW YORK},
copyright = {Copyright 2013 Elsevier B.V., All rights reserved.},
issn = {0004-5411},
journal = {Journal of the ACM},
language = {eng},
number = {6},
pages = {1-22},
publisher = {ACM},
title = {The effectiveness of lloyd-type methods for the k-means problem},
volume = {59},
year = {2012},
}

@INPROCEEDINGS{Onan2017,
  author={Onan, Aytu{\u g}},
  booktitle={2017 International Conference on Computer Science and Engineering (UBMK)}, 
  title={A K-medoids based clustering scheme with an application to document clustering}, 
  year={2017},
  volume={},
  number={},
  pages={354-359},
  doi={10.1109/UBMK.2017.8093409}
}

@article{Yu2018,
title = {An improved K-medoids algorithm based on step increasing and optimizing medoids},
journal = {Expert Systems with Applications},
volume = {92},
pages = {464-473},
year = {2018},
issn = {0957-4174},
doi = {https://doi.org/10.1016/j.eswa.2017.09.052},
author = {Donghua Yu and Guojun Liu and Maozu Guo and Xiaoyan Liu}
}

@article{Rodriguez2019,
author = {Rodriguez, Mayra Z. and Comin, Cesar H. and Casanova, Dalcimar and Bruno, Odemir M. and Amancio, Diego R. and Costa, Luciano da F. and Rodrigues, Francisco A.},
address = {SAN FRANCISCO},
copyright = {Copyright 2019 Elsevier B.V., All rights reserved.},
issn = {1932-6203},
journal = {PloS One},
language = {eng},
number = {1},
pages={e0210236},
publisher = {Public Library Science},
title = {Clustering algorithms: A comparative approach},
volume = {14},
year = {2019},
}

@incollection{Murugesan2021,
author = {Murugesan, Nivedha and Cho, Irene and Tortora, Cristina},
address = {Cham},
booktitle = {Data Analysis and Rationality in a Complex World},
copyright = {Springer Nature Switzerland AG 2021},
isbn = {9783030601034},
issn = {1431-8814},
language = {eng},
pages = {175-185},
publisher = {Springer International Publishing},
series = {Studies in Classification, Data Analysis, and Knowledge Organization},
title = {Benchmarking in Cluster Analysis: A Study on Spectral Clustering, DBSCAN, and K-Means},
volume = {5},
year = {2021},
}

@techreport{Verma2003,
  title={A comparison of spectral clustering algorithms},
  author={Verma, Deepak and Meila, Marina},
  institution={University of Washington Tech Rep UWCSE030501},
  volume={1},
  pages={1--18},
  year={2003}
}

@book{Kaufman1990,
address = {New York (N.Y.)},
author={Kaufman, L. and Rousseeuw, P.J.},
booktitle = {Finding groups in data an introduction to cluster analysis.},
isbn = {0-471-87876-6},
language = {eng},
publisher = {Wiley},
series = {Wiley-interscience publications},
title = {Finding groups in data: an introduction to cluster analysis.},
year = {1990},
}

@article{Jung2014,
  title={Clustering performance comparison using K-means and expectation maximization algorithms},
  author={Jung, Yong Gyu and Kang, Min Soo and Heo, Jun},
  journal={Biotechnology \& Biotechnological Equipment},
  volume={28},
  number={sup1},
  pages={S44--S48},
  year={2014},
  publisher={Taylor \& Francis}
}

@article{Costa2004,
  title={Comparative analysis of clustering methods for gene expression time course data},
  author={Costa, Ivan G and de Carvalho, Francisco de AT and de Souto, Marc{\'\i}lio CP},
  journal={Genetics and Molecular Biology},
  volume={27},
  pages={623--631},
  year={2004},
  publisher={SciELO Brasil}
}

@article{Kinnunen2011,
  title={Comparison of clustering methods: A case study of text-independent speaker modeling},
  author={Kinnunen, Tomi and Sidoroff, Ilja and Tuononen, Marko and Fr{\"a}nti, Pasi},
  journal={Pattern Recognition Letters},
  volume={32},
  number={13},
  pages={1604--1617},
  year={2011},
  publisher={Elsevier}
}

@article{DeSouto2008,
  title={Clustering cancer gene expression data: a comparative study},
  author={{de Souto}, Marcilio Cp and Costa, Ivan G and {de Araujo}, Daniel Sa and Ludermir, Teresa B and Schliep, Alexander},
  journal={BMC Bioinformatics},
  volume={9},
  number={1},
  pages={1--14},
  year={2008},
  publisher={BioMed Central}
}

@article{Kou2014,
  title={Evaluation of clustering algorithms for financial risk analysis using {MCDM} methods},
  author={Kou, Gang and Peng, Yi and Wang, Guoxun},
  journal={Information Sciences},
  volume={275},
  pages={1--12},
  year={2014},
  publisher={Elsevier}
}

@article{Mangiamelo1996,
  title={A comparison of {SOM} neural network and hierarchical clustering methods},
  author={Mangiameli, Paul and Chen, Shaw K and West, David},
  journal={European Journal of Operational Research},
  volume={93},
  number={2},
  pages={402--417},
  year={1996},
  publisher={Elsevier}
}

@article{Hennig2015,
author = {Hennig, Christian},
address = {AMSTERDAM},
copyright = {2015 The Authors},
issn = {0167-8655},
journal = {Pattern Recognition Letters},
language = {eng},
pages = {53-62},
publisher = {Elsevier B.V},
title = {What are the true clusters?},
volume = {64},
year = {2015},
}

@book{McNicholas2016a,
  title={Mixture Model-Based Classification},
  author={McNicholas, P.D.},
  isbn={9781482225679},
  url={https://books.google.be/books?id=txcNDgAAQBAJ},
  year={2016},
  publisher={CRC Press}
}

@article{McNicholas2016b,
author = {McNicholas, Paul D.},
address = {New York},
copyright = {The Author(s) 2016},
issn = {0176-4268},
journal = {Journal of Classification},
language = {eng},
number = {3},
pages = {331-373},
publisher = {Springer US},
title = {Model-Based Clustering},
volume = {33},
year = {2016},
}

@book{Timm2002,
address = {Pittsburgh},
isbn = {0-387-95347-7},
language = {eng},
publisher = {Springer},
author = {Timm, Neil H.},
series = {Springer texts in statistics},
title = {Applied multivariate analysis},
year = {2002},
}

@book{Wierzchon2019,
  title={Modern Algorithms of Cluster Analysis},
  author={Wierzcho{\'n}, S. and K{\l}opotek, M.},
  isbn={9783319887524},
  series={Studies in Big Data},
  url={https://books.google.be/books?id=YkaVwAEACAAJ},
  year={2019},
  publisher={Springer International Publishing}
}

@article{Halkidi2001,
author = {Halkidi, Maria and Batistakis, Yannis and Vazirgiannis, Michalis},
address = {New York},
copyright = {COPYRIGHT 2001 Springer},
issn = {0925-9902},
journal = {Journal of Intelligent Information Systems},
language = {eng},
number = {2},
pages = {107-145},
publisher = {Springer},
title = {On Clustering Validation Techniques},
volume = {17},
year = {2001},
}

@article{Govender2020,
title = {Application of k-means and hierarchical clustering techniques for analysis of air pollution: A review (1980-2019)},
journal = {Atmospheric Pollution Research},
volume = {11},
number = {1},
pages = {40-56},
year = {2020},
issn = {1309-1042},
doi = {https://doi.org/10.1016/j.apr.2019.09.009},
author = {P. Govender and V. Sivakumar}
}

@article{Vendramin2010,
author = {Vendramin, Lucas and Campello, Ricardo J. G. B. and Hruschka, Eduardo R.},
title = {Relative clustering validity criteria: A comparative overview},
journal = {Statistical Analysis and Data Mining: The ASA Data Science Journal},
volume = {3},
number = {4},
pages = {209-235},
doi = {https://doi.org/10.1002/sam.10080},
eprint = {https://onlinelibrary.wiley.com/doi/pdf/10.1002/sam.10080},
year = {2010}
}

@Manual{clusterPackage,
    title = {cluster: Cluster Analysis Basics and Extensions},
    author = {Martin Maechler and Peter Rousseeuw and Anja Struyf and Mia Hubert and Kurt Hornik},
    year = {2022},
    url = {https://CRAN.R-project.org/package=cluster},
    note = {R package version 2.1.4},
  }

@article{Franti2019,
title = {How much can k-means be improved by using better initialization and repeats?},
journal = {Pattern Recognition},
volume = {93},
pages = {95-112},
year = {2019},
issn = {0031-3203},
doi = {https://doi.org/10.1016/j.patcog.2019.04.014},
author = {Pasi Fr{\"a}nti and Sami Sieranoja}
}

@misc{BERT,
  doi = {10.48550/ARXIV.1810.04805},
note = {arXiv: 1810.04805. Available at: \url{https://arxiv.org/abs/1810.04805}},
  author = {Devlin, Jacob and Chang, Ming-Wei and Lee, Kenton and Toutanova, Kristina},
  title = {{BERT}: Pre-training of Deep Bidirectional Transformers for Language Understanding},
  publisher = {arXiv},
  year = {2018},
  copyright = {arXiv.org perpetual, non-exclusive license}
}

@misc{ECB,
  author = {{European Central Bank}},
  year = {2021},
  title = {Loans from euro area monetary financial institutions to non-financial corporations by economic activity: Explanatory notes},
  howpublished = {\url{https://www.ecb.europa.eu/stats/pdf/money/explanatory_notes_nace-en_sdw_dissemination_en.pdf?993f98fe6b628ebc6ff44b0af3d2e362}},
  note = {Accessed: 2023-02-03}
}

@incollection{Glue,
title = {8 - Characteristics of Adhesive Materials},
editor = {Sina Ebnesajjad},
booktitle = {Handbook of Adhesives and Surface Preparation},
publisher = {William Andrew Publishing},
address = {Oxford},
pages = {137-183},
year = {2011},
series = {Plastics Design Library},
isbn = {978-1-4377-4461-3},
doi = {https://doi.org/10.1016/B978-1-4377-4461-3.10008-2},
author = {Sina Ebnesajjad}
}

@misc{Guo2016,
  url = {arXiv: 1604.06737. Available at https://arxiv.org/abs/1604.06737},
  author = {Guo, Cheng and Berkhahn, Felix},
  title = {Entity Embeddings of Categorical Variables},
  publisher = {arXiv},
  year = {2016},
}

@inproceedings{Argyrou2009,
language = {eng},
pages = {19-27},
publisher = {Springer Berlin Heidelberg},
series = {Lecture Notes in Computer Science},
title = {Clustering Hierarchical Data Using Self-Organizing Map: A Graph-Theoretical Approach},
volume = {5629},
year = {2009},
author = {Argyrou, Argyris},
address = {Berlin, Heidelberg},
booktitle = {Advances in Self-Organizing Maps},
copyright = {Springer-Verlag Berlin Heidelberg 2009},
isbn = {3642023967},
issn = {0302-9743},
}

@book{Kohonen1995,
language = {eng},
publisher = {Springer},
series = {Springer series in information sciences 30},
title = {Self-organizing maps},
year = {1995},
author = {Kohonen, Teuvo},
address = {Berlin},
isbn = {3540586008},
}
			\clearpage
			\endgroup
			
			\bookmarksetup{startatroot}
			\appendix
			\section*{Appendices}
\addcontentsline{toc}{section}{Appendices}
\renewcommand{\thesection}{\Alph{section}}
\renewcommand{\thesubsection}{\Alph{section}.\arabic{subsection}}

\section{Distance and (dis)similarity metrics}\label{App:DistanceMeasures}
Using clustering algorithms, we aim to divide a set of data points $\boldsymbol{x}_{1}, \dots, \boldsymbol{x}_{J}$ into $J'$ homogeneous groups such that observations in each cluster $j'$ are more similar to each other compared to observations of other clusters $\mb{j'} \neq j'$. Consequently, most clustering algorithms rely on distance or (dis)similarity metrics between all pairwise observations. 

The most commonly used distance metric between two vectors $\bs{x}_j$ and $\bs{x}_{\mb{j}}$ is the squared Euclidean distance
\begin{equation}
	\begin{aligned}
		d_e(\bs{x}_j, \bs{x}_{\mb{j}}) = \|\bs{x}_j - \bs{x}_{\mb{j}}\|^2_2.
	\end{aligned}
\end{equation}
\noindent
Here, $\| \bs{x}_j \|_2 \coloneqq \sqrt{x_{j1}^2 + \dots + x_{jn_f}^2}$ and $n_f$ denotes the number of features considered. The squared Euclidean distance can be converted to the Gaussian similarity measure
\begin{equation}
	\begin{aligned}
		s_g(\bs{x}_j, \bs{x}_{\mb{j}}) = \exp\left( \frac{-\|\bs{x}_j - \bs{x}_{\mb{j}}\|^2_2}{\sigma^2} \right)
	\end{aligned}
\end{equation}
\noindent
which ranges from 0 (i.e. dissimilar) to 1 (i.e. identical). $\sigma$ is a scaling parameter set by the user \citep{Ng2001, Poon2012}. When $\sigma$ is small, the distance needs to be close to 0 to result in a high similarity measure. Conversely, for high $\sigma$, even large distances will result in a value close to 1. 

Ideally, vectors that lie close to each other are characterized by a low distance $d(\cdot, \cdot)$ and a high similarity measure $s(\cdot, \cdot)$. Euclidean based distance/similarity measures, however, are not appropriate to capture the similarities between embeddings \citep{Kogan2006}. Within NLP, the cosine similarity
\begin{equation}
	\begin{aligned}
		s_c(\bs{x}_j, \bs{x}_{\mb{j}}) = \frac{\bs{x}_{j}^\top \bs{x}_{\mb{j}}}{\|\bs{x}_{j}\|_2 \cdot \|\bs{x}_{\mb{j}}\|_2}
	\end{aligned}
\end{equation}
\noindent
is therefore most often used to measure the similarity between embeddings \citep{Mohammed2012,Schubert2021}. The cosine similarity ranges from -1 (opposite) to 1 (similar). In cluster analysis, however, we generally require the (dis)similarity measure to range from 0 to 1 \citep{Everitt2011, Kogan2006, Hastie2009}. In this case, we can use the angular similarity
\begin{equation}
	\begin{aligned}
		\textcolor{red}{s_a(\bs{x}_j, \bs{x}_{\mb{j}}) = 1 - \frac{\cos^{-1}(s_c(\bs{x}_j, \bs{x}_{\mb{j}}))}{\pi}}
	\end{aligned}
\end{equation}
which is restricted to [0, 1]. Hereto related is the angular distance
\begin{equation}
	\begin{aligned}
		\textcolor{red}{d_a(\bs{x}_j, \bs{x}_{\mb{j}}) = \frac{\cos^{-1}(s_c(\bs{x}_j, \bs{x}_{\mb{j}}))}{\pi}}.
	\end{aligned}
\end{equation}
The angular distance is a proper distance metric since it satisfies the triangle inequality $d(\bs{x}_j, \bs{x}_{\mb{j}}) \leq d(\bs{x}_j, \bs{x}_{z}) + d(\bs{x}_z, \bs{x}_{\mb{j}})$ for any $z$ \citep{Schubert2021, Phillips2021}. Conversely, the distance measure based on the cosine similarity does not satisfy this inequality.

\section{Clustering algorithms}\label{App:ClusteringAlgorithms}
\subsection{K-means clustering} With k-means clustering \citep{kmeans}, we group the $J$ categories into $J'$ clusters $(C_1, \dots, C_{J'})$ by minimizing
\begin{equation}\label{eq:kmeans}
	\begin{aligned}
		\argminA_{(C_1, \dots, C_{J'})} \sum_{j' = 1}^{J'} \sum_{\bs{x}_j \in C_{j'}} d(\bs{x}_j, \bs{c}_{j'}) = \argminA_{(C_1, \dots, C_{J'})} \sum_{j' = 1}^{J'} \sum_{\bs{x}_j \in C_{j'}} \| \bs{x}_j - \bs{c}_{j'}\|_2^2
	\end{aligned}
\end{equation}
\noindent
where $\bs{c}_{j'}$ denotes the cluster centre or centroid of cluster $C_{j'}$. $\bs{c}_{j'}$ is the sample mean of all $\bs{x}_j \in C_{j'}$
\begin{equation}
	\begin{aligned}
		\bs{c}_{j'} = \frac{1}{n_{j'}} \sum_{\bs{x}_j \in C_{j'}} \bs{x}_j.
	\end{aligned}
\end{equation}
\noindent
where $n_{j'}$ denotes the number of observations in cluster $C_{j'}$. Hence, with \eqref{eq:kmeans} we minimize the within-cluster sum of squares.

K-means is only suited for numeric features, is sensitive to outliers, has several local optima and the results are sensitive to the initialization \citep{Kogan2006, Everitt2011, Ostrovsky2012, Hastie2009}.

\subsection{K-medoids clustering} Contrary to k-means, k-medoids clustering \citep{PAM} uses an existing data point $\bs{x}_j$ as cluster centre. In addition, the distance measure in k-medoids clustering is not restricted to the Euclidean distance \citep{Rentzmann2019,Hastie2009}. It can be used with any distance or dissimilarity measure. With k-medoids clustering we minimize
\begin{equation}
	\begin{aligned}
		\argminA_{(C_1, \dots, C_{J'})} \sum_{j' = 1}^{J'} \sum_{\bs{x}_j \in C_{j'}} d(\bs{x}_j, \bs{c}_{j'})
	\end{aligned}
\end{equation}
where $\bs{c}_{j'}$ is the observation for which $\sum_{\bs{x}_j \in C_{j'}} d(\bs{x}_j, \bs{c}_{j'})$ is minimal \citep{Struyf1997}. This observation is most central within cluster $C_{j'}$ and is called a medoid.

K-medoids is applicable to any feature type and is less sensitive to outliers. Nonetheless, it still suffers from local optima and is sensitive to the initialization \citep{Onan2017, Yu2018}. 

\subsection{Spectral clustering} In spectral clustering, we represent the data using an undirected similarity graph $G = \ev{V, E}$, where $V = (v_1, \dots, v_j, \dots, v_J)$ stands for the set of vertices and $E$ denotes the set of edges \citep{Hastie2009, Luxburg2007, Wierzchon2019}. The weight of the edges are represented using a $J \times J$ similarity matrix $\mathcal{S}$ which contains all pairwise similarities $s(\cdot, \cdot) \geq 0$ between the observations. The diagonal entries in the $\mathcal{S}$ matrix are equal to zero. Vertices $v_j$ and $v_{\mb{j}}$ are connected if $s(\bs{x}_j, \bs{x}_{\mb{j}}) > 0$. Hereby, we reformulate clustering as a graph-partitioning problem. We want to partition the graph such that edges within a group $j'$ have high weights and edges between different groups $j' \neq \mb{j'}$ have low weights.

To represent the degree of the vertices, we set up a diagonal matrix $\mathcal{D}$ with diagonal elements $(j, j) = \sum_{\mb{j} = 1}^{J} s(\bs{x}_j, \bs{x}_{\mb{j}})$. We use $\mathcal{D}$ to transform the similarity matrix to the Laplacian matrix $\mc{L} = \mc{D} - \mc{S}$. Next, we compute the $J$ eigenvectors $(\bs{u}_1, \dots, \bs{u}_{J})$ of $\mc{L}$. To cluster the observations in $J'$ groups, we use the $J'$ eigenvectors $(\bs{u}_1, \dots, \bs{u}_{J'})$ corresponding to the smallest eigenvalues and stack these in columns to form the matrix $U \in \mathbb{R}^{J \times J'}$. In $U$, the $j^{th}$ row corresponds to the original observation $\bs{x}_j$. In the ideal case of $J'$ true clusters, $\mc{L}$ is a block diagonal matrix
\begin{equation}
	\begin{aligned}
		\mc{L} = 
		\begin{bmatrix}
			\mc{L}_1 & & &\\
			& \mc{L}_2 & &\\
			&& \ddots &&\\
			&&& \mc{L}_{J'}\\ 
		\end{bmatrix}
	\end{aligned}
\end{equation}
\noindent
when the vertices are ordered according to their cluster membership. Here, $\mc{L}_{j'}$ is the block corresponding to cluster $j'$. In this situation, $\mc{L}$ has $J'$ eigenvectors with eigenvalue zero and these eigenvectors are indicator vectors (i.e. the values are 1 for a specific cluster $j'$ and 0 for clusters $\mb{j}' \neq j'$) \citep{Hastie2009, Poon2012, Luxburg2007}. This allows us to easily identify the $J'$ groups. Consequently, in a second step, we apply k-means to $U$ and the results hereof determine the clustering solution.

We commonly refer to $\mc{L}$ as the unnormalized Laplacian. It is, however, preferred to use a normalized version of $\mc{L}$ \citep{vonLuxburg2004, vonLuxburg2008}. One way to normalize $\mc{L}$ is by applying the following transformation to $\mc{L}$
\begin{equation}
	\begin{aligned}
		\mc{D}^{-1 / 2} \mc{L} \mc{D}^{-1 / 2} = I - \mc{D}^{-1 / 2} \mc{S} \mc{D}^{-1 / 2}.
	\end{aligned}
\end{equation}
\noindent

Spectral clustering can be used with any feature type and is less sensitive to initialization issues, outliers and local optima \citep{Verma2003, Luxburg2007}. Moreover, spectral clustering is specifically designed to identify non-convex clusters (for every pair of points inside a convex cluster, the connecting straight line segment is within this cluster). Conversely, k-means, k-medoids and HCA generally do not work well with non-convex clusters \citep{Hastie2009, Luxburg2007}.
When compared to other clustering algorithms, spectral clustering often has a better overall performance \citep{Murugesan2021, Rodriguez2019}. Notwithstanding, spectral clustering is sensitive to the employed similarity metric \citep{Haberman1996, Luxburg2007, DeSouto2008}.

\subsection{Hierarchical clustering analysis} Contrary to the other clustering methods, hierarchical clustering analysis (HCA) does not start from a specification of the number of clusters. Instead, it builds a hierarchy of clusters which can be either top-down or bottom-up \citep{Hastie2009}. In the agglomerative or bottom-up approach, each observation is initially assigned to its own cluster and we recursively merge clusters into a single cluster. When merging two clusters, we select the pair for which the dissimilarity is smallest. Conversely, in the divisive or top-down approach, we start with all observations assigned to one cluster and at each step, the algorithm recursively splits one of the existing clusters into two new clusters. Here, we split the cluster that results in the largest between-cluster dissimilarity. Consequently, in both approaches we need to define how to measure the dissimilarity between two clusters. With complete-linkage, for example, the distance between two clusters $C_{j'}$ and $C_{\mb{j}'}$ is defined as the maximum distance $d(\cdot, \cdot)$ between two observations in the separate clusters
\begin{equation}
	\begin{aligned}
		d_{CL}(C_{j'}, C_{\mb{j}'}) = \max_{\substack{x_j \in C_{j'},\\ x_{\mb{j}} \in C_{\mb{j}'}}} (d(x_{j}, x_{\mb{j}})).
	\end{aligned}
\end{equation}
\noindent
Conversely, with single-linkage we define the distance between two clusters as
\begin{equation}
	\begin{aligned}
		d_{SL}(C_{j'}, C_{\mb{j}'}) = \min_{\substack{x_j \in C_{j'},\\ x_{\mb{j}} \in C_{\mb{j}'}}} (d(x_{j}, x_{\mb{j}})).
	\end{aligned}
\end{equation}
\noindent
Several other methods are available and we refer the reader to \citet{Hastie2009} for an overview. The visualization of the different steps in HCA is often referred to as a dendrogram. It plots a tree-like structure which shows how the clusters are formed at each step in the algorithm. To partition the data into $J'$ clusters, we cut the dendrogram horizontally at the height that results in $J'$ clusters.

Due to its design, HCA is less sensitive to initialization issues and local optima in comparison to k-means. In addition, we can employ HCA with any type of feature and HCA with single-linkage is more robust to outliers \citep{Everitt2011, Timm2002}. The disadvantage of HCA is that divisions or fusions of clusters are irrevocable \citep{Kaufman1990, Kogan2006}. Once a cluster has been split or merged, it cannot be undone.

\section{Internal cluster evaluation criteria}\label{App:EvaluationCriteria}
In the aforementioned clustering techniques, $J'$ can be considered a tuning parameter that needs to be carefully chosen from a range of different (integer) values. Hereto, we require a cluster validation index to select that $J'$ which results in the most optimal clustering solution. We divide the cluster validation indices into two groups, internal and external \citep{Liu2013, Everitt2011, Wierzchon2019, Halkidi2001}. Using external validation indices, we evaluate the clustering criterion with respect to the true partitioning (i.e. the actual assignment of the observations to different groups is known). Conversely, we rely on internal validation indices when we do not have the true cluster label at our disposal. Here, we evaluate the compactness and separation of a clustering solution. The compactness indicates how dense the clusters are and compact clusters are characterized by observations that are similar and close to each other. Clusters are well separated when observations of different clusters are dissimilar and far from each other. Consequently, we employ internal validation indices to choose that $J'$ which results in compact clusters that are well separated \citep{Liu2013, Everitt2011, Wierzchon2019}. 

Several internal validation indices exist and each index formalizes the compactness and separation of the clustering solution differently. An extensive overview of internal (and external) validation indices is given in \citet{Liu2013} and \citet{Wierzchon2019}. \citet{Vendramin2010} conducted an extensive comparison of the performance 40 internal validation criteria using 1080 data sets. These data sets were grouped into four categories: a) a low number of features (i.e. $\in (2, 3, 4)$); b) a high number of features (i.e. $\in (22, 23, 24)$); c) a low number of true clusters (i.e. $\in (2, 4, 6)$) and d) a high number of true clusters (i.e. $\in (12, 14, 16)$). The authors concluded that the silhouette and Cali{\'n}ski-Harabasz indices are superior compared to other validation criteria. These indices are well-known within cluster analysis \citep{Wierzchon2019, Govender2020, Vendramin2010}. Nonetheless, the results of \citet{Vendramin2010} do no necessarily generalize to our data set. We therefore include two additional, commonly used criteria: the Dunn-index and Davies-Bouldin index.

\subsection{Cali{\'n}ski-Harabasz index} The Cali{\'n}ski-Harabasz (CH) index \citep{Calinski1974,Liu2013} is defined as the ratio of the average between- to the within-sum of squares
\begin{equation}
	\begin{aligned}
		\frac{\sum_{j' = 1}^{J'} n_{j'} \|\bs{c}_{j'} - \bs{c}\|^2_2 / (J' - 1)}{\sum_{j' = 1}^{J'} \sum_{\bs{x}_j \in C_{j'}} \|\bs{x}_j - \bs{c}_{j'}\|^2_2 / (J - J')}
	\end{aligned}
\end{equation}
\noindent
where $\bs{c}$ denotes the global centre of all observations. We compute $\bs{c}$ as
\begin{equation}
	\begin{aligned}
		\bs{c} = \frac{1}{J} \sum_{j = 1}^{J} \bs{x}_j.
	\end{aligned}
\end{equation}
\noindent
The higher this index, the more compact and well-separated the clustering solution. This index is also known as the Pseudo F-statistic. The results of \citet{Vendramin2010} suggest that the CH index performs better when the number of features is high and the number of true clusters is low.

\subsection{Davies-Bouldin index} The Davis-Bouldin index is defined as \citep{Davies1979}

\begin{equation}\label{eq:DBindex}
	\begin{aligned}
		\frac{1}{J'} \sum_{j' = 1}^{J'} \max_{\mb{j}', \ \mb{j}' \neq j'} \left( \frac{\frac{1}{n_{j'}} \sum_{\bs{x}_j \in C_{j'}} d(\bs{x}_j, \bs{c}_{j'}) + \frac{1}{n_{\mb{j}'}} \sum_{\bs{x}_{\mb{j}} \in C_{\mb{j}'}} d(\bs{x}_{\mb{j}}, \bs{c}_{\mb{j}'})}{d(\bs{c}_{j'}, \bs{c}_{\mb{j}'})} \right).
	\end{aligned}
\end{equation}
\noindent
The numerator in \eqref{eq:DBindex} captures the compactness of clusters $C_{j'}$ and $C_{\mb{j}'}$ and dense clusters are characterized by low values. With the denominator, we measure the distance between the centroids of clusters $C_{j'}$ and $C_{\mb{j}'}$ and this signifies how well separated the clusters are. Hence, low ratios are indicative of dense clusters that are well separated. By taking the maximum ratio for a specific cluster $C_{\mb{j}'}$ we take the worst scenario possible. 

According to \citet{Vendramin2010}, the Davis-Bouldin index performs better for data sets with fewer features and this finding was more pronounced for data sets with a low number of true clusters.

\subsection{Dunn-index} The Dunn-index \citep{Dunn1974} is defined as the ratio of the minimum distance between the clusters to the maximum distance within clusters
\begin{equation}\label{eq:DunnIndex}
	\begin{aligned}
		\min_{1 \leq j' \leq J'} \left(\min_{\substack{1 \leq \mb{j}' \leq J'\\ j \neq \mb{j}}} \left( \frac{\min\limits_{\substack{x_j \in C_{j'}\\x_{\mb{j}} \in C_{\mb{j}'}}} d(x_{j}, x_{\mb{j}})}{\max\limits_{1 \leq \kappa \leq J'} \left\lbrace \max\limits_{\substack{x_j, x_{\mb{j}} \in C_{\kappa}}} d(x_{j}, x_{\mb{j}}) \right\rbrace } \right) \right)
	\end{aligned}
\end{equation}
\noindent
The higher this index, the better the clustering solution. Several variants exist of the Dunn-index (see, for example, \citet{Vendramin2010}) and we focus on the original formulation as given in \eqref{eq:DunnIndex}. In \citet{Vendramin2010}, the Dunn-index performed reasonably when focusing on the difference between the true and selected number of clusters. Notwithstanding, of all four indices considered in our paper, it has the lowest performance.
\clearpage
\subsection{Silhouette index} For a specific observation $\bs{x}_j \in C_{j'}$, we define the average dissimilarity of $\bs{x}_j$ to all other observations in cluster $C_{j'}$ as
\begin{equation}
	\begin{aligned}
		a(\bs{x}_j) = \frac{1}{n_{j'} - 1} \sum_{\bs{x}_{\mb{j}} \in C_{j'}, \mb{j} \neq j} d(x_{j}, x_{\mb{j}})
	\end{aligned}
\end{equation}
\noindent
and the average dissimilarity of $\bs{x}_j$ to all observations in cluster $C_{\mb{j}'}$ as
\begin{equation}
	\begin{aligned}
		e(\bs{x}_j) = \frac{1}{n_{\mb{j}'}} \sum_{\bs{x}_{\mb{j}} \in C_{\mb{j}'}} d(\bs{x}_{j}, \bs{x}_{\mb{j}}).
	\end{aligned}
\end{equation}
\noindent
We compute $e(\bs{x}_j)$ for all clusters $C_{j'} \neq C_{\mb{j}'}$ and calculate
\begin{equation}
	\begin{aligned}
		b(\bs{x}_j) = \min_{C_{j'} \neq C_{\mb{j}'}} e(\bs{x}_j).
	\end{aligned}
\end{equation}
\noindent
We call $b(\bs{x}_j)$ the neighbour of $\bs{x}_j$ as it is the closest observation of another cluster. We calculate the silhouette value $s(\bs{x}_j)$ as 
\begin{equation}
	\begin{aligned}
		s(\bs{x}_j) = \frac{b(\bs{x}_j) - a(\bs{x}_j)}{\max(a(\bs{x}_j), b(\bs{x}_j))}.
	\end{aligned}
\end{equation}
\noindent
$s(\bs{x}_j)$ indicates how well an observation $\bs{x}_j$ is clustered and from the definition, it follows that $-1 \leq s(\bs{x}_j) \leq 1$. For clusters with a single observation, we set $s(\cdot) = 0$. Values close to one indicate that the observation has been assigned to the appropriate cluster, since the smallest between dissimilarity $b(\bs{x}_j)$ is much larger than the within dissimilarity $a(\bs{x}_j)$ \citep{Rousseeuw1987}. Conversely, when $s(\bs{x}_j)$ is close to -1, $\bs{x}_j$ lies on average closer to the neighbouring cluster than to its own cluster and this suggests that this observation is not assigned to the appropriate cluster. We calculate the average silhouette width
\begin{equation}
	\begin{aligned}
		\tilde{s} = \frac{1}{J} \sum_{j = 1} s(\bs{x}_j).
	\end{aligned}
\end{equation}
\noindent
to evaluate how good the clustering solution is. Higher $\tilde{s}$'s are associated with a better clustering solution.

In the study of \citet{Vendramin2010}, the silhouette index had the most robust performance with regard to the different evaluation scenarios. The other evaluation criteria were more sensitive to the dimensionality of the data and the true number of clusters.

\clearpage

\section{Empirical distribution of the category-specific weighted average damage rates and expected claim frequencies}\label{App:EmpiricalDistrDRCF}
\begin{figure}[!h]
	\centering
	\caption{Empirical distribution of the category-specific weighted average damage rate at different levels in the hierarchy. One large value is removed to obtain a better visualization.}
	\makebox[\textwidth][c]{\includegraphics[width = \textwidth]{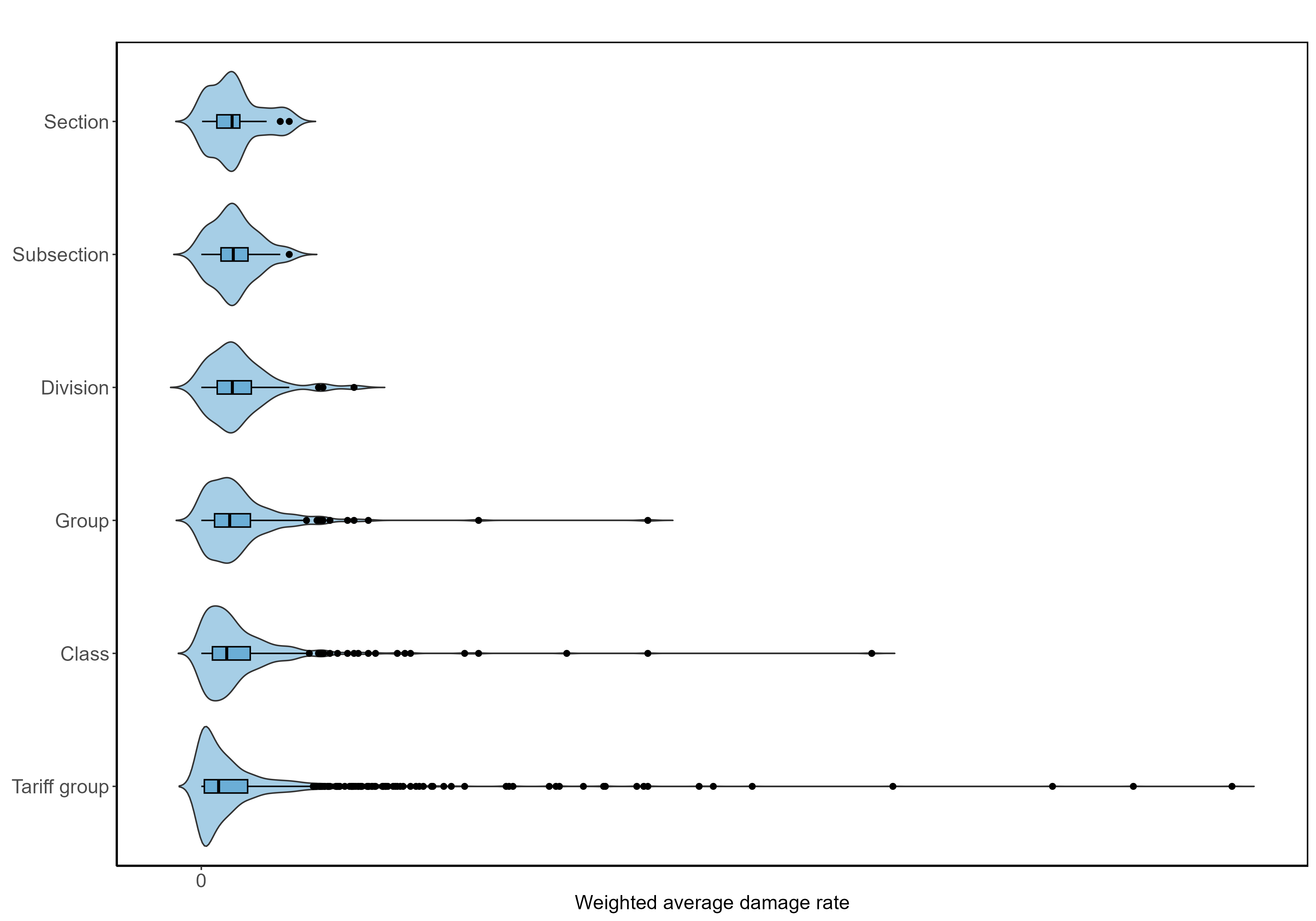}}
\end{figure}
\clearpage
\begin{figure}[H]
	\centering
	\caption{Empirical distribution of the category-specific expected claim frequency at different levels in the hierarchy. Two large values are removed to obtain a better visualization.}
	\makebox[\textwidth][c]{\includegraphics[width = \textwidth]{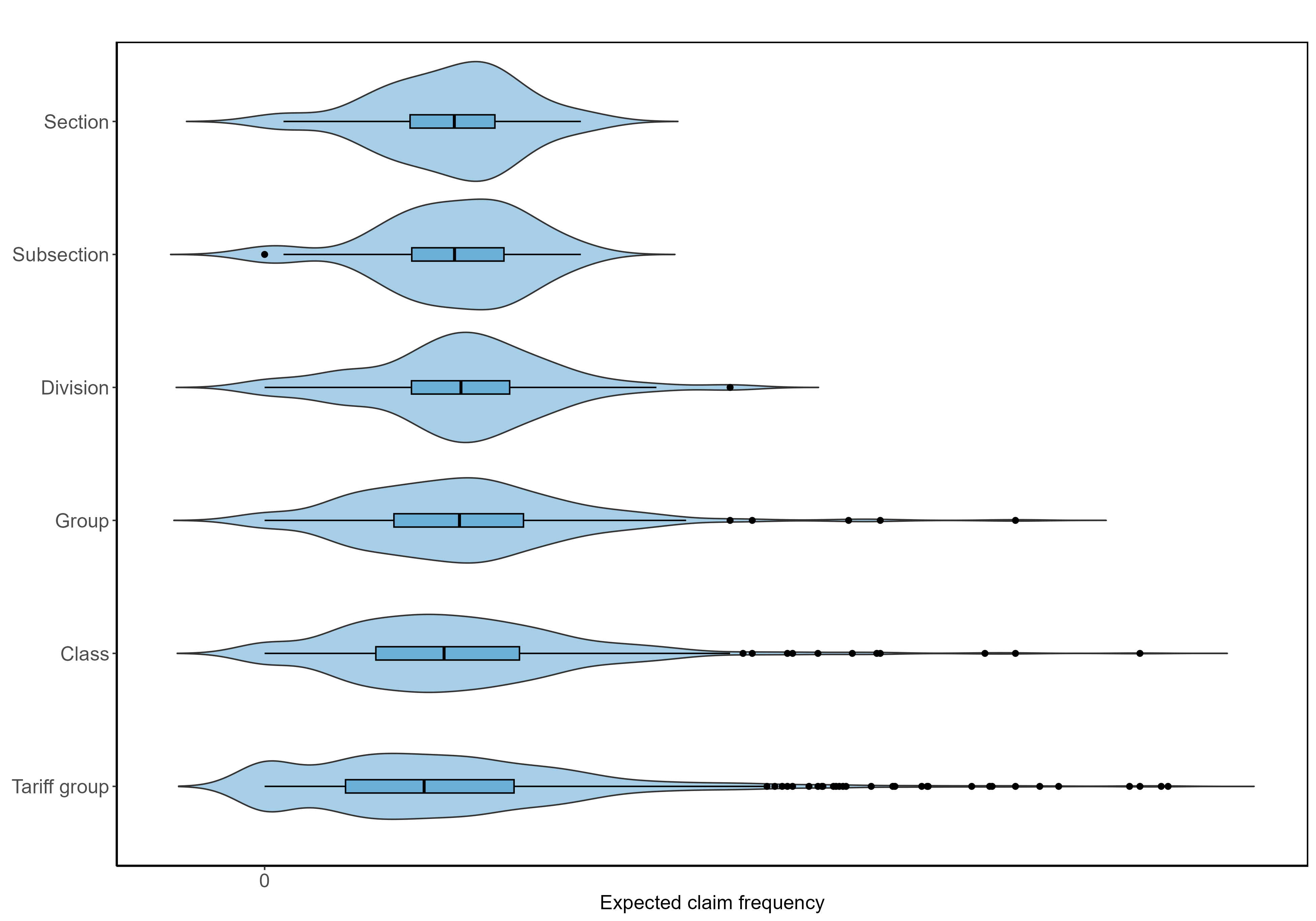}}
\end{figure}

\section{Low-dimensional representation of the embedding vectors}\label{App:tSNEFig}

\begin{figure}[H]
	\centering
	\caption{\label{fig:tSNE2}Low-dimensional visualization of all embedding vectors, resulting from the pre-trained USE v4 encoder, constructed for different categories at the \texttt{subsection} level. The text boxes display the textual labels. The blue dots connected to the boxes depict the position in the low-dimensional representation of the embeddings.}
	\makebox[\textwidth][c]{\includegraphics[width = \textwidth]{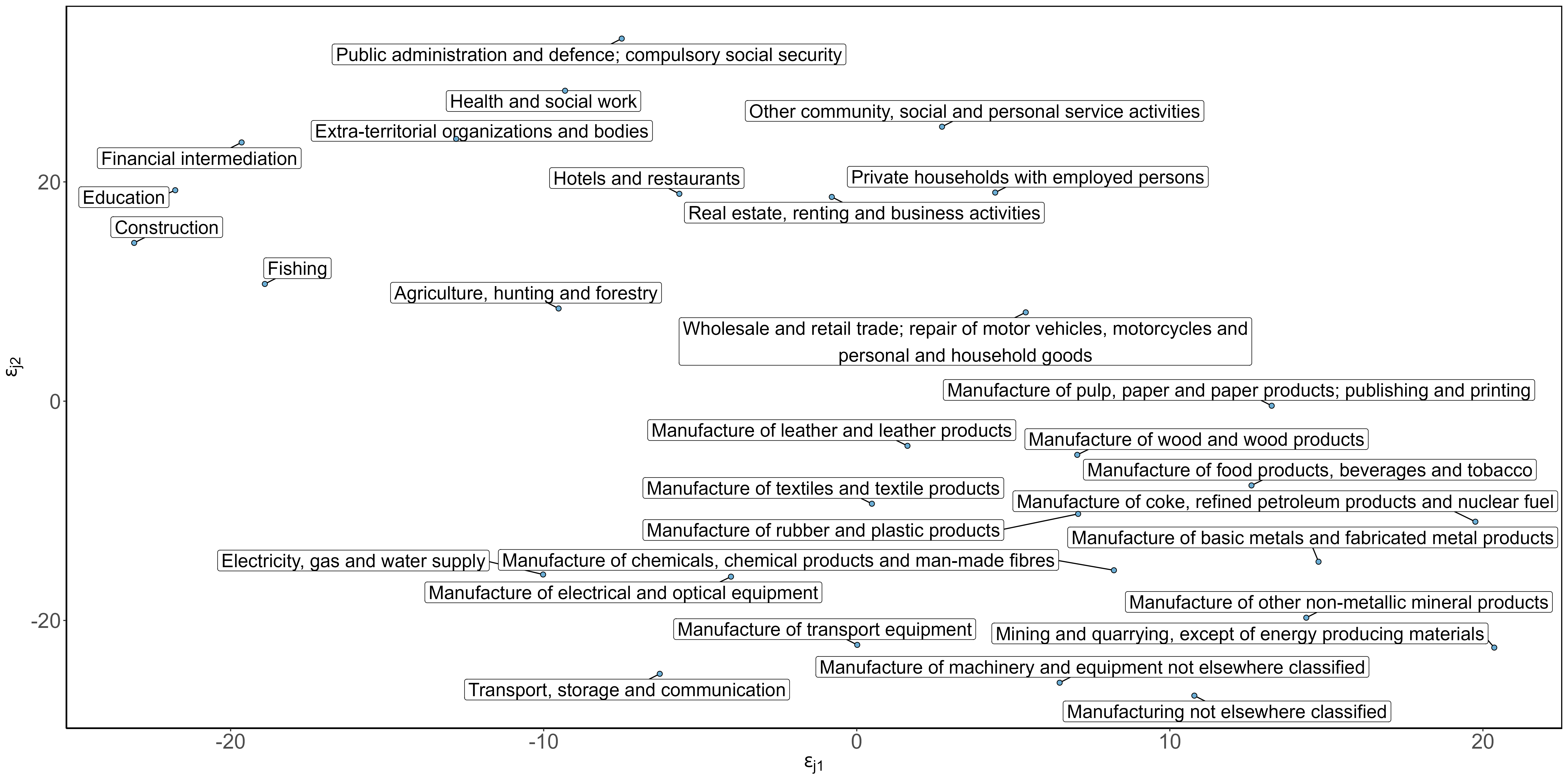}}
\end{figure}

\begin{figure}[H]
	\centering
	\caption{\label{fig:tSNE3}Low-dimensional visualization of all embedding vectors, resulting from the pre-trained USE v5 encoder, constructed for different categories at the \texttt{subsection} level. The text boxes display the textual labels. The blue dots connected to the boxes depict the position in the low-dimensional representation of the embeddings.}
	\makebox[\textwidth][c]{\includegraphics[width = \textwidth]{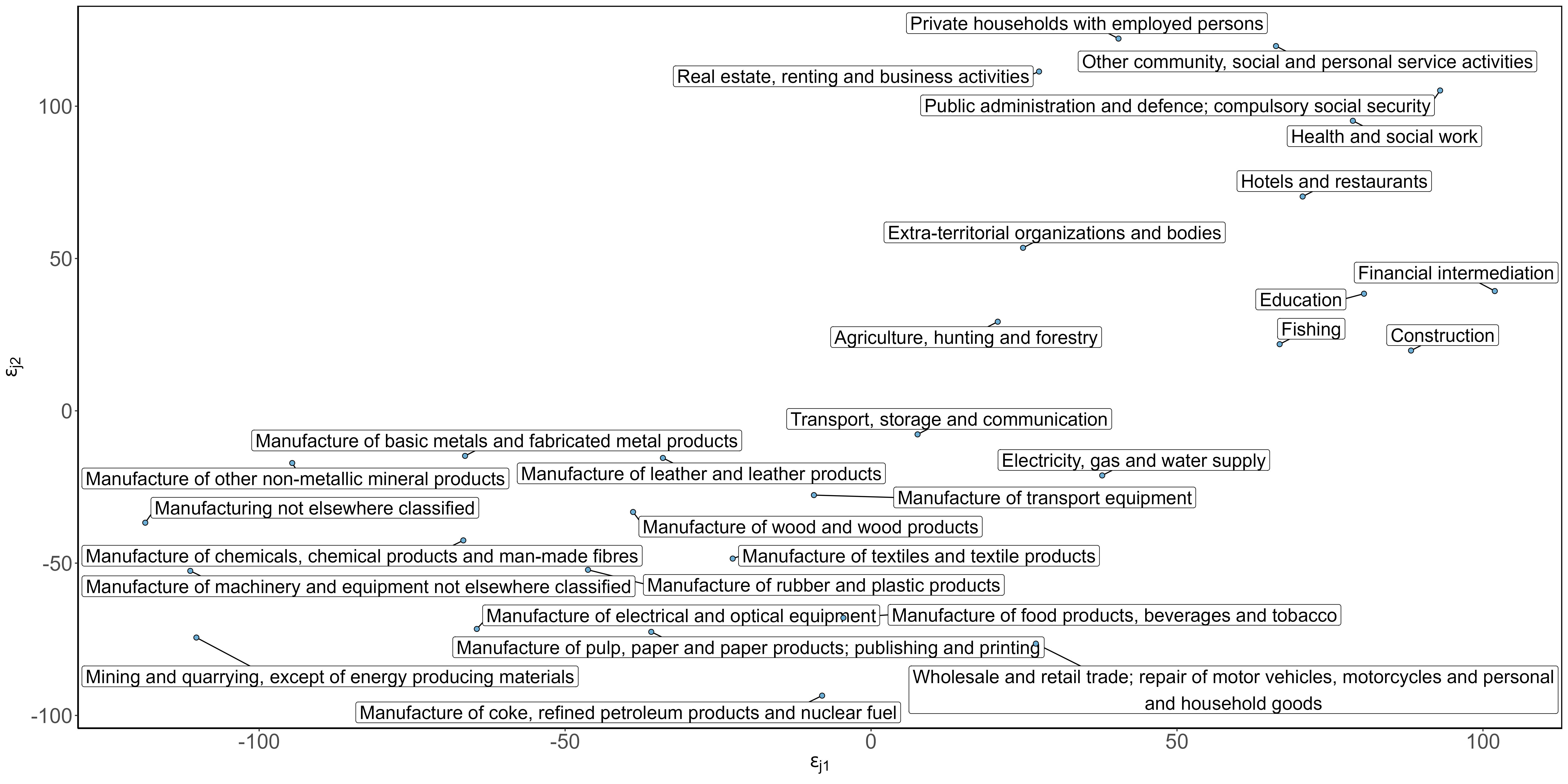}}
\end{figure}

\section{Predictive performance when using the angular distance matrix $\mc{D}$ for the cluster evaluation criteria}\label{App:PredPerfAngular}

\begin{table}[H]
	\centering
	\caption{Predictive performance on the training and test set, when the internal evaluation criterion is calculated using the angular distance and the complete feature vector.} 
	\label{tab:PredPerfAng}
	\begin{tabular}{@{\extracolsep{4pt}}lccccc@{}}
		\hline
		& & & Development & \multicolumn{2}{c}{Validation}\\
		\cline{4-4} \cline{5-6}
		& $J'$ & $\sum_{j' = 1}^{J'} K'_{j'}$ & Gini-index & Gini-index & Loss ratio\\ 
		\hline
		Benchmark &   18 &  641 & 0.658 & 0.585 & 1.006\\ 
		HCA:&&&&&\\
		\hspace{1mm} Silhouette index & 7 &  137 & 0.669 & 0.625 & 1.008   \\ 
		\hspace{1mm} Dunn index &   14 &  322 & 0.662 & 0.604 & 1.008 \\ 
		\hspace{1mm} Davies-Bouldin index &  14 &  316 & 0.670 & 0.613 & 1.007\\ 
		\hspace{1mm} CH index &  13 &  157 & 0.658 & 0.604 & 1.011\\ 
		
		k-medoids:&&&&&
		\\\hspace{1mm} Silhouette index &  6 &  135 & 0.655 & 0.593 & 1.008\\ 
		\hspace{1mm} Dunn index &  14 &  315 & 0.656 & 0.576 & 1.006\\ 
		\hspace{1mm} Davies-Bouldin index &   14 &  303 & 0.674 & 0.621 & 1.011\\ 
		\hspace{1mm} CH index &  10 &  125 & 0.658 & 0.627 & 1.010\\ 
		
		Spectral clustering:&&&&&\\
		\hspace{1mm} Silhouette index & 8 &  150 & 0.670 & 0.599 & 1.009\\ 
		\hspace{1mm} Dunn index & 14 &  299 & 0.663 & 0.585 & 1.008\\  
		\hspace{1mm} Davies-Bouldin index & 16 &  323 & 0.668 & 0.583 & 1.006 \\  
		\hspace{1mm} CH index &    16 &  224 & 0.671 & 0.619 & 1.008\\  
		\hline
	\end{tabular}
\end{table}
			
		\end{document}